\documentclass[]{emulateapj}
\usepackage{graphicx,natbib,latexsym}
\newcommand{\zi}{\v{Z}eljko Ivezi\'{c}}
\newcommand{\dao}{{\tt DAOPhot}}
\newcommand{\dop}{{\tt DoPhot}}
\newcommand{\sex}{{\tt SExtractor}}
\newcommand{\photo}{{\tt Photo}}

\newcommand{\all}{{\tt allframe}}

\begin{document} 

\title{In Pursuit of LSST Science Requirements: A Comparison of Photometry Algorithms}

\shorttitle{Photometry Comparison}
\shortauthors{Becker et~al.}

\author{
  Andrew C. Becker\altaffilmark{1},
  Nicole M. Silvestri\altaffilmark{1}, 
  Russell E. Owen\altaffilmark{1}, 
  \zi\altaffilmark{1}, 
  Robert H. Lupton\altaffilmark{2}
}

\altaffiltext{1}{Department of Astronomy, Box 351580, University of
  Washington, Seattle, WA 98195, U.S.A.; becker@astro.washington.edu, 
  nms@astro.washington.edu,
  rowen@u.washington.edu,
  ivezic@astro.washington.edu}

\altaffiltext{2}{Princeton University Observatory, Peyton Hall,
  Princeton, NJ 08544, U.S.A.; rhl@astro.princeton.edu}

\begin{abstract}

We have developed an end--to--end photometric data processing pipeline
to compare current photometric algorithms commonly used on
ground--based imaging data.  This testbed is exceedingly adaptable,
and enables us to perform many research and development tasks,
including image subtraction and co-addition, object detection and
measurements, the production of photometric catalogs, and the creation
and stocking of database tables with time--series information.  This
testing has been undertaken to evaluate existing photometry algorithms
for consideration by a next--generation image processing pipeline for
the Large Synoptic Survey Telescope (LSST).  We outline the results of
our tests for four packages: The Sloan Digital Sky Survey's (SDSS)
\photo\ package, \dao\ and {\tt allframe}, \dop, and two versions of
Source Extractor (\sex).  The ability of these algorithms to perform
point--source photometry, astrometry, shape measurements, star--galaxy
separation, and to measure objects at low signal--to--noise is
quantified.  We also perform a detailed crowded field comparison of
\dao\ and {\tt allframe}, and profile the speed and memory
requirements in detail for \sex.
We find that both \dao\ and \photo\ are able to perform aperture
photometry to high enough precision to meet LSST's science
requirements, and less adequately at PSF--fitting photometry.  \photo\
performs the best at simultaneous point and extended--source shape and
brightness measurements.  \sex\ is the fastest algorithm, and recent
upgrades in the software yield high--quality centroid and shape
measurements with little bias towards faint magnitudes.  {\tt
Allframe} yields the best photometric results in crowded fields.  

\end{abstract}


\section{Introduction}

The next generation of astronomical surveys will provide data rates
and volumes that dwarf those of current time--domain surveys
\citep[e.g.][]{2006AIPC..870...44T,2006IAUJD..13E...7K}, requiring
commensurate advances in astronomical image processing and data
management capabilities.  These surveys will enable synoptic study of
such diverse science aspects as the minor planets of the solar system
\citep{2006Icar..185..508J}, Galactic structure through
color--magnitude \citep{2005astro.ph.10520J} and proper motion
\citep{2004AJ....127.3034M} studies, time domain variability
\citep{2004ApJ...611..418B}, and the study of cosmological dark matter
and dark energy using type Ia supernovae \citep{2007ApJ...666..694W},
baryon acoustic oscillations \citep{2005ApJ...633..560E}, galaxy
clustering \citep{2004ApJ...603....1B}, and weak lensing
\citep{2006JCAP...08..008Z}.  These science goals require precision
astrometric and photometric measurements of both stars and galaxies.
The engineering challenge in these surveys is to design and
manufacture a system able to obtain data of requisite quality.  The
data management challenge is to reliably and rapidly transfer,
analyze, and store the raw data and data products, with the
algorithmic engineering challenge to realize the science goals through
precision analysis of the data.

The Science Requirements Document (SRD) for the Large Synoptic Survey
Telescope (LSST\footnote{http://www.lsst.org/}) includes constraints
on point--source photometry and astrometry, as well as on stellar and
galaxy shape measurements.  These requirements are not to be violated
in data or in software.  The goal of this research is to test the
latter, given a large set of input data.  In particular, the LSST SRD
requires that the root--mean--square (RMS) of the unresolved source
magnitude distribution around the mean value is not to exceed 0.005
magnitudes in the $g$, $r$, and $i$ passbands, when supported by
photon statistics.  The measured photometric errors shall not exceed
the quoted photometric errors by $10\%$.  The RMS of the distance
distribution for stellar pairs with separations of 5, 20, and
$200\arcmin$ shall not exceed 10, 10, and 15 milli--arcseconds in the
$g$, $r$ and $i$-bands, respectively.  Finally, for fields within 10
degrees of zenith, the $r$ and $i$-band point--source ellipticity
distribution will have a median value of no more than 0.04, and must
be correctable to a distribution with a median no larger than 0.002.

We compare here extant software packages in the context of these LSST
science requirements.  This includes 
\dao\ \citep{Stetson87},
\dop\ \citep{Schechter93}, 
{\tt allframe} \citep{Stetson94}, 
\sex\ \citep{Bertin96}, and
\photo\ \citep{2002SPIE.4836..350L}.  
We have established quality assessment metrics for comparing ensemble
measurements of stellar positions, shapes, and brightnesses.
Important algorithmic steps required to achieve this are the
separation of stars and galaxies, and the deblending of neighboring
objects.  Because the absolute ``truth'' is not known here, these
comparisons are by necessity relative.  We compare the times required
to reduce astronomical images, as well as memory consumption, when
possible.

While we have attempted to tune each package to obtain the best
results for the ensemble of data, it is very likely that better
results would emerge through individual study of each image.  As such,
this analysis reflects the results for a typical pipelined application
of each package.

We summarize the requirements for characterizing stellar and extended
sources in astronomical images in Section~\ref{sec-meas}.  We describe
the data used in the analysis in Section~\ref{sec-data}, our pipeline
infrastructure in Section~\ref{sec-pipeline}, and summarize the
algorithms we tested in Section~\ref{sec-alg}.
Our time--series database is outlined in Section~\ref{sec-db}, and the
algorithms used to ``cluster'' single detections into multiple
measurements of astronomical objects are described in
Section~\ref{sec-cluster}.  We discuss the methods used to select
objects from our database in Section~\ref{sec-method}.  We describe
the results of our analyses regarding star/galaxy separation,
photometry, shape measurements, centroiding, and photometric depth in
Sections~\ref{sec-phot}--\ref{sec-depth}.  We focus on a
crowded--field analysis of globular cluster M2 in
Section~\ref{sec-m2}, and on algorithm timing and scaling tests in
Section~\ref{sec-valgrind}.  We conclude with an overall summary in
Section~\ref{sec-summary}.

\section{Source Measurements in Astronomy}
\label{sec-meas}

The problem of point source photometry is a well--studied one, with
various solutions whose algorithms differ in their methods and
implementation \citep[e.g.][]{Howell,Thomson,Handler,Ivezic,Pin}.  The
problem requires the correct modeling of an image's point spread
function (PSF), the transfer function of point sources though the
atmosphere and the optics of the telescope.  This solution typically
includes an analytic model and an ``aperture correction'' that
compensates for the limitations of the model
\citep[e.g.][]{Tanvir,Handler,Kui}.  In practice, the aperture and PSF
fluxes are determined in a small aperture that is a small multiple of
the PSF full--width at half maximum (FWHM).  The aperture flux is an
unweighted measurement, while the PSF flux is derived using the PSF as
the weight.  The aperture fluxes of bright stars are next measured out
to a very large radius, where one is reasonably certain that all the
light has been collected.
%
%
The ratio of the bright star flux in the large and small apertures
yields a multiplicative flux correction to the small aperture
measurements.
In general, these aperture corrections need to vary across an
astronomical image because of spatial variation in the PSF.  For very
bright stars, aperture photometry yields a more accurate measurement
of the flux than PSF photometry, due to limitations of the analytic
model.  However, for faint stars near the sky limit, PSF photometry
yields a more precise measurement of the flux, since aperture
photometry includes many contributions from sky pixels.

Galaxy photometry is a much less studied issue, with a variety of
pitfalls.  Because of color changes in a galaxy's light profile, the
correct aperture to use before becoming sky--noise dominated is a
function of the passband one is observing in.  Galaxies are also
irregular in shape and may be deblended non--uniquely \citep{Kushner}.
Typically, a basic symmetric model (deVaucouleurs, exponential) is
fitted to the light profile.  For weak lensing science, which requires
precision measurement of the shapes of galaxies
\citep[e.g.][]{2002AJ....123..583B}, adaptive second moments of the
light profile are used to quantify the ellipticity of galaxies.
Photometric redshift measurements require the consistent accounting of
flux in a variety of passbands, and thus ideally requires a
simultaneous ensemble measurement of images taken through different
filters \citep{Collister}.

\section{The Data}
\label{sec-data}

One of the algorithms under study is the photometric reduction
pipeline used by the Sloan Digital Sky Survey (SDSS) : \photo.
\photo\ is one of the few packages, and the only one analyzed here,
that consistently performs both stellar PSF and extended source
photometry, and represents a solid precursor pipeline for future
surveys.  However, \photo\ has been designed to operate solely on data
from SDSS; testing of this algorithm requires that we operate on data
from SDSS.

SDSS uses a dedicated 2.5m telescope \citep{2006AJ....131.2332G} to
provide simultaneous 5--band imaging
\cite[$u,g,r,i,z$;][]{1996AJ....111.1748F}.  The imaging camera
contains 30 photometric CCDs arranged in 6 columns
\citep{1998AJ....116.3040G}.  The images are obtained in drift--scan
mode, and ``fields'' are defined corresponding to a scan length of
$9\arcmin$ (36 seconds of drift--scanning), with a field width of
$14\arcmin$.  The five images corresponding to a given field, obtained
in the order $r-i-u-z-g$, are simultaneously processed by \photo.

We have chosen to use data from two photometric runs of SDSS
equatorial Strip 82N for these comparisons.  These are runs 3437
(obtained MJD 52578) and 4207 (MJD 52936).  The data for run 3437
extend from $311\deg <$ RA $< 23\deg$ (J2000), with median $g, r$, and
$i$--band PSF FWHMs of $1.3\arcsec, 1.1\arcsec$, and $1.1\arcsec$,
respectively, and a median $r$--band sky brightness of $20.8$ mag
arcsec$^{-2}$.  The data for run 4207 extend from $305\deg <$ RA $<
60\deg$ (J2000), have a median seeing of $1.4\arcsec, 1.3\arcsec$
and $1.2\arcsec$ in the $g, r$ and $i$--band data, and median sky
brightness of $r = 20.7$ mag arcsec$^{-2}$.  There are approximately 27k
objects per square degree detected by \photo\ in these images.


Because \photo\ determines the PSF model for a given image by using
neighboring images (along the direction of the scan), the other
algorithms would be at a disadvantage when trying to measure the PSF
from a single frame.  For this reason, we ``stitch'' together 3 images
along the direction of the scan into a $14\arcmin$ by $27\arcmin$
image, with the frame of interest being in the middle.  The algorithms
operate on the entire stitched frame, but we accept only photometry
from the central section.

\section{The Analysis Pipeline}
\label{sec-pipeline}

To control the application of each algorithm to the data, we require a
form of middleware that records progress and distributes jobs.  For
this we have chosen to use the {\tt Photpipe} software developed by the
SuperMACHO and ESSENCE collaborations \citep{2002SPIE.4836..395S}.

The majority of {\tt Photpipe} is written in the {\tt Perl} language.
This provides the internal glue that strings together the various
processing steps.  In general, the image--level computations are
written in the {\tt C} language.  These applications are called by the
{\tt Perl} scripts.  

As a programmatic summary, the {\tt Photpipe} pipeline consists of a
series of {\it stages}, each of which has {\it actions} which it
undertakes, as well as {\it dependencies} on the successful completion
of previous stages.  By default, an ensemble of images is passed from
stage to stage using input and output lists.  We have added a stage
for \dao, \dop, and \sex, whose actions are merely to reduce each
image using the algorithm.  Results of the analysis are ingested into
our time--series database (Section~\ref{sec-db}).

We made an effort to explore the response of \dao, \dop, and \sex\ to
different input parameters.  However, because of the number of degrees
of freedom available to each (of order 100 for both \dop\ and \sex; of
order 10 for \dao\ and 60 for the {\tt Perl}--language scripts that
control its application) it was unfeasible to find which combination
of parameters yielded the optimal results for every analysis presented
here.  We did vary the obvious tuning parameters, such as the input
FWHM and significance threshold for object detection, degree of
variation and complexity in the PSF model, and clustering size for
matching up the ensemble of detections, ingesting the results of each
analysis into our database as a separate dataset.  In total we
ingested 112 permutations of dataset, algorithm, and algorithm input
parameters, and report here on those results that reflect our best
pipelined application of each algorithm.

\section{The Algorithms}
\label{sec-alg}

In the following sections, we briefly summarize the photometry
algorithms used in this analysis : \photo, \dao\ and {\tt allframe},
\dop, and two versions of \sex.  More complete descriptions of each
algorithm are given in the Appendix. 

The SDSS photometric pipeline \photo\ contains a complete suite of
data reduction tools that take the raw data stream, apply reduction
and calibration stages, and extract photometry from the calibrated
images.  Because the images we are using have been pre--processed by
\photo, we expect that \photo\ has a distinct advantage in the quality
of its photometric measurements.  The SDSS imaging point spread
function (PSF) is modeled heuristically in each band using a
Karhunen--Loeve (K--L) transform.  Objects are measured
self--consistently across all bands, and their positions and
brightnesses are fit using a variety of models, including PSF and
extended source models.

The \dao\ package contains a set of algorithms primarily designed to
do stellar photometry and astrometry in crowded fields.  The tools are
included as either subroutines in the executable program {\tt daophot}
or as independent executable programs.  \dao\ builds its PSF using
multiple iterations of source detection, PSF modeling, and source
subtraction.  The PSF model includes an analytic form as well as a
lookup table of corrections.  While {\tt daophot} operates on single
images, \all\ performs simultaneous measurements of all sources from a
stack of images.  \dao\ does not attempt to fully characterize
extended sources.  We designed a set of {\tt Perl}--language scripts
to automate the application of the \dao\ package.  While the scripts
have proven to be robust in the iterative building of PSFs \citep{AB},
they are also relatively slow.  A significant fraction of the
computing time spent running \dao\ is due to this implementation
choice, and not necessarily intrinsic to the \dao\ source code.

The \dop\ package is designed to robustly produce a catalog of stellar
positions, magnitudes and star/galaxy classifications for detections
from astronomical images.  \dop\ was designed to work on a large
number of images quickly with little to no interaction with the user.
However, the version of \dop\ tested here is not the original software
implementation, but instead a version that has been extensively
modified to operate robustly in the {\tt Photpipe} environment.  \dop\
uses a single PSF model that is {\it not} allowed to vary spatially,
in contrast to \photo\ and \dao, whose PSF models are allowed to vary
across the image.

\sex\ is designed to quickly produce reliable aperture photometry
catalogs on a large number of astronomical sources.  \sex\ has been
used to produce object catalogs for a variety of astronomical imaging
surveys to date such as the NOAO Deep Wide--Field Survey
\citep{NOAODeep}, GOODS--N Survey \citep{GOODS}, Deep Lens Survey
\citep{DLS}, IRAC Shallow Survey \citep{IRAC}, and the MAST Survey
\citep{MAST}.  Aside from the ease of installation, \sex\ is also
notable for its speed and versatility.  It is one of the few packages
that aspires to distinguish and photometer both stars and galaxies,
although its lack of a PSF model limits the accuracy of faint
point--source photometry.  Newer versions of the software include
adaptive windowing functions to provide more accurate centroids and
shapes than the default (isophotal) measurements.

\section{The Database}
\label{sec-db}

To enable the following analysis, we installed a MYSQL client and
server on our local computers and constructed a database to store our
test results (both science and performance benchmarking).  

We developed a variety of {\tt Python}--language scripts to help
properly ingest data (pipeline versions, parameter files, file
locations, etc.)  into the database in an organized manner.  We
ingested metadata on over 1000 SDSS images processed through \photo\
in five colors ($ugriz$) resulting in over 10 million detections in
our {\tt Objects} table.  The main tables of our database are {\tt
Image}, {\tt Object} and {\tt AlgRun}.

\begin{itemize}

\item {\bf \tt Image}: Metadata about images including data source
(e.g. SDSS), date, exposure time, filter and a pointer to World
Coordinate System (WCS) information for the image.

\item {\bf \tt Object}: Data for sources (detections from an image)
and objects (clusters of sources), including position (x,y and
RA/Dec), classification and various measures of intensity. In
addition, sources are linked to the image on which they were detected.

\item {\bf \tt AlgRun}: Information about a particular run of a
component, including the input parameters used for that run.  All
told, 112 instances of pipeline runs were ingested into the database,
representing different combinations of input data, photometry
algorithm, and input parameters.  Both the {\tt Object} and {\tt
Image} tables link to the {\tt AlgRun} table.

\end{itemize}

\section{Clustering of Sources into Objects}
\label{sec-cluster}

After ingest of sources and images into the database, we require a
method to associate sources into objects.  This allows us to collate
the $ugriz$ data for a single astronomical object, as well as to match
up the reductions from different algorithms or from different nights.
We use the {\tt OPTICS} algorithm to do this clustering.

The {\tt OPTICS} algorithm \cite[Ordering Points To Identify the
Clustering Structure;][]{optics} is a density--based method to
identify clusters of points in databases.  In this ordering, a {\it
reachability distance} is defined between neighboring points.  When
this distance is exceeded for neighboring points, the boundary of a
cluster is defined.  {\tt OPTICS} is an improvement of the {\tt
DBSCAN} algorithm \citep{dbscan}.

The user provides a minimum number of points to define the cluster
core.  In our case, for a given object we have 4 algorithms operating
on 5 filters and 2 nights of data, meaning we ideally expect 40 points
in a cluster.  We run {\tt OPTICS} requiring a minimum of 5 points to
include objects missed in some filters due to their color, missed on
some nights due to different image depths, or missed in different
algorithms due to the vagaries of the software.  Since we only have 3
algorithms besides \photo\ running on these data, an artifact in one
image and in one filter should not lead to a spurious cluster.  We do
however find spurious clusters in the wings of bright stars, where
multiple algorithms may detect signal in multiple passbands on
multiple nights.

The user also defines reachability distance $\epsilon$ for a given
core set of points.  For all points in this neighborhood, all points
within $\epsilon$ of it are searched, repeating until no more points
can be added to the cluster.  The data are stored in a tree--based
spatial index.  A search in the neighborhood $\epsilon$ of a given
object scales with the number of points {\it N} as {\it N log(N)}.  We
chose a clustering distance of 1 pixel ($0.4\arcsec$).

\begin{figure*}[htbp]
  \begin{center}
    \leavevmode
    \includegraphics[width=3.5in]{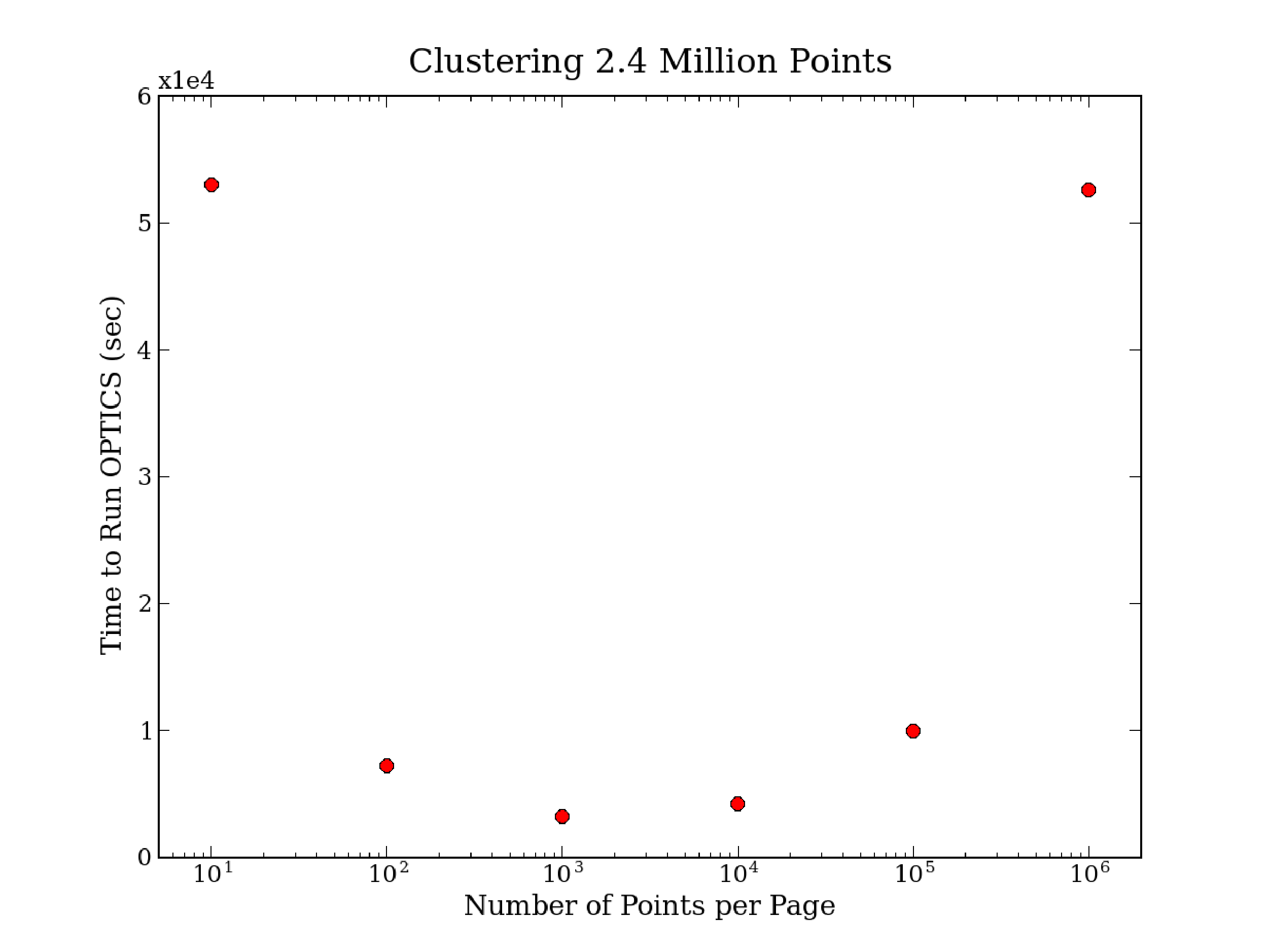}
  \end{center}
  \caption[Run Time for {\tt OPTICS} as a Function of Leaf Size]{Run
  time for clustering 2.4 million points as a function of leaf size in
  the internal lean--tree database used by {\tt OPTICS}.  Note the
  y-axis in units of $10^4$ seconds.}
  \label{fig-optics}
\end{figure*}

One way we found to optimize the clustering was to relate the size of
each page in the database to the length of the input list to be
clustered.  We found that too large (or too small) a page size would
impact the computation of the clustering by an order of magnitude.
Figure~\ref{fig-optics} demonstrates the {\tt OPTICS} run time as a
function of the number of points per page (or ``leaf'') in the
database.

\section{Methodology}
\label{sec-method}

In this section and those below, we describe the practical methods
used to quantify \dao, \dop, \photo, and \sex.

Our analyses are designed to ascertain the level of systematics
inherent to each photometry algorithm by comparing the measured
properties of objects on multiple nights.  We also compare brightness,
shape, and centroiding measurements by the different algorithms on the
same imaging data.  We start with the assumption that \photo's
star--galaxy classification is ``truth'', and use this information to
derive similar classification boundaries for the other algorithms.  We
then repeat our analyses using these new algorithm--derived
boundaries.

Our initial queries to the {\tt Object} table select {\it all} objects
from the comparison algorithms, but only a subset of detections from
\photo.  We only include \photo\ detections where the {\tt
objc\_flags}\footnote{http://www.sdss.org/dr5/products/catalogs/flags.html}
suggest that it is not {\tt SATURATED}, {\tt BLENDED}, or {\tt
BRIGHT}, was found in the {\tt BINNED1} image, and was not {\tt
DEBLENDED\_AS\_MOVING}.  These objects essentially serve as the
``seed'' objects that we use for clustering.  

We start this process by selecting only clusters where \photo\ has
detections in both runs that it thinks are stars.  This criterion is
used to select measurements from other algorithms to be used for
magnitude zero--pointing, determination of star--selection criteria,
and comparison of shape measurements and photometric depth.  We use
PSF magnitudes when available, and aperture magnitudes
otherwise\footnote{Aperture photometry is performed at a radius of
$7.4\arcsec$}.

\dao, \dop, and \sex\ report their results in instrumental magnitudes,
and we have to derive zero--point offsets if we want to directly
compare their data to \photo.  For each algorithm, filter, and run
combination, we take all \photo--selected stars and find the 3--sigma
clipped average difference in magnitudes between \photo\ and the
algorithm (we use aperture magnitudes for \sex; PSF magnitudes for
\dao\ and \dop).


\section{Star/Galaxy Separation}
\label{sec_ana-stargal}

\begin{figure*}[htbp]
  \begin{center}
    \leavevmode
    \includegraphics[width=3.5in]{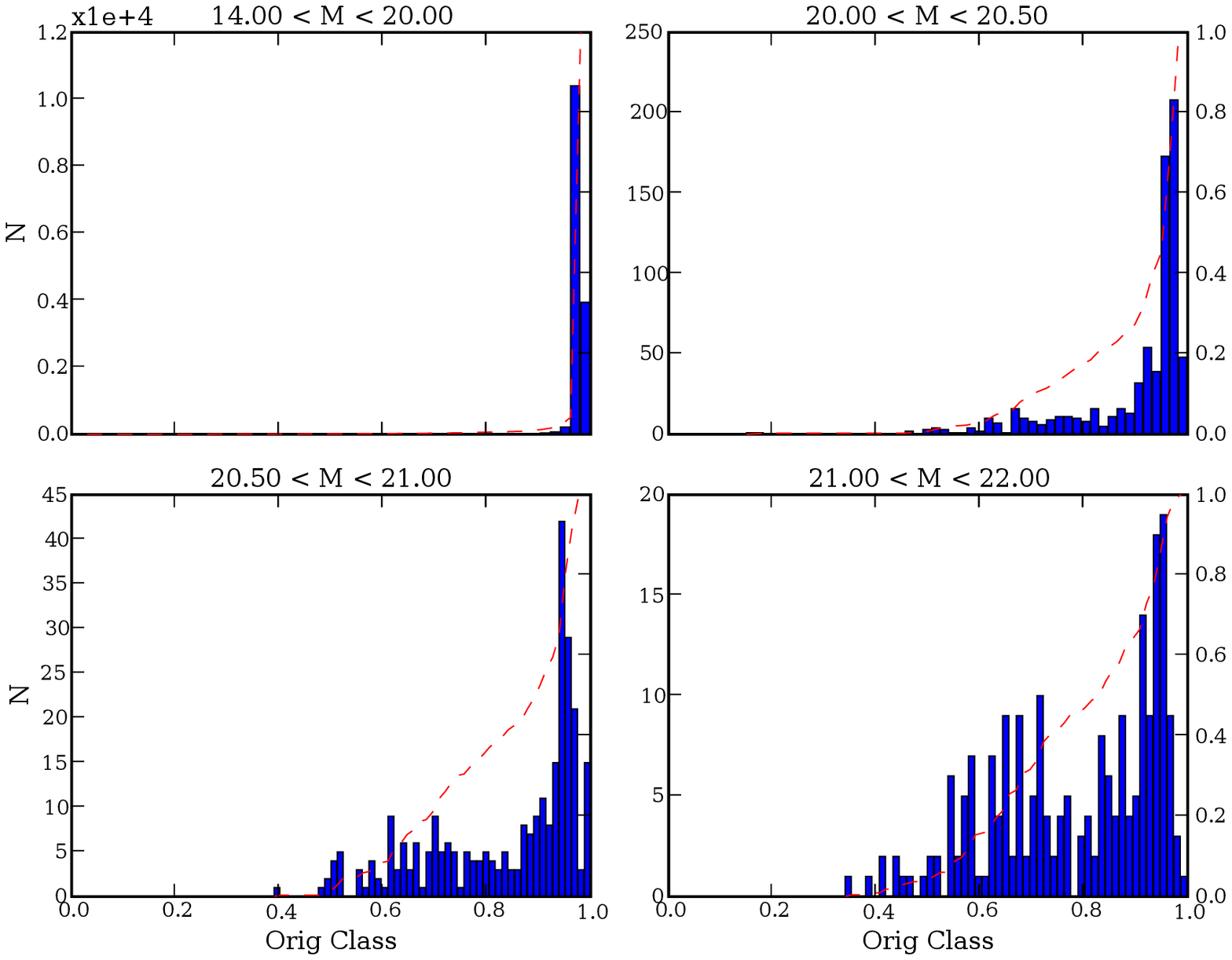}
    \includegraphics[width=3.5in]{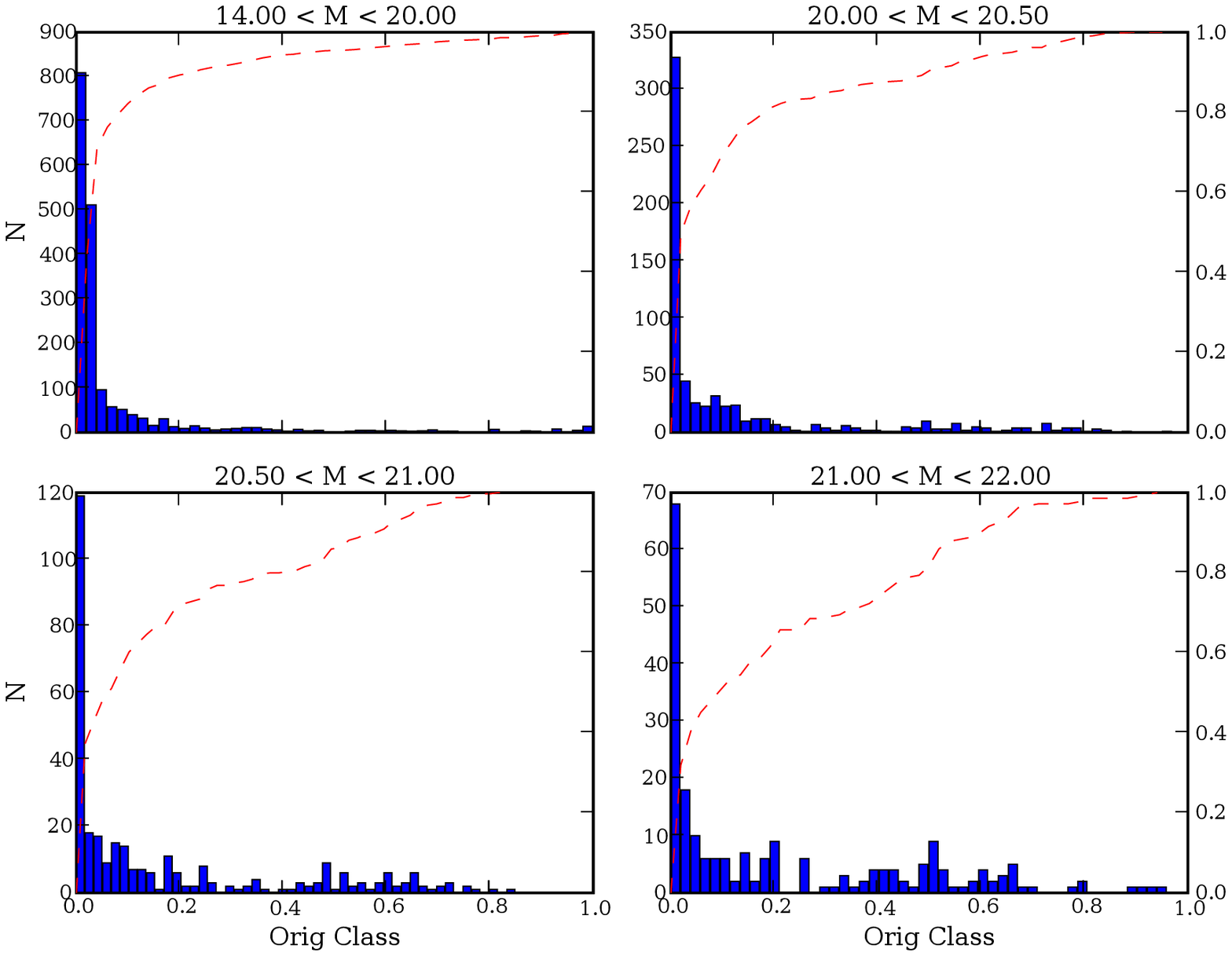}
  \end{center}
  \caption[Example {\tt CLASS\_STAR} for \sex]{Distribution of the
  {\tt CLASS\_STAR} parameter for \sex\ reductions of $r$--band data
  from run 4207.  The {\it top} panel shows objects that \photo\
  classifies as stars, and the {\it bottom} panel objects that \photo\
  classifies as galaxies.  The data are split by magnitude into 4
  bins.  The dashed line shows the cumulative fraction.  Note the
  highly skewed distributions.}
  \label{fig-class_sex}
\end{figure*}

The initial step in this analysis is to define star/galaxy boundaries
for each algorithm.  To do this, we select all objects that \photo\
classifies as stars and galaxies, and plot the distribution of the
star/galaxy separation metrics from each algorithm.  In particular, we
have chosen to use {\tt Sharp} for \dao, {\tt Type} for \dop, and {\tt
CLASS\_STAR} for \sex.  By studying the distribution of these
parameters, we can derive star/galaxy classification schemes for each
algorithm.  For all \photo--selected stars and galaxies, we plot each
algorithm's star/galaxy parameter in 4 magnitude bins : $14 < r < 20;
20 < r < 20.5; 20.5 < r < 21; 21 < r < 22$.
Each window contains a histogram and the cumulative distribution of
that parameter plotted as a dashed line.  We show example results for
\dao\ in Figure~\ref{fig-class_dao}, and \sex\ in
Figure~\ref{fig-class_sex}.  

\subsection{Results Using \photo's Classification}
\begin{figure*}[htbp]
  \begin{center}
    \leavevmode
    \includegraphics[width=3.5in]{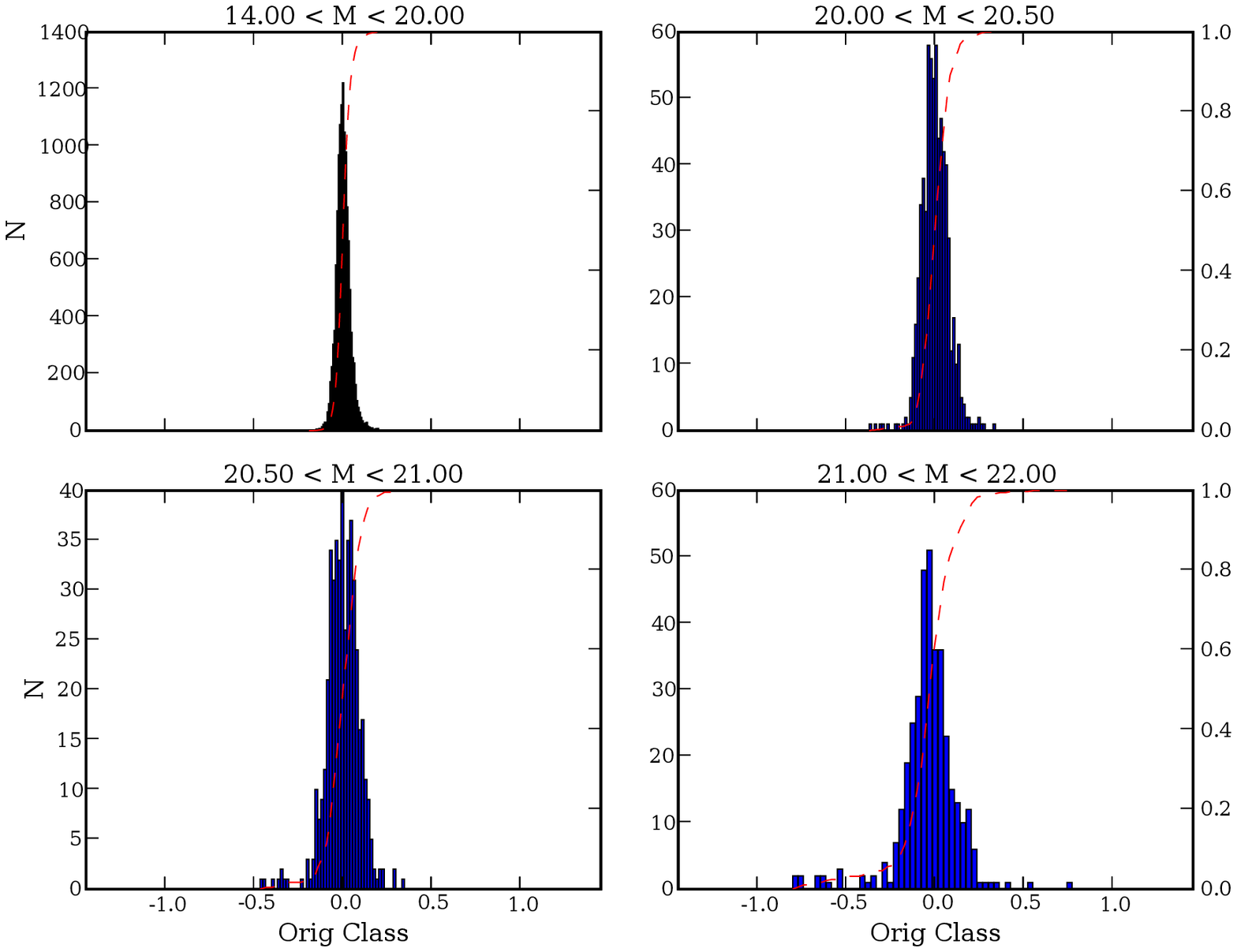}
    \includegraphics[width=3.5in]{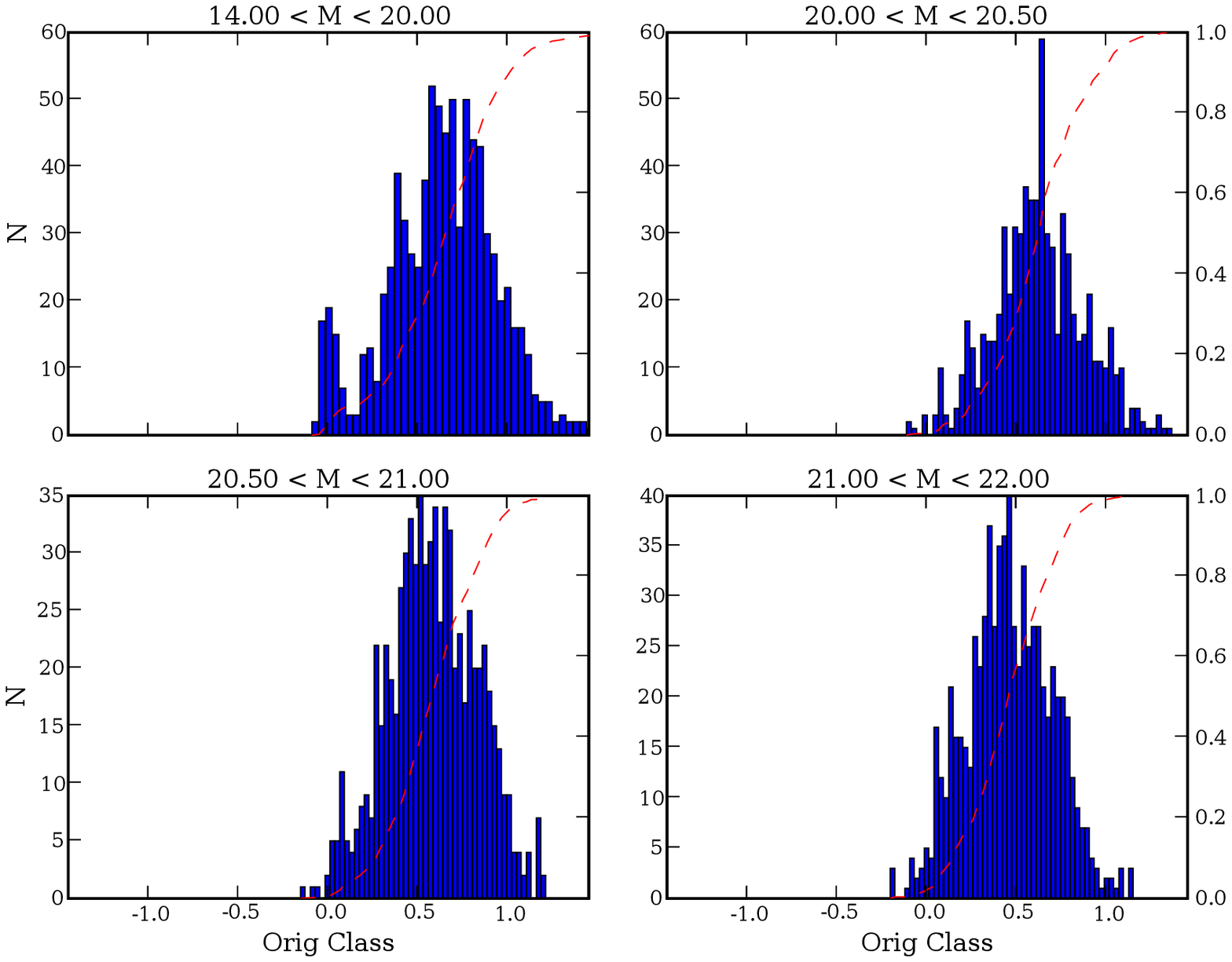}
  \end{center}
  \caption[Example {\tt Sharp} for \dao]{Distribution of the {\tt
  Sharp} parameter for \dao\ reductions of $r$--band data from run
  4207.  The {\it left} figure shows objects that \photo\ classifies
  as stars, and the {\it right} figure objects that \photo\ classifies
  as galaxies.  The data are split by magnitude into 4 bins.  The
  dashed line shows the cumulative fraction.  Note the distribution is
  symmetric around value 0.0 for stars and biased towards values
  greater than 0.0 for galaxies.}
  \label{fig-class_dao}
\end{figure*}

In \dao, {\tt Sharp} for stars is distributed in a near Gaussian that
is centered on value 0.0 with a characteristic width.
Figure~\ref{fig-class_dao} shows the $r$-band distribution from run
4207.  The data are split into 4 magnitude bins.  The distribution for
stars are plotted in the left figure; for galaxies on the right.  As
expected, the width of the stellar {\tt Sharp} distribution widens as
you go to fainter objects, from 0.04 at the bright end to 0.17 at the
faint end.  The parameter distribution for galaxies remains relatively
constant with magnitude.  We have combined the analyses from runs 3437
and 4207, and calculated the width of the stellar distribution in the
brightest bin.  The mean and width of this distribution is listed in
Table~\ref{tab-daosharp}.  We define our filter--dependent \dao\
star--selection criterion as anything having {\tt Sharp} within
$3\sigma$ of the mean in the brightest bin.  We define galaxies as
those objects with {\tt Sharp} larger than $+3\sigma$ from the mean.
Anything with {\tt Sharp} less than $-3\sigma$ from the mean is
sharper than the PSF and likely to be an image artifact.  We note that
other selection criteria are possible and may lead to better results,
such as using parameters {\tt Sharp} and {\tt Chi} in combination.
However, {\tt Sharp}'s highly symmetric distribution for stars and
highly skewed distribution for galaxies in Figure~\ref{fig-class_dao}
suggests that it is appropriate, although not necessarily optimal, to
use it as the sole criterion.  The same is true for the other metrics
defined below.

\dop\ returns a {\tt Type} parameter for each object it measures.  A
{\tt Type = 1} object is considered a ``perfect'' star, and is used in
the computation of the weighted PSF.  A {\tt Type = 3} object is not
as peaked as a single star, and is assumed to be a blend.  It is
however photometered with a single PSF.  A {\tt Type = 7} object is
too faint to do a full 7--parameter fit, so a 4--parameter fit was
undertaken.  We found that stars in our data had almost exclusively
{\tt Type = 1}, with very few having {\tt Type = 7}.  We found that
galaxies tended to have {\tt Type = 3} or {\tt Type = 1}, with a small
fraction of {\tt Type = 7}.  Since this is our only selection
criterion, we select stars as all objects with {\tt Type} = 1 and
galaxies as all objects with {\tt Type} = 3, recognizing that our
stars will have non-zero contamination by galaxies.

In \sex, {\tt CLASS\_STAR} is designed to be a star/galaxy
classification toggle, where a value of 1 represents an object highly
likely to be a star.  This requires that the correct input FWHM be
applied for the filtering to work optimally.  Therefore we use the
FWHM as derived by \photo\ as inputs to \sex.  As
Figure~\ref{fig-class_sex} shows, this parameter tends to work well.
The top panel shows the distribution for stars, and the bottom for
galaxies.  For all filters except for $u$-band, we chose a cutoff of
{\tt CLASS\_STAR} = 0.8 as the line separating stars from galaxies.
In the $u$-band, many of the stars are also distributed near {\tt
CLASS\_STAR} = 0, and we lowered our delineation to {\tt CLASS\_STAR}
= 0.2.

The extent of galaxy contamination in these algorithms is summarized
in Table~\ref{tab-sg1} and Table~\ref{tab-sg2}.  We list in
Table~\ref{tab-sg1} the total fraction of objects that were classified
as stars by both the algorithm and \photo\ (S--S); as stars in the
algorithm and galaxies in \photo\ (S--G); as galaxies in the algorithm
and stars in \photo\ (G--S); and galaxies in both algorithms (G--G).
We make a similar comparison in Table~\ref{tab-sg2}, which lists the
fraction of all objects that each algorithm (mis)classified in both
runs.  We limit this selection to objects brighter than 21$^{st}$
magnitude, where \photo's star--galaxy separation has been tested
extensively and is considered ``truth'' for the purposes of these
comparisons.

From Table~\ref{tab-sg1}, we see that \dop\ and \photo\ disagree on
anywhere from $1$ to $10\%$ of all bright objects (increasing to $\sim
20\%$ when looking at all brightnesses).  In general, \dop\ is more
likely to classify something as a star that \photo\ thinks is a
galaxy.  The fraction of detected \photo-classified galaxies is also
lowest in \dop, suggesting that this algorithm is very inefficient at
detecting galaxies, and biased towards classifying galaxies it does
find as stars.  \sex\ tends to disagree with \photo\ in the opposite
sense -- \sex\ is likely to call something a galaxy that \photo\
classifies as a star.  Run 3437 is particularly egregious in this
regard.  The most obvious cause is that we fed the wrong initial
estimate of the stellar FWHM (derived from the \photo\ analysis) to
the package, and it was therefore making poorly informed choices for
star/galaxy separation.  However, runs 3437 and 4207 were treated
equally in this regard, so this is likely not the culprit.

\begin{figure*}[htbp]
  \begin{center}
    \leavevmode
    \includegraphics[width=3.5in]{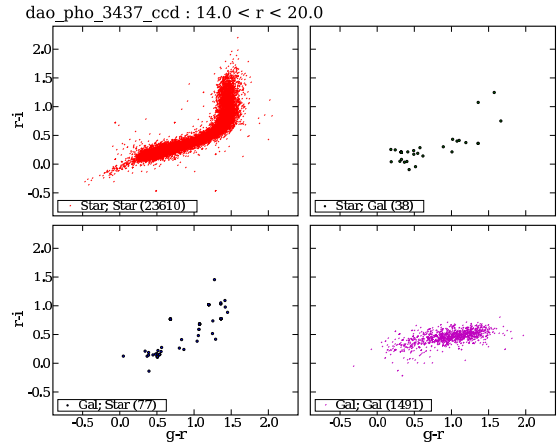}
  \end{center}
  \caption[Color--Color Diagram; \dao\ vs. \photo]{These panels show
  $g-r$, $r-i$ diagrams (derived from the \photo\ magnitudes) for
  objects with $14 < r < 20$.  These are the subset of objects that
  had detections in $g$, $r$, and $i$ in \dao\ and \photo\ from run
  3437.  In the upper left is the set of objects that both \dao\ and
  \photo\ called stars; in the upper right, \dao\ classified as a star
  and \photo\ classified as a galaxy; in the lower left, \dao\
  classified as a galaxy and \photo\ classified as a star; in the
  lower right, both algorithms classified as galaxies.}
  \label{fig-ccd_dao}
\end{figure*}

\dao\ agrees with \photo\ a large fraction of the time, and is
slightly more likely to call a \photo-classified star a galaxy than a
\photo-classified galaxy a star.  We have created plots such as
Figure~\ref{fig-ccd_dao} to investigate each permutation of
(mis)classifications.  These depict color--color diagrams of objects
classified in $g$, $r$, and $i$ as either stars or galaxies.  We plot
here only the bright objects ($14 < r < 20$) classified by both \dao\
and \photo\ in run 3437 (the figure for run 4207 is very similar).  To
yield a point on this diagram, the object must be classified the same
by each algorithm in all 3 passbands.  Thus the fraction of objects in
each window will slightly disagree with the entries in
Table~\ref{tab-sg1}.  Its clear that the misclassifications (the
off--diagonal plots) are drawn more from the stellar than the galactic
locus, thus we conclude that \dao\ correctly calls some objects stars
that \photo\ incorrectly calls galaxies, and vice versa.

\subsection{Results Using Each Algorithm's Classification}
We also investigate the consistency {\it within} a given algorithm by
looking at the classifications of the same object detected in both
runs.  This is listed in Table~\ref{tab-sg2}.  As discussed above,
\dop\ is biased towards calling objects stars, but shows here that it
is very self consistent in that regard.  \sex\ classifies a higher
fraction of objects as galaxies than do the other algorithms, and
apparently had difficulty with objects classified as stars in 4207 and
galaxies in 3437.  \dao\ disagrees with itself for $12\%$ of objects,
while \photo\ is the most consistent ($\sim 2\%$) with regards to
misclassifications of these bright objects.
We note that if we examine the {\it entire} sample of clustered
objects, including objects fainter than 21$^{st}$ magnitude, the
misclassification rates in Table~\ref{tab-sg2} degrade worst for
\photo, increasing from $\sim 2\%$ to $\sim 12\%$.  The ratios for the
other algorithms tend to remain constant at fainter magnitudes.

\subsection{Classification Conclusions}
Both \dop\ and \sex\ have inadequacies in their star/galaxy
classification schemes as derived in this experiment.  It is very
likely that improvements can be made to \sex\ using the non--linear
filters from {\tt Enhance Your Extraction
(EyE)}~\footnote{http$://$terapix$.$iap$.$fr$/$soft$/$eye}, and it
should be carefully considered as an option with the potential to
contribute to LSST algorithm development.  Surprisingly, \dao\ does a
better job at classification than these algorithms, although its
galaxy {\it characterization} methods are limited.  \photo\ is the
best all--around package in this regard due to its extensive analysis
and characterization of each object.


\section{Photometry}
\label{sec-phot}

\begin{figure*}[htbp]
  \begin{center}
    \leavevmode
    \includegraphics[width=3.5in]{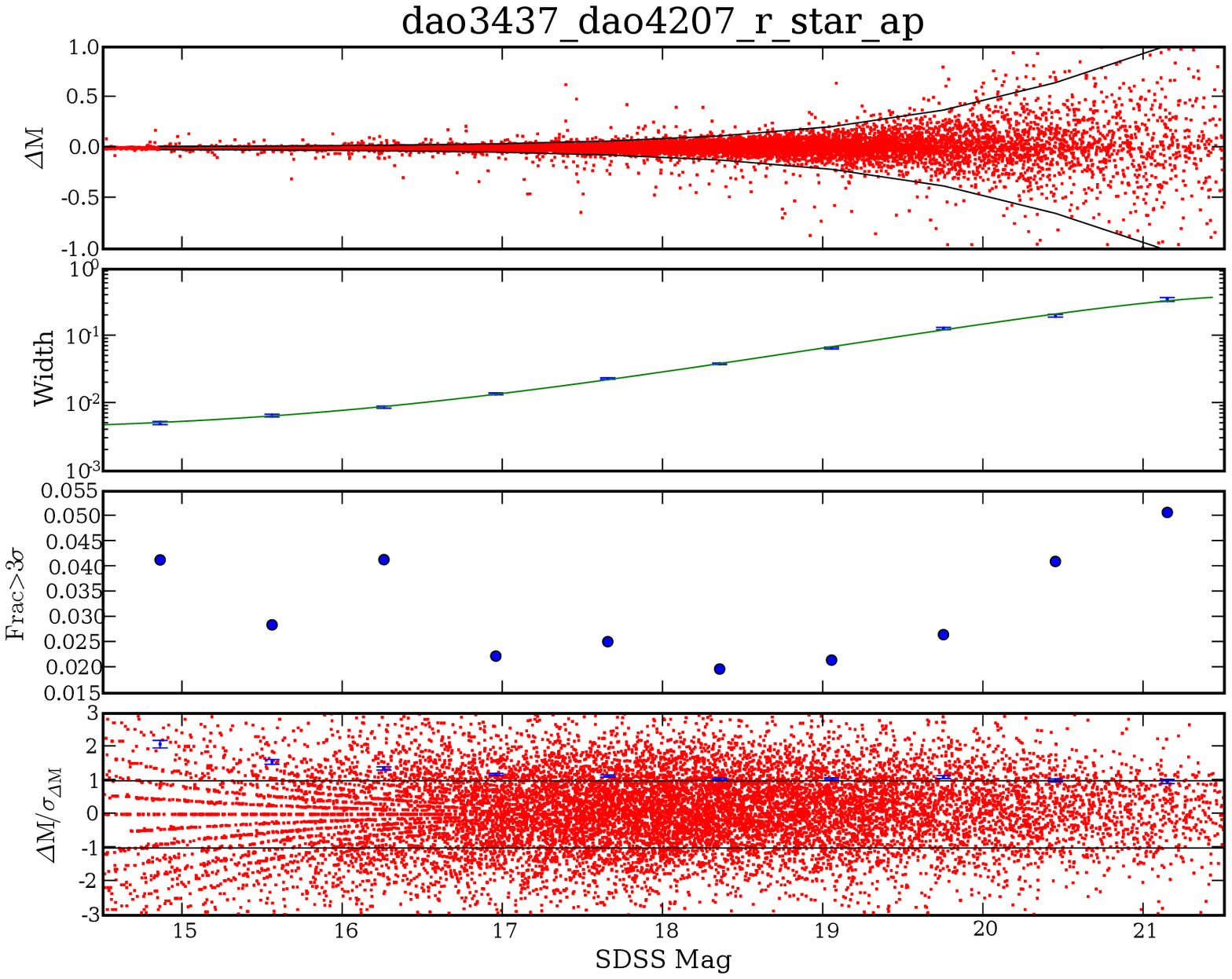}
    \includegraphics[width=3.5in]{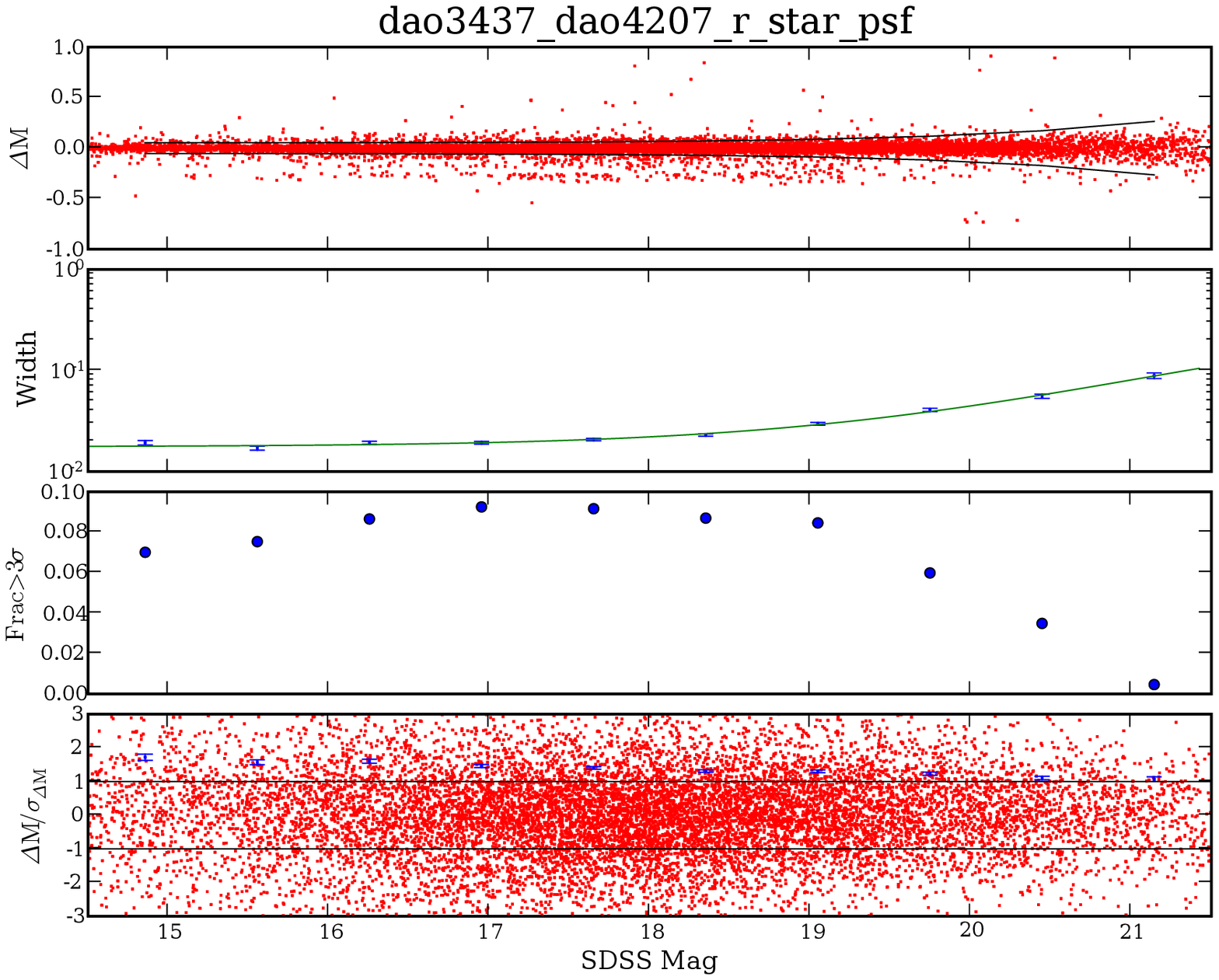}
  \end{center}
  \caption[$r$--band $\Delta$M for \dao]{Figure
  described in Section~\ref{sec_ana-centroid} for \dao's $r$--band 
  photometry of stars.  Figure on the left is for aperture photometry, on the right is PSF photometry.}
  \label{fig-dm_dao}
\end{figure*}

For \photo--selected stars and galaxies, we calculate the difference
of an object's magnitude as measured by a algorithm {\tt alg1} in {\tt
run1} and {\tt alg1} in {\tt run2}, or by algorithm {\tt alg1} in {\tt
run1} and algorithm {\tt alg2} in {\tt run1}.  We plot these
distributions as a function of magnitude.  We do this for both
aperture and PSF (when available) magnitudes, and for stars and
galaxies.  Example $r$-band results for \dao\ are shown in
Figure~\ref{fig-dm_dao} for both aperture and PSF photometry.  Each
figure contains four panels, described below.

\subsection{Panel 1}
\label{subsec-panel1}

The differences in measured magnitudes ($\Delta$M = M1 - M2) are
plotted as a function of \photo's magnitude.  The median $\Delta$M of
objects brighter than $18^{\rm th}$ magnitude (or the brightest
magnitude plus one if no objects brighter than $18^{\rm th}$ are
present; typically this uses thousands of objects) was subtracted off
of the entire distribution, so that it is centered on $y = 0$.  We cut
out the brightest and dimmest $0.5\%$ of the data to avoid outliers.
At the bright end, the width stops following Poisson statistics and
levels off at a characteristic width indicative of systematics in the
analysis.  It is this width that we choose to characterize our
algorithms.

For aperture magnitudes, the systematic floor is smaller at the bright
end because there is no reliance on any PSF model, and aperture
measurements are ideally Poisson limited.  This distribution shows a
characteristic broadening at fainter magnitudes as measurements become
sky--noise dominated.  We naively expected most algorithms to perform
similarly well in aperture magnitude measurements.  However, there are
enough degrees of freedom in centroiding and in treating the
brightness of neighboring objects that these results in actuality are
significantly different.

For PSF magnitudes, the bright--end systematic floor is much larger
due to reliance on a PSF model which is certain to be incomplete at
some level.  Ideally, gross errors in the PSF model come out in the
aperture correction, and this systematic floor is then indicative of
the degree of spatial variation in the aperture corrections.  At
fainter magnitudes, the distribution remains much tighter than for
aperture measurements since sky noise does not contribute as much in a
PSF--weighted measurement.

\subsection{Panel 2}
\label{subsec-panel2}

We divide the $\Delta$M distribution into 10 bins.  The points in each
bin are sorted by $\Delta$M and the first (Q1) and third (Q3) quartile
are determined (the indices corresponding to 0.25 and 0.75 the length
of the sorted array, respectively).  The value of the points
associated with Q1 and Q3 are used to determine the interquartile
range (IQR) of these data.  We choose to use the IQR to lessen our
sensitivity to outliers (such as variable stars).

We find the uncertainty in this width by assuming the data are
normally distributed, where $\sigma_{\rm mean} = 0.74 * IQR$ and
$\sigma_{\rm median} = \sqrt{\pi/2} * \sigma_{\rm mean}$.  The
standard deviation in the IQR is $\sigma_{\rm IQR} = \sqrt{\pi} * 0.55
* IQR$.  The uncertainty in the IQR is $\sigma_{\rm IQR} /
\sqrt{N-1}$.

We plot $\sigma_{\rm mean}$ and its uncertainty (as derived from the
IQR) in each bin.  These data are then fit with the functional form $A
+ B z + C z^2$, where $z = 10^{0.4 * M}$, which describes well the
growth of this envelope with magnitude.  This best fit is plotted as a
solid line.  We evaluate this equation one magnitude below the
brightest data point, and use this single number to characterize the
systematics inherent in the comparison.  The $3-\sigma$ envelope
allowed by this relationship is plotted in Panel~1.  These results are
summarized in Table~\ref{tab-deltaM1} for \photo-selected stars.

We note that the LSST Science Requirement Document states that
photometry should be reproducible to 0.005 magnitudes.  That
translates into a systematic bin width at the bright end of $\sqrt{2}$
* 0.005, or 0.007 magnitudes.

\subsection{Panel 3}
\label{subsec-panel3}

We evaluate and plot the fraction of stars in Panel~1 that are more
than $3-\sigma$ from the mean.  For night--to--night comparisons, this
is very sensitive to the level of variability in the sample.  For
algorithm--to--algorithm comparisons on a given set of data, it allows
us to uncover differences in the algorithms.

\subsection{Panel 4}
\label{subsec-panel4}

We add in quadrature the uncertainties associated with each component
M1 and M2 and plot the distribution of $\Delta$M / $\sigma_{\Delta{\rm
M}}$.  These data are binned, and we derive each bin's IQR and its
uncertainty and overplot these points.  If the photometry packages
accurately quantify the measurement uncertainties, these binned points
should all lie near 1.0.

\subsection{Results Using \photo--Selected Stars}
We have designed three variants of the tests described above to
characterize the algorithms' photometric performance : comparing
photometry of data taken on different nights as an overall
characterization of each algorithm; comparing different algorithms'
photometry of the same data, providing a relative characterization
that is insensitive to stellar variability; and comparing aperture and
PSF magnitudes from the same algorithm on the same data, yielding an
estimate of the scatter introduced by spatial variation of the
aperture corrections.

We first characterize the photometric accuracy of each algorithm by
comparing the brightness of \photo-selected stars measured in both
SDSS runs.  Figure~\ref{fig-dm_dao} shows example $r$-band summary
plots for \dao\ in both aperture and PSF photometry for
\photo-selected stars.  The width of the $\Delta$M distributions are
summarized in Table~\ref{tab-deltaM1}.  We note that both \photo\ and
\dao\ produce $g$, $r$, and $i$-band aperture photometry that meets
LSST's SRD on photometric accuracy.  No other algorithms are able to
meet this requirement, failing to reach the benchmark of 0.007
magnitudes.  We note that {\it no} algorithms are able to meet the SRD
in PSF photometry -- the numbers consistently fall short by a factor
of 2--3.  \dop\ performs worst in terms of PSF photometry.

Most algorithms tend to underestimate aperture magnitudes errors of
bright objects compared to the empirical scatter, with the exception
of \photo\ which tends to {\it overestimate} the aperture errors of
bright objects by as much as a factor of 2.  \sex\ underestimates the
aperture errors of {\it all} objects by a factor of 2--3.  \photo's
PSF magnitude errors represent the empirical scatter very faithfully.
\dop\ and \dao\ underestimate their PSF error uncertainties by $\sim
20\%$.  

We next look at the width of the $\Delta$M distribution for different
algorithms running on the exact same data.  This is insensitive to
stellar variability, and allows us to localize any differences to the
algorithms themselves.  The results for the $r$-band are listed in
Tables~\ref{tab-deltaM5} and \ref{tab-deltaM7} for PSF and aperture
photometry, respectively.  The aperture results are very similar for
all pairs of algorithms, while the PSF photometry comparison of
\photo\ to \dao\ is superior to any comparison using \dop.

Finally, we compare aperture and PSF magnitudes from the algorithms,
yielding an estimate of the additional scatter coming from spatial
variation in the aperture corrections (Table~\ref{tab-deltaM9}).  We
limit our comparison to \photo-selected stars.  {\it A-priori}, we
expect \photo\ to outperform all other algorithms here, since its PSF
magnitudes have already been aperture corrected.  Ideally, the scatter
here should be very close to the aperture photometry results in
Table~\ref{tab-deltaM1}.  Table~\ref{tab-deltaM9} indicates that
\photo's results are equivalent to \dao's, and closer to the PSF
photometry scatter than the aperture photometry scatter.  This
suggests that \photo's aperture corrections have {\it not}
successfully accounted for spatial variation in the PSF.  The numbers
in Table~\ref{tab-deltaM9} do tend to bridge the difference between
the aperture and PSF scatter in Table~\ref{tab-deltaM1}, verifying
that the PSF photometry scatter contains a baseline contribution from
the aperture photometry and an additional contribution from aperture
corrections.

\subsection{Results With Algorithm--Selected Stars}

We repeat this analysis using objects each algorithm selects as a
star.  These results are listed in Table~\ref{tab-deltaM3}, and are
very similar to the \photo-selected analysis.  The largest difference
is that the fraction of $3\sigma$ outliers increases by a factor of
2--3, indicating that the star--galaxy classification schemes for the
algorithms are inferior to \photo's.  Some fraction of this additional
scatter comes from not knowing exactly which pixels in the images have
been interpolated over by \photo\ due to cosmic rays or bad pixels.

\subsection{Photometry Conclusions} 
The aperture and PSF photometry from \dao\ and \photo\ are clearly
superior.  In particular, \dao\ performed as well as \photo, which is
encouraging as \photo\ was designed and commissioned with this SDSS
data set in mind.

No algorithms were able to meet the LSST SRD in terms of PSF
photometry.  The ideal aperture corrections to the PSF photometry
should bring the PSF scatter in--line with that from the aperture
photometry.  The only algorithm for which this degree of calibration
has been done is \photo.  However, it appears that \photo\ has not
sufficiently compensated for spatial variations in its aperture
corrections to PSF magnitudes, since its aperture vs. PSF scatter are
commensurate with \dao's.

As far as calculating uncertainties, the PSF magnitude errors from
\photo\ most closely track the empirical uncertainties.  Aperture
photometry uncertainties are either over or underestimated in {\it
all} algorithms.

It is clear that the task of PSF photometry still requires significant
research and development if LSST is to meet its SRD in terms of
photometric accuracy.  


\section{Shape Measurements} 

For the \photo--selected stars and galaxies, we extract the algorithm
shape parameters {\tt Ixx, Iyy}, and {\tt Ixy} (\dao\ does not report
these values on an object--by--object basis).  We calculate the
ellipticities derived from these moments

\begin{eqnarray}
  e1 = \frac{Ixx - Iyy}{Ixx + Iyy} & ~~;~~ & e2 = \frac{2 Ixy}{Ixx + Iyy} 
\end{eqnarray}

and generate figures comparing each algorithm's shape measurements to
\photo's, dividing the data into 4 magnitude bins.  We plot a linear
relationship between \photo's shape and that from the algorithm.  The
RMS of the scatter about this line is calculated and listed in
Table~\ref{tab-e_star_sex} for \photo-selected stars, and
Table~\ref{tab-e_gal_sex} for \photo-selected galaxies.
Figure~\ref{fig-e_sex} shows a representative set of figures comparing
$r$--band \photo\ and \sex\ ellipticity parameters from run 3437.

\subsection{Shape Measurement Results}
\sex\ is the only algorithm that we tested which reliably calculates
the shapes of galaxies, thus we have limited our comparison of shape
measurements to \photo\ and \sex.  In addition, for ease of tabulation
and interpretation, we present only the results of the $r$--band
analyses.  We note that the $g$ and $i$-band results are
quantitatively similar.

We compare the ellipticities derived from both the ``isophotal'' shape
measurements from \sex\ 2.3.2 and the ``windowed'' measurements from
\sex\ 2.4.4.  The linear relationships between \photo's and \sex's
$r$-band measurements, in the form {\tt e$_{\photo}$ = A + B
e$_{\sex}$}, are shown in Table~\ref{tab-e_star_sex} for stars, and
Table~\ref{tab-e_gal_sex} for galaxies.  We report these numbers for
the brightest magnitude bin ($14 < r < 20$).  We also list the RMS
scatter about this line.

We first note the significantly reduced scatter from the best--fit
linear relationships when using the ``windowed'' shape measurements
from \sex\ 2.4.4.  In particular, this yields up to an order of
magnitude less scatter in the stellar shape measures
(Table~\ref{tab-e_star_sex}), suggesting that \sex\ 2.3.2 is {\it not}
to be used for determining stellar shapes and ellipticities.  The
improvement for galaxies is a more modest factor of 3
(Table~\ref{tab-e_gal_sex}), but still very significant.

The ellipticities of galaxies in \sex\ 2.3.2 is similar to in \photo\
(slope $\sim 1$); the ellipticities of both stars and galaxies in
\sex\ 2.4.4 is different than in \photo\ (slope $\sim$ 2.0 for stars,
$\sim 1.8$ for galaxies).  Figure~\ref{fig-e_sex} shows an example
plot of ellipticity comparisons for \photo-selected galaxies.  The
left panel shows this relationship for \sex\ 2.3.2, and the right
panel for \sex\ 2.4.4.  The isophotal measurements clearly lead to a
tighter relationship.

\subsubsection{Shape Measurement Conclusions} 
Adaptive second moments are more reliable than isophotal moments.  We
recommend that all \sex\ analyses relying upon shape measurements use
``windowed'' shape measures.  Non--windowed shape measures should {\it
not} be used for stars.


\section{Centroiding}
\label{sec_ana-centroid}

We also compare centroiding offsets between objects as measured in the
same images by different algorithms.  To do this accurately, we must
first determine the conventions used to describe the image array.  For
both \dao\ and \sex, the center of the lower--left hand corner pixel
(LLHC) is coordinate (1.0, 1.0).  In \photo\ and \dop, the LLHC is at
coordinate (0.5, 0.5).

We perform an analysis similar to that described in
Section~\ref{sec-phot} but describing the distribution of {\it pixel}
offsets as a function of magnitude.  This should reveal any
centroiding biases as a function of magnitude.  Example
Figure~\ref{fig-centroid} includes the three panels described in
Section~\ref{subsec-panel1}, Section~\ref{subsec-panel2}, and
Section~\ref{subsec-panel3}.  Here the width of the bright end of the
distribution in Panel~1 reflects centroiding systematics.

We also plot in each Panel~1 a quadratic fit to the median value of
the $X,Y$--coordinate pixel offsets of the form $\Delta_{X,Y} = A + B
z + C z^2$, where $z = M - M0$, $M0$ is the magnitude of the first
(brightest) bin and $M$ the central magnitude for each bin.  We plot
the median values and their uncertainties, and the functional fit as a
solid line.  Any shape to this distribution ($B \neq C \neq 0$)
suggests systematics in object centroiding as a function of magnitude.
These results are summarized in Table~\ref{tab-centroid2a} for
\photo-selected stars.  Table~\ref{tab-centroid2c} shows the width of
this distribution, evaluated 1 magnitude below the brightest
unsaturated star, comparing algorithm to algorithm for $r$-band
centroids in run 3437 (upper triangular matrix) and run 4207 (lower
triangular matrix).

\subsection{Centroiding Results}
We compare the measured positions of objects in each image as a
function of magnitude.  Accurate centroiding is required to deliver
the SRD relative astrometry requirement of $0.01 \arcsec$ (here 0.025
pixels).  We are unable to comment on the absolute astrometry
requirements since that involves knowledge of astrometric distortions
in the focal plane, which are different here than will be the case in
LSST.

We list the results of the quadratic fit in Table~\ref{tab-centroid2a}
for \photo-selected stars.  \sex\ 2.3.2 consistently has significant
offset, linear, and quadratic terms.  \dop\ rarely shows significant
quadratic terms, but tends to have significant zeropoint offsets at
$\sim 0.01$ pixels.  Both \dao\ and \sex\ 2.4.4 compare very well with
\photo's positional measurements, routinely having offsets below 0.005
pixels, linear terms below 0.003 pixels/magnitude, and quadratic terms
below 0.001 pixels/magnitude$^2$.

An example demonstrating the improvements between \sex\ 2.3.2 and
\sex\ 2.4.4 is shown in Figure~\ref{fig-centroid}.  Here we plot 2
figures containing the three panels described in
Section~\ref{sec_ana-centroid}.  The left panel shows the distribution
of $z$-band $\Delta$X pixel offsets between \sex\ 2.3.2 and \photo.
The right panel provides a comparison between \sex\ 2.4.4 and \photo.
It is clear there is a much smaller trend of the median pixel offset
with magnitude in \sex\ 2.4.4, as well as a smaller overall RMS to the
distribution.  

We use this RMS at the bright end to further characterize the
centroiding accuracy.  
This comparison of all algorithm centroids is shown in
Table~\ref{tab-centroid2c} for $r$-band $x$--coordinate centroids.
This table indicates that the algorithms are much more consistent with
each other than they are with \photo, as the RMS is consistently
highest in those comparisons including \photo.  Compared to RMSs of
order 0.02--0.03 pixels for comparisons with \photo, the other
algorithms are consistent to 0.01 pixels or better.  We trace this
back to \photo's astrometric corrections derived from the PSF behavior
\citep{2003AJ....125.1559P}, which the other algorithms do not account
for.  These corrections demonstrably produce better {\it absolute}
astrometry, since they account for biases in positions due to the
complex PSFs.  We thus expect relative astrometry to be accomplished
in software to better than 0.01 pixels, or more than 200 times smaller
than the image FWHM.  Absolute astrometry may require corrections
similar to what has been undertaken by SDSS.

\subsection{Centroiding Conclusions}
The LSST SRD relative astrometry requirement of $0.01 \arcsec$ (1/70
the median SRD r--band seeing of 0.7 $\arcsec$) is not likely to be
violated in software.  The ``windowed'' centroids of \sex\ 2.4.4 are
comparable to the PSF centroids of \dao\ and \photo, and a significant
improvement over \sex\ 2.3.2.


\section{Photometric Depth}
\label{sec-depth}

We select all clustered objects that have been classified as a star by
each algorithm for each run, and create star count histograms.
We find the bin with the maximum number of stars found by each
algorithm, as well as the cumulative fraction of the histogram as a
function of magnitude.  We characterize the photometric depth of each
algorithm by determining the magnitude bins below which 95\%
(M$_{95}$) and 99\% (M$_{99}$) of the objects have been detected.
These values, as well as the peak of the functions, are listed in
Table~\ref{tab-depth}.

\subsection{Photometric Depth Results}

Using M$_{99}$ as a proxy for photometric depth, \photo\ is
consistently deeper than \dao\ and \dop\ in PSF magnitudes, in many
cases significantly.
We can trace this back to the definition of ``significance'' in the
object detection stages.  For example, \dao\ triggers off the central
pixel of an object in the image convolved with its PSF, yielding a
weighted sum of neighboring pixels.  \photo\ does a similar smoothing,
but also grows the source by an amount approximately equal to the
radius of the seeing disk, and defines a source as a connected set of
pixels that are detected in at least one of the 5 passbands.
Unfortunately, it is not sufficient to merely lower \dao's object
detection threshold to compensate for these differences without also
enacting a change in how the algorithm evaluates the notion of
``significance''.  By lowering the threshold we would be allowing an
unacceptable number of artifacts through along with the fainter
astronomical objects.  The ideal object detection algorithm would
trigger off of medium significance pixels and determine the integrated
significance of all neighboring (e.g. 8-connected) pixels, comparing
the latter to the user--defined detection threshold.

The comparison between \photo\ and \sex\ is slightly more difficult,
since aperture photometry is not the ideal measurement to use in star
count comparisons.  For example, the peaks of \photo's aperture
photometry star--counts are frequently 2-3 magnitudes fainter than for
its PSF star--counts.  At least for the $g$ and $r$ passbands, the
metric M$_{99}$ is approximately the same for aperture and PSF
photometry, so we use these filters in our \sex\ comparison.  On both
nights, \sex\ stops more than 1 magnitude brighter than \photo\ in
$g$, and slightly less than 1 magnitude in $r$.

\subsection{Photometric Depth Conclusions} 
It is difficult to compare photometric depths in the context of
incomplete star/galaxy separation schemes.  The star counts of all
algorithms are contaminated to some degree by galaxies.  However,
because \photo\ measures and deblends stars and galaxies
simultaneously, we believe this yields the most accurate
classification criteria, and thus the most accurate star counts.

\dao\ is primarily designed to photometer stars, and while it does a
reasonable job of agreeing with \photo\ on object classification
(Table~\ref{tab-sg2}), it also is over--complete compared to \photo\
for brighter objects, where \photo\ is known to do well, and is also
incomplete for fainter objects.  The former is likely due to detection
of artifacts in the images, as well as misclassification of galaxies
as stars.


\section{Analysis of Globular Cluster M2}
\label{sec-m2}

Globular Cluster M2 (NGC 7089) is located in our imaging strip.  This
cluster contains approximately 150,000 stars, with a core radius of
$0.34\arcsec$.  This is a highly concentrated structure, and will test
the limits of any photometric software tasked to analyze it.  In fact,
the majority of \photo's attempts to reduce images containing this
cluster are unsuccessful, failing at the stage of deblending.

We have chosen to use this particular field to test {\tt daophot}'s
and {\tt allframe}'s abilities to do stellar photometry in crowded
fields.  With the vast majority of objects in these images being
cluster stars, we expect minimal contamination from background
galaxies.  We do however expect to encounter problems with the
brightest cluster stars ($13^{th}$ magnitude), which saturate in the
standard SDSS exposures.  In the images we are using, saturated pixels
and bleeds have been interpolated over by \photo, leaving the profiles
of these objects inconsistent with the PSF.  \dao\ is therefore
inclined to consider these objects extended, and will fit an ensemble
of PSFs to the object until enough have been added to ``vacuum'' up
all of its flux.

This analysis will also serve as a proxy for how close LSST can
observe to the Galactic plane and still maintain a given level of
photometric precision.  However, in such crowded fields, aperture
photometry is neigh impossible.  And as Section~\ref{sec-phot} has
shown, PSF photometry is unable to produce results with the required
accuracy.  It is unclear if it is possible, even in the most idealized
case, for the SRD requirements to be met in such crowded fields.

\subsection{Photometry}

Due to the degree of stellar crowding in this field, {\tt OPTICS}
clustering runs yielded marginal results with a clustering distance of
1 pixel ($0.4\arcsec$).  This was characterized by large scatter when
matching the centroids of objects in {\tt daophot} and {\tt allframe},
at the level of 0.8 pixel RMS in the $r$--band.  We instead chose to
cluster the data with a half pixel ($0.2\arcsec$) clustering distance,
which yielded much improved results (RMS scatter of 0.04 pixel in the
$r$--band).  Clustering at a quarter pixel ($0.1\arcsec$) did not
significantly alter the results.

The results for the $\Delta$M distribution measurements are listed in
Table~\ref{tab-M2deltaM}.  For both algorithms, we used the
star--galaxy classification schemes derived from the previous analyses
and described in Table~\ref{tab-daosharp}.

The results of this analysis are very encouraging.  We first note that
the first two sets of data ({\tt daophot} and {\tt allframe}) in
Table~\ref{tab-M2deltaM} correspond to objects classified by {\tt
daophot} as stars.  To have clustered with {\tt daophot} detections,
this subset of the data will not reach as deep as the full {\tt
allframe} reductions.  Therefore these numbers do not directly reflect
{\tt allframe}'s photometry of faint objects, but instead the fact
that {\tt allframe} is better able to deblend the stars used in this
analysis from faint objects that were missed in {\tt daophot}.  The
second set of {\tt allframe} results are for objects classified by
{\tt allframe} as stars, and thus also probes the distribution of
stars missed in {\tt daophot} because they were too faint or blended.
We emphasize that the PSFs used in the two analyses are exactly the
same, and any improvements may be directly attributed to better
deblending and centroiding.

The aperture photometry results are considerably worse here than as
reflected in the sparse--field analysis described in
Table~\ref{tab-deltaM1} and Table~\ref{tab-deltaM3}.  This is to be
expected, as the field is extraordinarily crowded and there is a very
steep and significant background sky gradient due to unresolved
cluster stars.  Both the $r$ and $i$--band aperture results are
considerably worse than in the other passbands, in this case due to
the extreme crowding conditions in these filters.

The PSF photometry shows a marked improvement over the aperture
photometry results, particularly in the $r$ and $i$--band data where
the images are most crowded.  The $g$-band PSF photometry is the most
problematic in the \dao\ reductions.  However, the magnitude scatter
for objects classified by {\tt daophot} as stars is reduced by
approximately $25\%$ when going to the stacked analysis of {\tt
allframe}.  In particular, the $g$-band photometry improves
significantly, suggesting that \dao\ did a poor job of selecting all
the stellar $g$-band objects, and a proper deblending was only
possible by using constraints from the $r$ and $i$-band data.  We also
note that the {\tt allframe} PSF photometry results are commensurate
with the sparse--field analyses described in Table~\ref{tab-deltaM1}
and Table~\ref{tab-deltaM3}.  This indicates that {\tt daophot}+{\tt
allframe} is indeed a powerful combination that is able to perform
consistent stellar PSF photometry across the range of crowding
conditions expected in LSST.

The final set of numbers in Table~\ref{tab-M2deltaM}, reflecting the
analysis of objects classified by {\tt allframe} as stars, shows a
slight increase in the scatter of photometric measurements.  The
degradation is likely due to the impact of {\tt allframe} detecting
fainter, more crowded objects, for which photometry is more difficult.
However, the PSF--photometry results are still better than {\tt
daophot}'s single--image analysis of this field, and essentially
equivalent to the sparse--field analysis results presented in
Table~\ref{tab-deltaM3}.

\subsubsection{Photometry as a Function of Crowding}

\begin{figure*}[htbp]
  \begin{center}
    \leavevmode
    \includegraphics[width=3.5in]{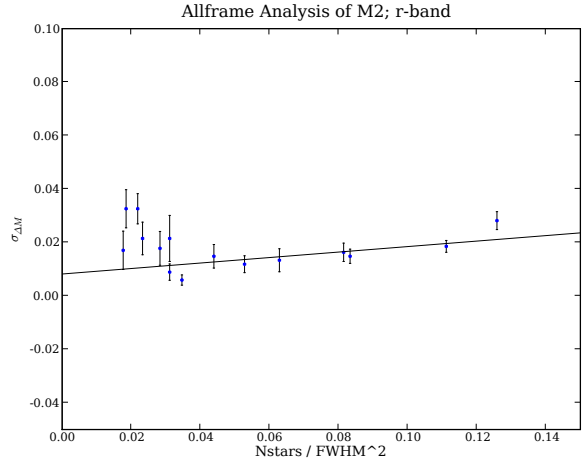}
  \end{center}
  \caption[$\sigma_{\Delta M}$ vs Crowding for M2]{$\sigma_{\Delta M}$
  plotted as a function of local crowding conditions, derived from
  {\tt allframe} analysis of globular cluster M2, for the $r$--band
  data.  We divided the image up into multiple regions and for each
  derived the width of the $\Delta$M distribution from the brightest 3
  magnitudes of stars.  We normalized the number of {\it all} stars in
  each region by the area of the region and the average FWHM of the
  two images.  The x--axis reflects the crowding conditions, and
  corresponds to the total number of stars per seeing disk.}

  \label{fig-M2deltaMvsCrowding}
\end{figure*}

Given the broad range of stellar densities in these images, we are
able to constrain how \dao's ability to do PSF photometry degrades as
a function of local crowding conditions.  To do this we have divided
the image up into 200 pixel by 200 pixel regions, and select only
those objects that {\tt allframe} classifies as stars in both runs.
We count the total number of such objects in this region, as well as
the total number of ``bright'' objects in this region, where we define
``bright'' as the brightest 3 magnitudes of objects.  We calculate the
$\sigma_{\rm mean}$ from the interquartile range of $\Delta$M for the
bright objects, and plot this against the total number of stars in the
bin.  We normalize this by the area of the box, yielding the local
number of stars per pixel, and then multiply by the averaged FWHM$^2$
of the two images, yielding the approximate number of stars per seeing
disk.  We fit a line to the relationship of $\Delta$M vs number of
stars per FWHM$^2$.  These results are summarized in
Table~\ref{tab-M2deltaMvsCrowding}.  We show the plots for the
$r$--band data in Figure~\ref{fig-M2deltaMvsCrowding}.  Extrapolation
back to an empty field (number of stars = 0) yields numbers that are
very close to the SRD requirement on photometric accuracy.

\subsection{Photometric Depth}

We select stars on an algorithm--by--algorithm basis, and find the
peak of the star count histograms are the same for both {\tt daophot}
and {\tt allframe}, approximately $r = 20.5$, $g = 21.0$, $i = 20.2$
for run 4207.  However, {\tt allframe} finds approximately 1.5 times
the total number of objects in the $g$-band data, 1.3 in the $r$-band,
and 1.4 in the $i$-band.  This is due to {\tt allframe}'s ability to
resolve and photometer blended neighbors that contaminate an object's
{\tt Sharp}-ness in {\tt daophot}, as well as its extra photometric
depth.  Table~\ref{tab-M2depth} characterizes the depth per run and
passband.  For both algorithms, we list the peak of the histogram
(M$_{max}$), the magnitude bin below which 95\% of the stars are
contained (M$_{95}$), and the bin below which 99\% of the stars are
contained (M$_{99}$).  Using M$_{99}$ as our proxy, {\tt allframe}
accurately photometers objects nearly a magnitude deeper than in {\tt
daophot} in the $g$-band, 0.3 magnitudes in the $r$-band, and 0.5
magnitudes in the $i$-band.  This is a remarkable improvement
considering that we only have 2 images per passband to work with.  The
fact that we can combine the constraints from images in different
filters into a global analysis allows us to make such improvements in
depth.

\begin{figure*}[htbp]
  \begin{center}
    \leavevmode
    \includegraphics[width=3.5in]{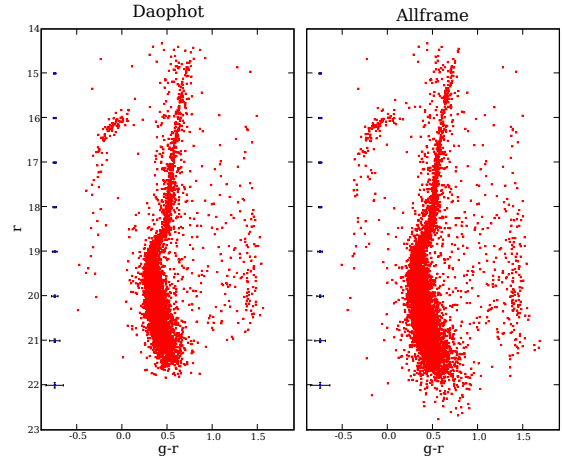}
  \end{center}
  \caption[Color--magnitude Diagram of M2]{Color--magnitude diagram
  (CMD) of M2 reconstructed from {\tt daophot} and {\tt allframe}
  analysis.  All clustered objects classified by each algorithm as
  stars in both runs and in both the $r$ and $g$-bands were used.  We
  also plot typical error bars in 8 magnitude bins.  The {\tt
  allframe} CMD contains 70\% more points than the {\tt daophot} CMD,
  and reaches approximately 0.3 magnitudes deeper in the $r$--band.}
  \label{fig-M2_cmd}
\end{figure*}

Figure~\ref{fig-M2_cmd} shows a $r$ vs. $g-r$ color--magnitude diagram
(CMD) of all stars in the SDSS images containing M2.  We have not
selected against field stars, which contaminate the cluster CMD.  For
each algorithm, we query for all clustered objects that were
classified as stars in both runs {\it and} in both passbands to yield
the final ensembles of points.  {\tt Allframe} finds 1.7 times the
number of stars as {\tt daophot}.
We plot the averaged magnitudes and colors of the objects, as well as
typical error bars on each point in 8 magnitude bins.

\subsection{Conclusions from Study of M2}

The {\tt allframe} analysis has shown that it is an encouraging
precursor to LSST's envisioned Deep Detection Pipeline ensemble
analysis of imaging data \citep{2005AAS...207.2631R}.  We are able to
use {\it all} images of a given part of the sky to attain extra depth
and precision in the measurements of {\it all} objects in the field.
Potential improvements to this process include regeneration of the PSF
during the ensemble analysis, as well as characterization of extended
objects.

\section{Processing Time and Scalability}
\label{sec-valgrind}

During processing, we recorded the total elapsed time to run each
algorithm on all images.  However, during testing we noticed severe
degradations in performance during periods of heavy disk access.  This
is a known problem with the Redundant Array of Independent Drives
(RAID) controller on the host machine, and makes the {\it absolute}
numbers in this section inaccurate.  The {\it relative} numbers are
likely to be less affected.

We do not have information for \dop\ on run 4207 because the file
containing the times for this run was corrupted.  We emphasize that
the \dao\ results are not entirely localizable to the internal
algorithms, but are also due to inefficiencies in our controlling {\tt
Perl} scripts (Section~\ref{sec-alg}).  We fit the trend of processing
time with the number of detections, and present these results in
Table~\ref{tab-timefit}.  \sex\ is the fastest algorithm, with version
2.3.2 slightly faster than version 2.4.4, primarily due to the
overhead in calculating windowed quantities in the latter.  There
appears to be a minimum threshold of at least 4 seconds necessary for
\sex\ to process an individual stitched image regardless of the number
of detections found, due to overhead associated with the reading and
writing of data products.  \dao\ shows a significant trend with number
of detections and has the steepest scaling laws.  The \dop\ entry in
Table~\ref{tab-timefit} is a bit misleading, as \dop\ tends to be
relatively insensitive to the number of objects ultimately detected in
the image.  This suggests that much of the processing time is spent on
common--mode items such as the PSF generation.

\subsection{Additional Testing \label{sec-darkstar}}

In an effort to eliminate the influence of the RAID controller, we
also ran time trials on a new computer.  We selected four images (two
from each run) covering the range of total detections per image found
by \sex\ in the $r$-filter.  The ``stitched'' images are approximately
2k $x$ 4k in size.  We decided to examine the scaling of resource
usage with image size by chopping each image into a 2k $x$ 2k image.
We also produce an LSST--sized image by placing a copy of each image
next to itself to yield a 4k $x$ 4k image.  We store a copy of each
image with a variety of bit depths to determine how this might effect
\sex's behavior.  We store a copy of each image as 16 and 32--bit
integers (BITPIX=16,32), and as 32 and 64--bit floats
(BITPIX=-32,-64).  In summary, we have 4 images with different numbers
of objects; we have 3 copies of each image in different sizes; and we
store each of these with 4 different bit depths.  In total, this
yields 48 different configurations.

Each of these images was {\tt SExtracted} 50 times in a row to
determine the average elapsed time per image, averaging over any
extraneous system load.  \sex\ was run while there were no other tasks
queued on the machine for the duration of each run.  We monitored the
memory usage of each process as a function of time by scanning the
file {\tt /proc/PID/status} every half second.  We extract the values
{\tt VmSize} and {\tt VmRSS}.  {\tt VmSize} is the total amount of
memory required by this program, and {\tt VmRSS} is the "Resident Set
Size" (the amount actually in memory at a given moment).  We extracted
the total processing time by using the executable {\tt /usr/bin/time}
and summing the {\tt user} CPU and {\tt system} CPU times -- each
process had $98\%$ or greater of the CPU.  Table \ref{sexds} lists the
results of these trials.

We first examine the profiling as a function of image bit depth.  The
maximum memory used by \sex\ is {\it not} a function of image bit
depth for a given--sized image.  This suggests that \sex\ translates
an image into a ``native'' bit depth before processing.  The total
processing times for BITPIX of 16, 32, and -32 are very similar; the
BITPIX = 64 images take on average $10\%$ longer to process,
suggesting significant overhead in translating from 64--bit images.
We restrict our analysis henceforth to 32--bit float images.

\begin{figure*}[htbp]
  \begin{center}
    \leavevmode
    \includegraphics[width=3.5in]{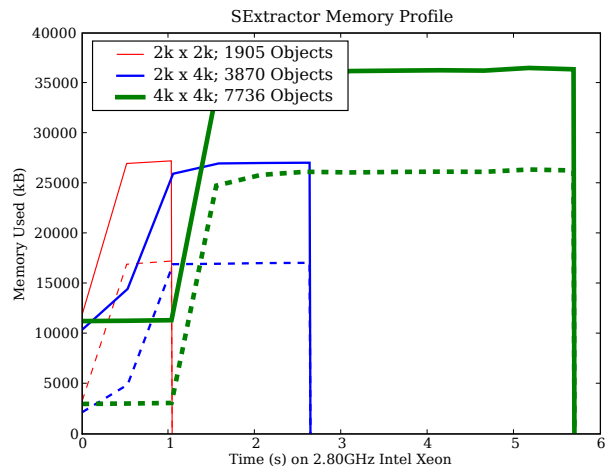}
  \end{center}
  \caption[\sex\ Memory Requirements vs. Time] {Detailed look at the
  average memory required by \sex\ as a function of time.  The {\it
  solid} lines correspond to {\tt VmSize}, while the {\it dashed}
  lines correspond to {\tt VmRSS}. }
  \label{fig-mem1}
\end{figure*}

We next look at the memory consumed as a function of time for a given
run.  Since we only sample the memory usage in 0.5 second intervals,
this will be somewhat poorly determined for the short analyses.  We
choose to make representative plots using the last image in
Table~\ref{sexds}.  Figure~\ref{fig-mem1} shows the average memory
usage as a function of time for the 3 image sizes.  Note that the
total processing time shown here can be up to 0.5 seconds smaller than
the values listed in Table~\ref{sexds} due to our coarse sampling.

It is interesting to note the memory consumption profiles generally
differ due to the different processing times, but the {\it maximum}
memory used does not scale directly with the image size or the total
number of objects.  The memory requirements grow only marginally more
expensive, suggesting that \sex\ undertakes an effective degree of
intelligent memory management.  For example, the 4k $x$ 4k image
consumes less than twice as much memory as the 2k $x$ 4k image.

\begin{figure*}[htbp]
  \begin{center}
    \leavevmode
    \includegraphics[width=3.5in]{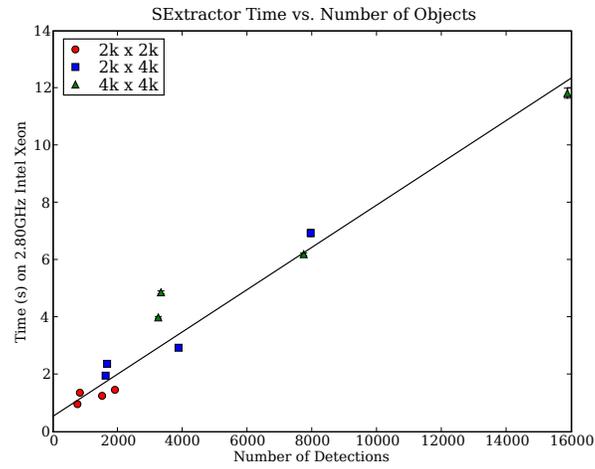}
  \end{center}
  \caption[\sex\ Processing Time vs. Number of Detections]
  {Plot of total \sex\ processing time as a function of number of
  detections in the image.  The {\it red} circles are from the 2k $x$
  2k images, {\it blue} squares from the 2k $x$ 4k images, and {\it
  green} triangles from the 4k $x$ 4k images.  A joint fit to all the
  data is shown in black, with the functional form $y = 0.5468~x +
  0.0007$.}
  \label{fig-sexds}
\end{figure*}

We next examine the total processing time as a function of the number
of objects in the image.  These data are plotted in
Figure~\ref{fig-sexds}.  We plot the data from the 2k $x$ 2k images as
circles, 2k $x$ 4k as squares, and 4k $x$ 4k as triangles.  A linear
regression yields the relationship $y = 0.5468~x + 0.0007$.  Comparing
this to the entries in Table~\ref{tab-timefit} is instructive.  The
zero--point processing time of 0.5 seconds is much shorter than
previous results of $\sim 4$ seconds, almost certainly due to the
aforementioned RAID issues impeding disk I/O.  The slope is similar :
every $\sim 1300$ objects being measured adds an additional second of
processing time.  We regard these tests on this machine to yield the
most reliable timing results.

\subsection{Processing Time Conclusions} 
\sex\ version 2.3.2 was the fastest of these algorithms.  However,
with slightly longer processing time we gain a considerable amount of
accuracy in the position and shapes of detected objects by using the
``windowed'' parameters from \sex\ 2.4.4.

Disk access is a fundamental issue that can significantly impede image
processing tasks.

The timing tests in Section~\ref{sec-darkstar} produce the most
reliable {\it absolute} numbers.  If we assume that the LSST focal
plane is populated with 4k $x$ 4k devices, than we expect that a
single detector may be photometered in (0.5 s) * (2.8 GHz) = 1.4 GHz
s, with an additional overhead of 1.4 GHz s for every 1300 objects in
the image.  We have not tested how these numbers scale with processor
speed.

\section{Summary of Results}
\label{sec-summary}

\subsection{Star/Galaxy Separation}

Each package undertakes some measure of object classification.  In all
cases, the benchmark profile is the PSF.  \dao\ and \dop\ compare each
object to the PSF profile.  \sex\ compares the width of each object
with the input PSF FWHM.  In comparison, \photo\ compares the flux
measured using the PSF to the flux from galaxy model fits.

Both \dop\ and \sex\ fared poorly compared to \dao\ and \photo\
(Tables~\ref{tab-sg1} and \ref{tab-sg2}).  However, \sex\ has the
option to use neural--network filters to enhance its performance.
\dao\ does a good job at object classification, but does not
explicitly compute object moments.  Objects where \dao\ and \photo\
disagree tend to be drawn from the stellar locus
(Figure~\ref{fig-ccd_dao}).

\photo\ is the most advanced package in this task, with \sex\ having
the most potential for improvement through add--on software like {\tt
EyE}.

\subsection{Photometry}

Both \dao\ and \photo\ are able to satisfy LSST's science requirements
on photometric accuracy (0.005 magnitudes unless precluded by photon
statistics) for aperture measurements only.  This is realized in the
$g$, $r$, and $i$-band datasets.  PSF photometry is unable to reach
this accuracy, and consistently falls short by a factor of $\sim 2-3$.
\dao\ provides marginally better results than \photo\ in both aperture
and PSF photometry in our normal analysis.  \dop\ consistently
under-performs in both aperture and PSF photometry.  \sex\ provides
adequate aperture photometry, but does not yet have the capability to
easily build and use a PSF model.  These results are summarized in
Table~\ref{tab-deltaM1} (for \photo-selected stars) and
Table~\ref{tab-deltaM3} (for algorithm--selected stars).

The additional scatter in the PSF magnitudes can be traced back to
inadequate aperture corrections to the PSF flux.  We highlight that
the determination of this quantity, as well as its spatial variation
across an image, is a crucial issue in LSST algorithm development.

From our analysis of globular cluster M2, we find that \dao\ is able
to provide PSF magnitudes in a crowded field with an accuracy similar
to a sparse field analysis.  A stacked analysis of the data using {\tt
allframe} yields an improvement of approximately $25\%$
(Table~\ref{tab-M2deltaM}) in photometric accuracy, and a
passband--dependent increase in photometric depth
(Table~\ref{tab-M2depth}).  We find a marginal degradation in
photometric accuracy with local crowding conditions
(Table~\ref{tab-M2deltaMvsCrowding}).  {\tt Allframe} is able to
maintain $2\%$ accuracy in $r$-band PSF photometry in crowding of up
to 0.12 stars per PSF FWHM$^2$ ($\sim 880$ stars arcmin$^{-1}$ in
0.7\arcsec seeing).

\subsection{Shape Measurements}

\begin{figure*}[htbp]
  \begin{center}
    \leavevmode
    \includegraphics[width=3.5in]{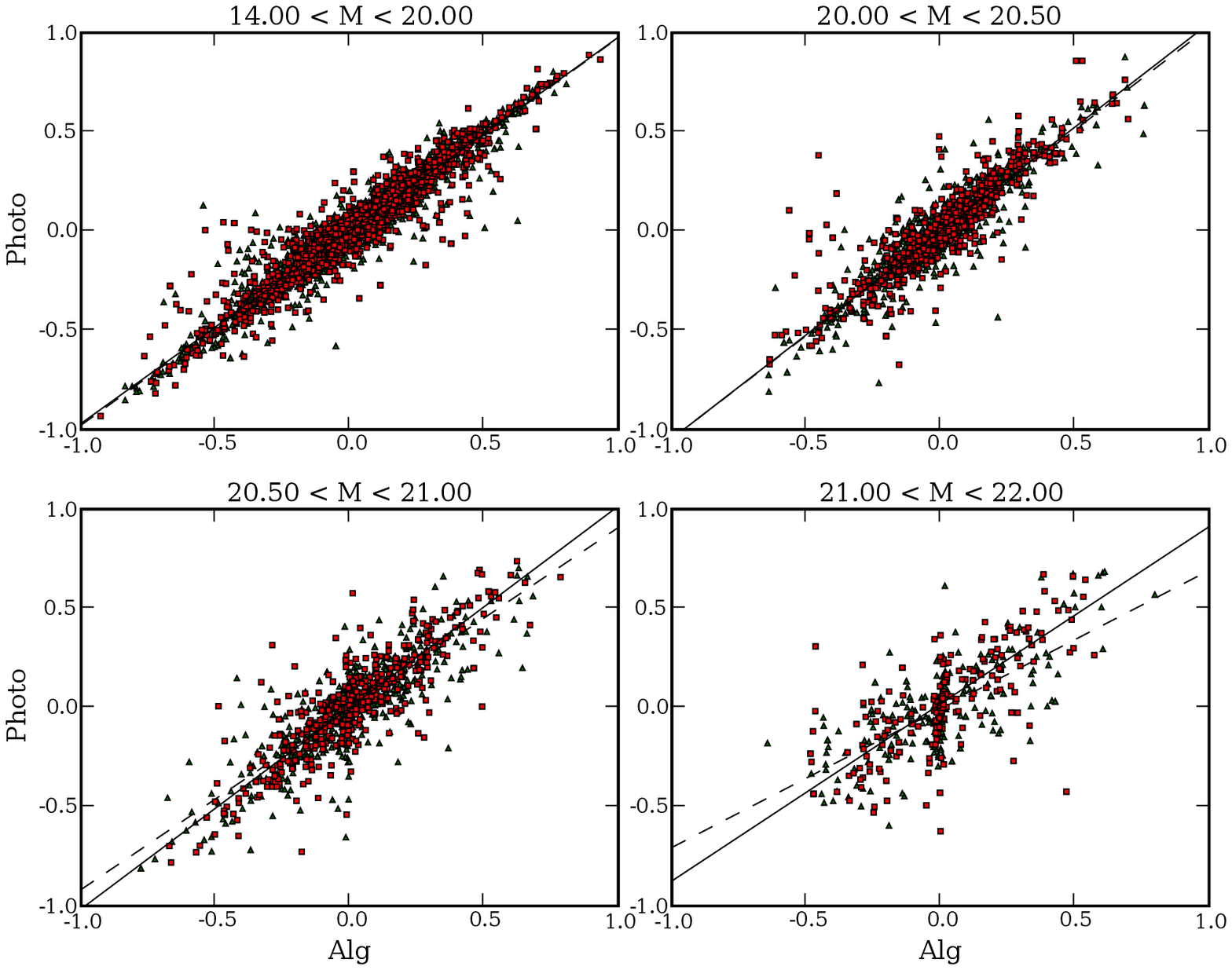}
    \includegraphics[width=3.5in]{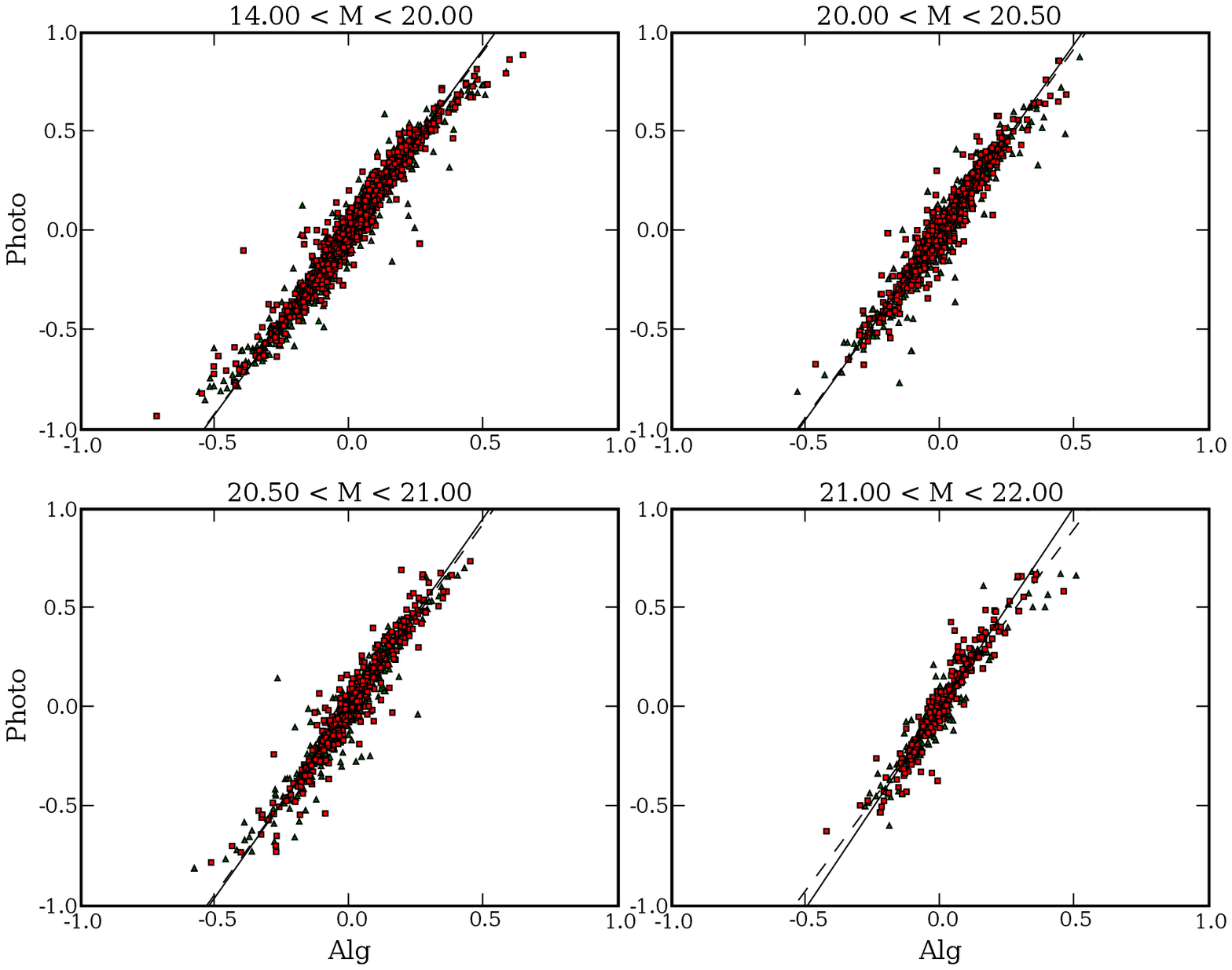}
  \end{center}
  \caption[Ellipticity Measurements; \sex\ 2.3.2. vs. \sex\ 2.4.4
    vs. \photo]{Comparison of run 3437 $r$-band galaxy ellipticity
    measurements in \sex\ and \photo.  {\tt e1} is plotted as green
    triangles, and {\tt e2} as red squares.  In each figure, the 4
    panels are for data in different $r$--band magnitude bins, and
    compare the shape measured in \sex\ on the $x$--axis, and \photo\
    on the $y$--axis.  The {\it left} figure shows results from \sex\
    2.3.2 and the {\it right} figure \sex\ 2.4.4.  The lines show the
    best fits given in Table~\ref{tab-e_gal_sex}, dashed for {\tt e1}
    and solid for {\tt e2}.}
  \label{fig-e_sex}
\end{figure*}

\sex\ and \photo\ are the only packages that provide reliable
estimates of object shapes, using second moment analysis.  \photo\ is
also the only package that also fits galaxy models (exponential, de
Vaucouleurs) to each object.  \sex\ version 2.3.2 uses isophotal
second moments, which degrade rapidly as a function of magnitude
compared to \photo's adaptive second moments (e.g. left panel of
Figure~\ref{fig-e_sex}).  These measurements should {\it not} be used
to measure the shapes of stars.  \sex\ versions 2.4.4 and greater use
``windowed'' second moments that yield ellipticities comparable to
\photo's (e.g. right panel of Figure~\ref{fig-e_sex}).  \photo\ and
\sex\ 2.4.4's stellar ellipticity measurements are extremely
consistent, their differences having an RMS of 0.001--0.004
(Table~\ref{tab-e_star_sex}).  This is more than a factor of 10
smaller than LSST's science requirement that the median of the
distribution be no larger than 0.04, indicating that the algorithmic
contribution to the stellar ellipticity distribution should be
negligible.

\subsection{Centroiding}

\begin{figure*}[htbp]
  \begin{center}
    \leavevmode
    \includegraphics[width=3.5in]{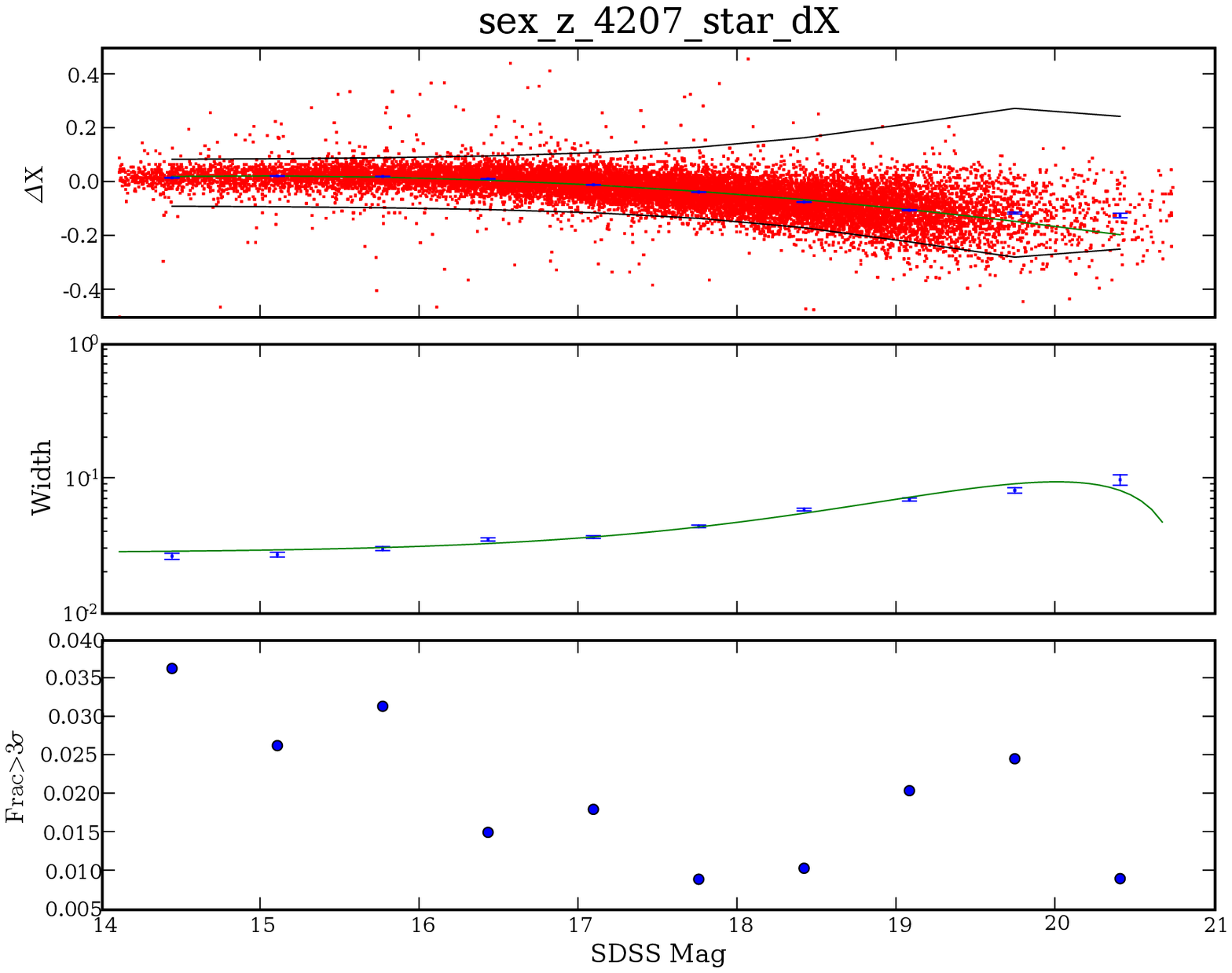}
    \includegraphics[width=3.5in]{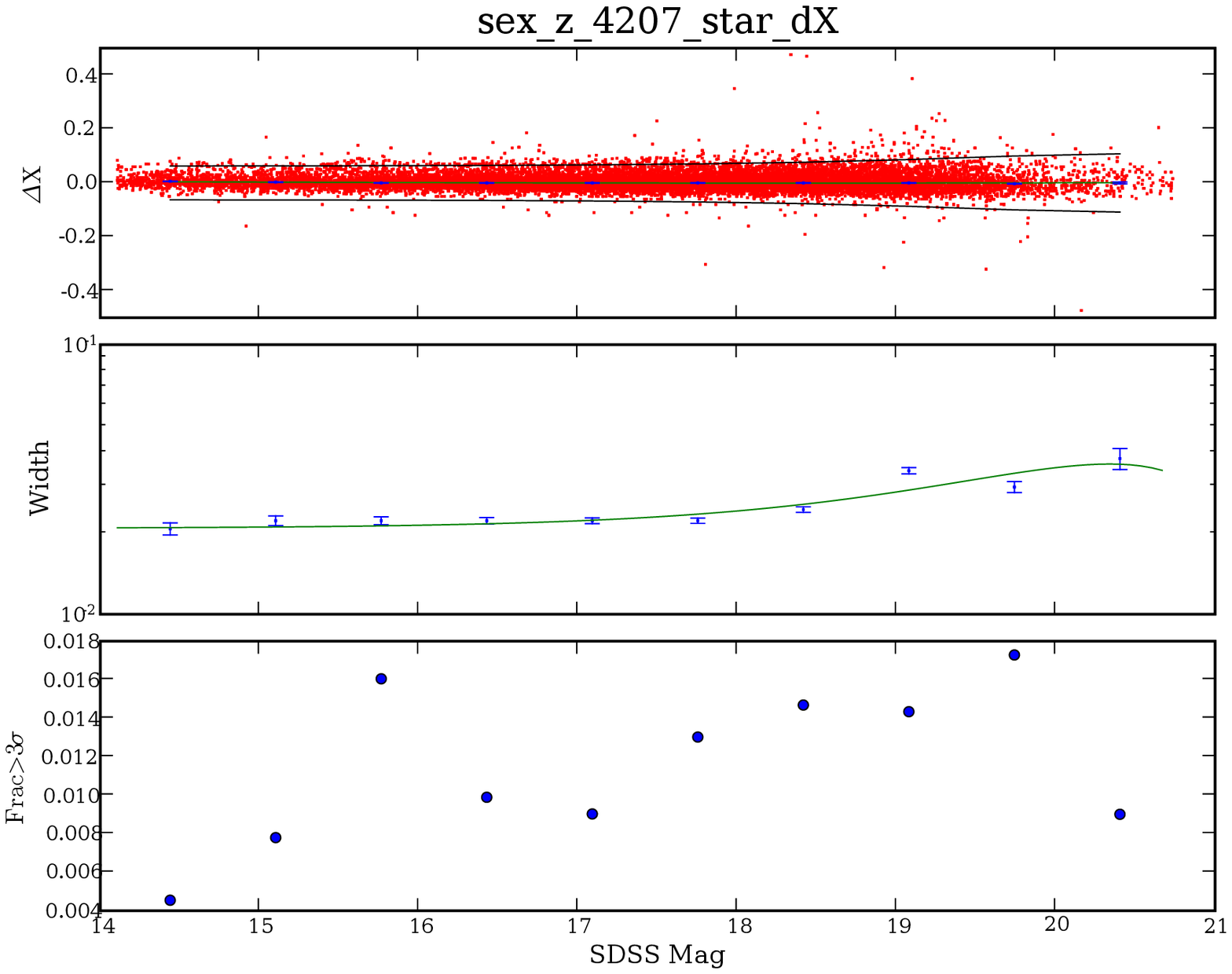}
  \end{center}
  \caption[$\Delta$XY Plots for \sex]{Differences in measured stellar
    positions between \sex\ and \photo\ plotted as a function of
    magnitude for $z$--band data from run 4207.  The $x$ coordinate is
    perpendicular to the scan direction in SDSS data.  The {\it left}
    panel shows these results for \sex\ 2.3.2, while the {\it right}
    panel shows the results for \sex\ 2.4.4.  These particular plots
    were chosen to demonstrate the improvements in centroiding between
    \sex\ versions 2.3.2 and 2.4.4.}
  \label{fig-centroid}
\end{figure*}

By comparing the calculated x,y centroids of objects to \photo's
centroids, we find very strong systematic trends in isophotal
centroiding accuracy as a function of magnitude for \sex\ version
2.3.2 (top panel of Figure~\ref{fig-centroid};
Table~\ref{tab-centroid2a}).  The windowed centroids in \sex\ version
2.4.4 and greater remedy this systematic (bottom panel of
Figure~\ref{fig-centroid}).  The centroiding RMS at the bright end
(compared to \photo) for most algorithms is 1/100 the PSF FWHM.  An
algorithm--to--algorithm comparison yields a typical centroiding RMS
of better than 1/200 the FWHM, with \photo\ the clear outlier due to
its absolute astrometry corrections (Tables~\ref{tab-centroid2c}).

The LSST relative astrometry requirement of $0.01\arcsec$ is not
likely to be violated in software.  The absolute astrometry
requirements of $0.05\arcsec$ may require corrections similar to
\photo's.

\subsection{Summary}

The one area where current algorithms do not clearly exceed the
constraints set out in LSST's SRD is in photometric accuracy.  \photo\
and \dao\ are able to deliver the requisite quality, but only in
aperture photometry, and then just at the threshold of acceptability.
Advances in PSF modeling and in wide--field aperture corrections and
sky subtraction are likely needed to ensure that the software can
deliver on the promise of LSST.
%

To summarize \photo's advantages : Its aperture photometry meets the
LSST science requirements; its PSF photometry is as good as \dao; it
is reliably able to discriminate stars from galaxies; it is the only
algorithm that does galaxy model fitting; the 5-band simultaneous
photometry is very similar to the envisioned LSST Deep Detection
analysis; and its star/galaxy deblender is robust under a variety of
conditions.  The disadvantages of \photo\ are : it is not very
flexible with respect to the format of input data, only operating on
SDSS images; the code as designed is not very portable; the deblender
is not designed for crowded fields.

To summarize \dao's advantages : Its PSF photometry is the best among
the algorithms considered here; star/galaxy separation is surprisingly
robust; it provides the best solution for point source photometry in
crowded fields; {\tt allframe} is also a useful Deep Detection
precursor algorithm.  Its disadvantages are : it is relatively slow,
and it does no galaxy characterization.

To summarize \dop's advantages : It is easily pipelined, and will take
almost any input data.  Its disadvantages are : its PSF does not vary
spatially, and it returns the poorest results with respect to both
photometry and astrometry (excluding \sex\ isophotal centroids).

Finally, to summarize \sex's advantages : It is very fast and the code
is very portable; its aperture photometry returns acceptable results;
its windowed shapes are as good as \photo's adaptive shapes; the
windowed centroids are as good as PSF centroids; the deblending model
is very extensible; and the inclusion of neural networking for object
classification is novel and potentially very powerful.  Its
disadvantages are : there is no easily accessible PSF modeling, and
the isophotal shape and positional measurements may be significantly
biased at faint magnitudes.

\acknowledgments 

We thank P. Stetson, E. Bertin, and A. Rest for valuable insights
regarding the photometry packages, and T. Axelrod and J. Kantor for
many and varied LSST Data Management discussions.  We also thank the
anonymous referee for suggestions on the content and format of this
manuscript.
LSST is a public-private partnership.  Design and development activity
is supported by in part the National Science Foundation under
Scientific Program Order No. 9 (AST-0551161) through Cooperative
Agreement AST-0132798.  Portions of this work are supported by the
Department of Energy under contract DE-AC02-76SF00515 with the
Stanford Linear Accelerator Center, contract DE-AC02-98CH10886 with
Brookhaven National Laboratory, and contract W-7405-ENG-48 with
Lawrence Livermore National Laboratory.  Additional funding comes from
private donations, grants to universities, and in-kind support at
Department of Energy laboratories and other LSSTC Institutional
Members.


\bibliographystyle{apj}
\bibliography{ms}

\clearpage


\clearpage

\begin{table}[htbp]
  \setlength{\abovecaptionskip}{0pt}
  \setlength{\belowcaptionskip}{10pt}
  \begin{center}
    \caption{\dao\ ``Sharp'' Distribution for \photo--Selected Stars}
    \label{tab-daosharp}
    \begin{tabular}{cll}
      \hline\hline
          {\em Filter}  & {\em Mean}  &  {\em RMS}  \\
          \hline
          u & 0.004 & 0.096 \\
          g & 0.001 & 0.062 \\
          r & 0.000 & 0.043 \\
          i & 0.003 & 0.045 \\
          z & 0.003 & 0.081 \\
          \hline
    \end{tabular}
  \end{center}
  Note. -- Distribution of \dao\ ``Sharp'' parameters for objects
  classified by \photo\ as stars.  We find these distributions by
  combining all data from runs 3437 and 4207.  These numbers were
  derived from the $3\sigma$ clipped distribution of {\tt Sharp}-ness
  parameters for all \dao\ measurements that were clustered with
  objects \photo\ classified as stars between $r = 14^{th}$ and $r =
  20^{th}$ magnitude.  \dao--selected stars are subsequently defined
  as anything having a sharpness within $\pm 3$ RMS of the mean.
  \dao--selected galaxies are objects with a sharpness larger than
  $+3$ RMS of the mean; objects with sharpness smaller than $-3$ RMS
  of the mean are likely cosmic rays or other defects.
\end{table}

\begin{table}[htbp]
  \setlength{\abovecaptionskip}{0pt}
  \setlength{\belowcaptionskip}{10pt}
  \begin{center}
    \caption{Object Classification; Algorithm vs. \photo}
    \label{tab-sg1}
    \begin{tabular}{ccccccc}
      \hline\hline
          {\em Algorithm} & {\em Run} & {\em Filter} & {\em S-S} & {\em S-G} & {\em G-S} & {\em G-G} \\
          \hline
          \dao\    & 3437     & g & 0.93 & 0.01 & 0.02 & 0.04 \\
          $\cdots$ & $\cdots$ & r & 0.82 & 0.01 & 0.05 & 0.12 \\
          $\cdots$ & $\cdots$ & i & 0.81 & 0.01 & 0.05 & 0.13 \\
          $\cdots$ & 4207     & g & 0.95 & 0.01 & 0.01 & 0.03 \\
          $\cdots$ & $\cdots$ & r & 0.87 & 0.01 & 0.02 & 0.09 \\
          $\cdots$ & $\cdots$ & i & 0.85 & 0.02 & 0.03 & 0.10 \\
          \hline
          \dop\    & 3437     & g & 0.93 & 0.04 & 0.00 & 0.03 \\
          $\cdots$ & $\cdots$ & r & 0.87 & 0.07 & 0.00 & 0.05 \\
          $\cdots$ & $\cdots$ & i & 0.83 & 0.13 & 0.00 & 0.04 \\
          $\cdots$ & 4207     & g & 0.96 & 0.01 & 0.00 & 0.03 \\
          $\cdots$ & $\cdots$ & r & 0.91 & 0.05 & 0.00 & 0.04 \\
          $\cdots$ & $\cdots$ & i & 0.87 & 0.09 & 0.00 & 0.03 \\
          \hline
          \sex\    & 3437     & g & 0.35 & 0.00 & 0.59 & 0.06 \\
          $\cdots$ & $\cdots$ & r & 0.57 & 0.00 & 0.28 & 0.15 \\
          $\cdots$ & $\cdots$ & i & 0.56 & 0.00 & 0.25 & 0.18 \\
          $\cdots$ & 4207     & g & 0.90 & 0.00 & 0.05 & 0.05 \\
          $\cdots$ & $\cdots$ & r & 0.83 & 0.01 & 0.02 & 0.14 \\
          $\cdots$ & $\cdots$ & i & 0.74 & 0.01 & 0.09 & 0.16 \\
          \hline
    \end{tabular}
  \end{center}
  Note. -- The fraction of total clustered objects brighter than 21st
  magnitude classified by the algorithm and \photo\ as a star (S-S);
  classified by the algorithm as a star and \photo\ as a galaxy (S-G);
  classified by the algorithm as a galaxy and \photo\ as a star (G-S);
  and classified by both the algorithm and \photo\ as a galaxy (G-G).
  This table indicates the degree of agreement between algorithms for
  a given set of data.
\end{table}

\begin{table}[htbp]
  \setlength{\abovecaptionskip}{0pt}
  \setlength{\belowcaptionskip}{10pt}
  \begin{center}
    \caption{Object Classification; Algorithm vs. Itself}
    \label{tab-sg2}
    \begin{tabular}{cccccc}
      \hline\hline
          {\em Algorithm} & {\em Filter} & {\em S-S} & {\em S-G} & {\em G-S} & {\em G-G} \\
          \hline
          \dao\    & g & 0.77 & 0.06 & 0.06 & 0.12 \\
          $\cdots$ & r & 0.65 & 0.07 & 0.06 & 0.22 \\
          $\cdots$ & i & 0.68 & 0.06 & 0.05 & 0.20 \\
          \hline
          \dop\    & g & 0.92 & 0.03 & 0.01 & 0.04 \\
          $\cdots$ & r & 0.93 & 0.02 & 0.01 & 0.03 \\
          $\cdots$ & i & 0.93 & 0.02 & 0.02 & 0.04 \\
         \hline
          \photo\  & g & 0.94 & 0.02 & 0.00 & 0.04 \\
          $\cdots$ & r & 0.90 & 0.01 & 0.00 & 0.08 \\
          $\cdots$ & i & 0.86 & 0.02 & 0.01 & 0.11 \\
          \hline
          \sex\    & g & 0.23 & 0.58 & 0.01 & 0.17 \\
          $\cdots$ & r & 0.43 & 0.35 & 0.01 & 0.21 \\
          $\cdots$ & i & 0.45 & 0.24 & 0.02 & 0.29 \\
          \hline
    \end{tabular}
  \end{center}
  Note. -- The fraction of total clustered objects brighter than 21st
  magnitude classified by the algorithm in both runs as a star (S-S);
  classified as a star in run 4207 and galaxy in 3437 (S-G);
  classified as a galaxy in run 4207 and star in 3437 (G-S); and as a
  galaxy in both runs (G-G).  This table indicates the degree of
  agreement within a given algorithm for a given set of objects.
\end{table}

\begin{table}[htbp]
  \setlength{\abovecaptionskip}{0pt}
  \setlength{\belowcaptionskip}{10pt}
  \begin{center}
    \caption{Width of $\Delta$M Distribution For \photo--Selected Stars}
    \label{tab-deltaM1}
    \begin{tabular}{llccccc}
      \hline\hline
          {\em Algorithm} & {\em Magnitude}  & {\em u}  & {\em g}  & {\em r}  & {\em i}  & {\em z}  \\
          \hline
          \dao\       & Aperture & 0.027   & {\bf 0.006}   & {\bf 0.006}   & {\bf 0.007}   & 0.015 \\
          $\cdots$    & PSF      & 0.032   & 0.018   & 0.018   & 0.017   & 0.017\\
          \hline
          \dop\       & Aperture & 0.024   & 0.009   & 0.008   & 0.008   & 0.011\\
          $\cdots$    & PSF      & 0.031   & 0.026   & 0.031   & 0.037   & 0.031\\
          \hline
          \photo\     & Aperture & 0.027   & {\bf 0.007}   & {\bf 0.006}   & {\bf 0.007}   & 0.015  \\
          $\cdots$    & PSF      & 0.029   & 0.019   & 0.019   & 0.021   & 0.019\\
          \hline
          \sex\ 2.3.2 & Aperture & 0.057   & 0.009   & 0.008   & 0.010   &  0.035 \\
          $\cdots$    & PSF      & $\cdots$ & $\cdots$ & $\cdots$ & $\cdots$ & $\cdots$  \\
          \hline
          \sex\ 2.4.4 & Aperture & 0.057   & 0.009   & 0.008   & 0.010   &  0.035 \\
          $\cdots$    & PSF      & $\cdots$ & $\cdots$ & $\cdots$ & $\cdots$ & $\cdots$  \\
          \hline
    \end{tabular}
  \end{center}
  Note. -- Characteristic widths of $\Delta$M , evaluated 1 magnitude
  below the brightest non--saturated object, representing the
  repeatability of photometric measurements of objects classified by
  \photo\ as stars, as described in Section~\ref{sec-phot}.
  Measurements compatible with LSST's science requirements (0.007
  magnitudes) are highlighted in bold.  
\end{table}

\begin{table}[htbp]
  \setlength{\abovecaptionskip}{0pt}
  \setlength{\belowcaptionskip}{10pt}
  \begin{center}
    \caption{Width of Stellar r--band $\Delta$M Distribution Algorithm to Algorithm; Aperture Magnitudes}
    \label{tab-deltaM5}
    \begin{tabular}{l|cccc}
      \hline\hline
                     & {\em \dao} & {\em \dop}  & {\em \photo} & {\em \sex 2.4.4} \\
          \hline
          \dao\      & $\cdots$   &  0.011      & 0.009        & 0.009      \\
          \dop\      & 0.007      & $\cdots$    & 0.007        & 0.010      \\
          \photo\    & 0.006      &  0.005      & $\cdots$     & 0.008      \\
          \sex\ 2.4.4& 0.007      &  0.008      & 0.005        & $\cdots$   \\
          \hline
    \end{tabular}
  \end{center}
  Note. -- Comparison of the characteristic width of $\Delta$M at the
  bright end of the distribution derived from comparisons of different
  algorithms on the same images.  The upper triangular matrix reflects
  r--band aperture measurements of \photo-selected stars seen in run
  3437, and the lower triangular for run 4207.
\end{table}

\begin{table}[htbp]
  \setlength{\abovecaptionskip}{0pt}
  \setlength{\belowcaptionskip}{10pt}
  \begin{center}
    \caption{Width of Stellar r--band $\Delta$M Distribution Algorithm to Algorithm; PSF Magnitudes}
    \label{tab-deltaM7}
    \begin{tabular}{l|ccc}
      \hline\hline
                     & {\em \dao} & {\em \dop}  & {\em \photo} \\
          \hline
          \dao\      & $\cdots$   &  0.033      & 0.018        \\
          \dop\      & 0.031      & $\cdots$    & 0.032        \\
          \photo\    & 0.018      &  0.025      & $\cdots$     \\
          \hline
    \end{tabular}
  \end{center}
  Note. -- Comparison of the characteristic width of $\Delta$M at the
  bright end of the distribution derived from comparisons of different
  algorithms on the same images.  The upper triangular matrix reflects
  r--band aperture measurements of \photo-selected galaxies seen in
  run 3437, and the lower triangular for run 4207.
\end{table}

\begin{table}[htbp]
  \setlength{\abovecaptionskip}{0pt}
  \setlength{\belowcaptionskip}{10pt}
  \begin{center}
    \caption{Width of $\Delta$M Distribution For \photo-Selected Stars; PSF vs. Aperture Magnitudes}
    \label{tab-deltaM9}
    \begin{tabular}{llccccc}
      \hline\hline
          {\em Algorithm} & {\em Run}  & {\em u}  & {\em g}  & {\em r}  & {\em i}  & {\em z}  \\
          \hline
          \dao\       & 3437 & 0.021 & 0.013 & 0.013 & 0.016 & 0.021 \\
          $\cdots$    & 4207 & 0.021 & 0.016 & 0.014 & 0.019 & 0.024 \\
          \hline
          \dop\       & 3437 & 0.022 & 0.018 & 0.024 & 0.028 & 0.030 \\
          $\cdots$    & 4207 & 0.017 & 0.018 & 0.027 & 0.034 & 0.024 \\
          \hline
          \photo\     & 3437 & 0.021 & 0.014 & 0.012 & 0.013 & 0.015 \\
          $\cdots$    & 4207 & 0.020 & 0.017 & 0.014 & 0.015 & 0.018 \\
          \hline
    \end{tabular}
  \end{center}
  Note. -- Characteristic widths representing the repeatability of
  photometric measurements of objects classified by \photo\ as stars,
  as described in Section~\ref{sec-phot}.  This table compares
  aperture vs. PSF magnitudes, and is primarily sensitive to spatial
  variations in the aperture corrections to PSF photometry.
\end{table}

\begin{table}[htbp]
  \setlength{\abovecaptionskip}{0pt}
  \setlength{\belowcaptionskip}{10pt}
  \begin{center}
    \caption{Width of $\Delta$M Distribution For Algorithm--Selected Stars}
    \label{tab-deltaM3}
    \begin{tabular}{llccccc}
      \hline\hline
          {\em Algorithm} & {\em Magnitude}  & {\em u}  & {\em g}  & {\em r}  & {\em i}  & {\em z}  \\
          \hline
          \dao\      & Aperture & 0.017   & {\bf 0.007}   & {\bf 0.007}   & 0.008   &  {\bf 0.006} \\
          $\cdots$   & PSF      & 0.030   & 0.020   & 0.018   & 0.016   &  0.017\\
          \hline
          \dop\      & Aperture & 0.017   & 0.009   & 0.009   & 0.009   &  {\bf 0.007} \\
          $\cdots$   & PSF      & 0.027   & 0.026   & 0.032   & 0.036   &  0.027\\
          \hline
          \photo\    & Aperture & 0.027   & {\bf 0.007}   & {\bf 0.006}   & {\bf 0.007}   & 0.015  \\
          $\cdots$   & PSF      & 0.029   & 0.019   & 0.019   & 0.021   & 0.019 \\
          \hline
          \sex\ 2.4.4& Aperture & 0.019   & {\bf 0.007}   & 0.009   &0.010   &  0.011 \\
          $\cdots$   & PSF      & $\cdots$ & $\cdots$ & $\cdots$ & $\cdots$ & $\cdots$  \\
          \hline
    \end{tabular}
  \end{center}
  Note. -- We repeat the analysis summarized in
  Table~\ref{tab-deltaM1} but instead use the algorithm's
  classification scheme instead of \photo's
  (Section~\ref{sec_ana-stargal}).  Objects must be classified as
  stars in both runs.  \photo\ results are the same as in
  Table~\ref{tab-deltaM1}.
\end{table}

\begin{table}[htbp]
  \setlength{\abovecaptionskip}{0pt}
  \setlength{\belowcaptionskip}{10pt}
  \begin{center}
    \caption{Comparison of Stellar $r$-band Ellipticities}
    \label{tab-e_star_sex}
    \begin{tabular}{cccrrr}
      \hline\hline
          {\em Ellipticity}     & {\em Algorithm} & {\em Run} & {\em RMS} & {\em Intercept} & {\em Slope} \\
          \hline
          e1 & \sex\ 2.3.2 & 3437 & 0.026 & 0.020  & 0.406 \\
          e2 & $\cdots$    & 3437 & 0.021 & -0.033 & 0.447 \\
          e1 & $\cdots$    & 4207 & 0.034 & -0.051 & 0.393 \\
          e2 & $\cdots$    & 4207 & 0.030 & 0.014  & 0.420 \\
          \hline
          e1 & \sex\ 2.4.4 & 3437 & 0.002 & -0.003 & 2.046 \\
          e2 & $\cdots$    & 3437 & 0.001 & -0.000 & 2.060 \\
          e1 & $\cdots$    & 4207 & 0.004 & -0.016 & 2.141 \\
          e2 & $\cdots$    & 4207 & 0.002 & 0.001  & 2.181 \\
          \hline
    \end{tabular}
  \end{center}
  Note. -- Comparison of \photo\ and \sex\ $r$-band ellipticity
  measures for \photo-selected stars with $14 < r < 20$.  We fit a
  line to the relationship and evaluate the RMS perpendicular to the
  principal axis.  \sex\ 2.3.2 uses ``isophotal'' shape measures, and
  \sex\ 2.4.4 ``windowed'' shape measures.
\end{table}

\begin{table}[htbp]
  \setlength{\abovecaptionskip}{0pt}
  \setlength{\belowcaptionskip}{10pt}
  \begin{center}
    \caption{Comparison of Galaxy $r$-band Ellipticities}
    \label{tab-e_gal_sex}
    \begin{tabular}{cccrrr}
      \hline\hline
          {\em Ellipticity}     & {\em Algorithm} & {\em Run} & {\em RMS} & {\em Intercept} & {\em Slope} \\
          \hline
          e1 & \sex\ 2.3.2 & 3437 & 0.036 & 0.005  & 0.987 \\
          e2 & $\cdots$    & 3437 & 0.037 & -0.002 & 0.976 \\
          e1 & $\cdots$    & 4207 & 0.037 & -0.001 & 0.976 \\
          e2 & $\cdots$    & 4207 & 0.038 & 0.004  & 0.973 \\
          \hline
          e1 & \sex\ 2.4.4 & 3437 & 0.016 & 0.005  & 1.834 \\
          e2 & $\cdots$    & 3437 & 0.015 & -0.004 & 1.848 \\
          e1 & $\cdots$    & 4207 & 0.016 & -0.001 & 1.825 \\
          e2 & $\cdots$    & 4207 & 0.017 & 0.002  & 1.842 \\
          \hline
    \end{tabular}
  \end{center}
  Note. -- Same as Table~\ref{tab-e_star_sex}, but for \photo-selected galaxies.
\end{table}

\begin{table*}[htbp]
  \setlength{\abovecaptionskip}{0pt}
  \setlength{\belowcaptionskip}{10pt}
  \begin{center}
    \caption{Centroiding Offsets (in Pixels) for Stars as a Function of Magnitude}
    \label{tab-centroid2a}
    \begin{tabular}{cccrrrrrrr}
      \hline\hline
          {\em Algorithm} & {\em Run} & {\em Filter} & {\em M0} & {\em A$_x$} & {\em B$_x$} & {\em C$_x$} & {\em A$_y$} & {\em B$_y$} & {\em C$_y$}\\
          \hline
          \dao\    & 3437     & u & 16.41 & 0.000 & 0.002 & 0.000 & -0.001 & 0.000 & -0.000 \\
          $\cdots$ & $\cdots$ & g & 15.29 & 0.002 & -0.001 & 0.000 & -0.003 & 0.005 & -0.001  \\
          $\cdots$ & $\cdots$ & r & 14.79 & -0.000 & 0.000 & -0.000 & -0.005 & 0.004 & -0.001 \\
          $\cdots$ & $\cdots$ & i & 14.61 & -0.000 & 0.000 & -0.000 & -0.003 & 0.003 & -0.001 \\
          $\cdots$ & $\cdots$ & z & 14.44 & 0.003 & -0.003 & 0.000 & -0.001 & 0.002 & -0.000  \\
          \hline
    \end{tabular}
  \end{center}
  Note.  -- Table~\ref{tab-centroid2a} is published in its entirety in the
  electronic edition of the PASP. A portion is shown here for guidance
  regarding its form and content.

         -- Results of the analysis described in
  Section~\ref{sec_ana-centroid} for \photo-selected stars.
  Coefficients subscripted $x$ are for the x--axis offsets, $y$ are
  for the y--axis.  This analysis tests systematics in centroiding as
  a function of magnitude.

\end{table*}

\begin{table}[htbp]
  \setlength{\abovecaptionskip}{0pt}
  \setlength{\belowcaptionskip}{10pt}
  \begin{center}
    \caption{$r$--band Centroiding RMS$_x$ (in Pixels) for \photo-Selected Stars; Algorithm vs Algorithm}
    \label{tab-centroid2c}
    \begin{tabular}{l|cccc}
      \hline\hline
                     & {\em \dao} & {\em \dop}  & {\em \photo} & {\em \sex 2.4.4} \\
          \hline
          \dao\      & $\cdots$   &  0.008      & 0.029        & 0.007      \\
          \dop\      & 0.011      & $\cdots$    & 0.024        & 0.004      \\
          \photo\    & 0.030      &  0.021      & $\cdots$     & 0.024      \\
          \sex\ 2.4.4& 0.011      &  0.007      & 0.021        & $\cdots$   \\
          \hline
    \end{tabular}
  \end{center}
  Note. -- Width of the stellar positional offset distribution
  evaluated 1 magnitude below the brightest object.  The upper
  triangular matrix reflects r--band measurements of \photo-selected
  stars in run 3437, and the lower triangular for run 4207.
\end{table}

\begin{table}[htbp]
  \setlength{\abovecaptionskip}{0pt}
  \setlength{\belowcaptionskip}{10pt}
  \begin{center}
    \caption{Comparison of Photometric Depth}
    \label{tab-depth}
    \begin{tabular}{lllccc|cc}
      \hline\hline
          {\em Run} & {\em Filter} & {\em Magnitude} & {\em \dao} & {\em \dop} &{\em \photo} & {\em \sex} & {\em \photo$^*$} \\
          \hline
          3437     & u        & M$_{max}$ & 20.36 & 20.91 & 22.00 & 19.99 & 19.99  \\
          $\cdots$ & $\cdots$ & M$_{95}$  & 21.63 & 21.81 & 22.54 & 21.98 & 22.16 \\
          $\cdots$ & $\cdots$ & M$_{99}$  & 21.99 & 21.99 & 23.45 & 23.25 & 23.07 \\
          \hline
          $\cdots$ & g        & M$_{max}$ & 20.68 & 22.34 & 22.55 & 19.44 & 20.27  \\
          $\cdots$ & $\cdots$ & M$_{95}$  & 21.71 & 22.75 & 22.96 & 21.10 & 22.75 \\
          \hline
    \end{tabular}
  \end{center}
  Note. -- Table~\ref{tab-depth} is published in its entirety in the
  electronic edition of the PASP. A portion is shown here for guidance
  regarding its form and content.

        -- Comparison of the photometric depths of each algorithm.  We
  use three numbers to characterize this quantity.  M$_{max}$
  represents the maximum of the measured star count histogram; M$_{95}$
  is the bin below which 95\% of the stars are contained; M$_{99}$ is
  the bin below which 99\% of the stars are.  PSF magnitudes are used
  to compare \dao, \dop, and \photo.  For \sex, we use \photo's
  aperture magnitudes, listed as \photo$^*$, for comparison.
\end{table}

\begin{table}[htbp]
  \setlength{\abovecaptionskip}{0pt}
  \setlength{\belowcaptionskip}{10pt}
  \begin{center}
    \caption{Width of $\Delta$M Distribution For Algorithm--Selected Stars in M2}
    \label{tab-M2deltaM}
    \begin{tabular}{llccccc}
      \hline\hline
          {\em Algorithm} & {\em Magnitude}  & {\em u}  & {\em g}  & {\em r}  & {\em i}  & {\em z}  \\
          \hline
          {\tt daophot}  & Aperture & 0.015   & 0.013   & 0.036   & 0.029   & 0.016  \\
          $\cdots$       & PSF      & 0.024   & 0.032   & 0.018   & 0.015   & 0.016  \\
          {\tt allframe} & Aperture & 0.026   & 0.012   & 0.036   & 0.031   & 0.017 \\
          $\cdots$       & PSF      & 0.018   & 0.020   & 0.014   & 0.011   & 0.011  \\
          \hline
          {\tt allframe} & Aperture & 0.039   & 0.024   & 0.046   & 0.045   & 0.023 \\
          $\cdots$       & PSF      & 0.020   & 0.028   & 0.018   & 0.014   & 0.012  \\
          \hline
    \end{tabular}
  \end{center}
  Note. -- We repeat the analyses summarized in
  Section~\ref{sec-phot} for globular cluster M2.  We restrict
  our analyses to the algorithms {\tt daophot} and {\tt allframe}.  The
  first set of {\tt allframe} results correspond to objects classified
  by {\tt daophot} as stars.  The second set correspond to objects
  classified by {\tt allframe} as stars.
\end{table}

\begin{table}[htbp]
  \setlength{\abovecaptionskip}{0pt}
  \setlength{\belowcaptionskip}{10pt}
  \begin{center}
    \caption{Width of $\Delta$M Distribution in M2 as a Function of Crowding}
    \label{tab-M2deltaMvsCrowding}
    \begin{tabular}{lcc}
      \hline\hline
          {\em Filter} & {\em Intercept}  & {\em Slope}  \\
          \hline
          u & 0.020 & 0.121 \\
          g & 0.018 & 0.134 \\
          r & 0.008 & 0.103 \\
          i & 0.008 & 0.077 \\
          z & 0.007 & 0.050 \\
          \hline
    \end{tabular}
  \end{center}
  Note. -- We repeat the analyses summarized in Section~\ref{sec-phot}
  for globular cluster M2, this time plotting $\Delta$M as a function
  of crowding conditions in the image.  The $r$--band data are plotted
  in Figure~\ref{fig-M2deltaMvsCrowding}.
\end{table}

\begin{table}[htbp]
  \setlength{\abovecaptionskip}{0pt}
  \setlength{\belowcaptionskip}{10pt}
  \begin{center}
    \caption{Comparison of Photometric Depth in M2}
    \label{tab-M2depth}
    \begin{tabular}{lllcc}
      \hline\hline
          {\em Run} & {\em Filter} & {\em Magnitude} & {\em {\tt daophot}} & {\em {\tt allframe}} \\
          \hline
          3437     & u        & M$_{max}$ & 20.25 & 20.65 \\
          $\cdots$ & $\cdots$ & M$_{95}$  & 21.84 & 22.44 \\
          $\cdots$ & $\cdots$ & M$_{99}$  & 22.24 & 22.84 \\
          \hline
          $\cdots$ & g        & M$_{max}$ & 20.80 & 20.80 \\
          $\cdots$ & $\cdots$ & M$_{95}$  & 22.14 & 22.91 \\
          \hline
    \end{tabular}
  \end{center}
  
  Note.  -- Table~\ref{tab-M2depth} is published in its entirety in the
  electronic edition of the PASP. A portion is shown here for guidance
  regarding its form and content.

         -- Comparison of the photometric depths of each algorithm.
  We use three numbers to characterize this quantity.  M$_{max}$
  represents the maximum of the measured star count histogram;
  M$_{95}$ is the bin below which 95\% of the stars are contained;
  M$_{99}$ is the bin below which 99\% of the stars are.  PSF
  magnitudes are used in this comparison.
\end{table}

\begin{table}[htbp]
  \setlength{\abovecaptionskip}{0pt}
  \setlength{\belowcaptionskip}{10pt}
  \begin{center}
    \caption{Algorithm Processing Time as a Function of The Number of Sources}
    \label{tab-timefit}
    \begin{tabular}{llccc}
      \hline\hline
          {\em Algorithm} & {\em SDSS Run} & {\em Slope}       & {\em y-Intercept} \\
          {\em }          & {\em }         & {\em (sec/\#Det)} & {\em (sec)} \\
          \hline \hline
          \dao\ & 3437 & 0.260 & 10 \\
                & 4207 & 0.090 & 170 \\
          \hline
          \dop\ & 3437 & 0.025 & 101 \\
                & 4207 & \nodata & \nodata \\
          \hline
          \sex\ v2.3.2 & 3437 & 0.010 & 4.2 \\
                       & 4207 & 0.001 & 4.3 \\
          \hline
          \sex\ v2.4.4 & 3437 & 0.001 & 4.5 \\
                       & 4207 & 0.001 & 4.7 \\
          \hline \hline
    \end{tabular}
  \end{center}
  Note. -- Scaling of processing time with the number of sources in
  the images.  We determine the time it takes each algorithm to
  process one image versus the number of sources detected in that
  image.  We find the linear trend with source number, listing here
  the slope and intercept.  
\end{table}

\begin{table*}[htbp]
\setlength{\abovecaptionskip}{0pt}
\setlength{\belowcaptionskip}{10pt}
\begin{center}
\caption{\sex\ Profiling}
\label{sexds}
{\footnotesize
\begin{tabular}{llcrllrr}
\hline\hline
{\em Image} & {\em NObj} & {\em Size} & {\em BITPIX} & {\em VmSize kB} & {\em VmR
SS kB} & {\em Time (s)} & {\em RMS (s)} \\
\hline
         r-003437\_0170        & 750   & 2k x 2k &  16 & 26134 (3.1) & 16552 (2.0) & 0.95 & 0.01 \\
                               &       &         &  32 & 25943 (1.5) & 16435 (1.0) & 0.98 & 0.02 \\
                               &       &         & -32 & 26015 (1.5) & 16451 (1.0) & 0.97 & 0.01 \\
                               &       &         & -64 & 26107 (0.8) & 16488 (0.5) & 1.09 & 0.02 \\
                               & 1619  & 2k x 4k &  16 & 26724 (1.6) & 16698 (1.0) & 1.92 & 0.02 \\
\hline
\end{tabular}
}
\end{center}

   Note. -- Table~\ref{sexds} is published in its entirety in the
electronic edition of the PASP. A portion is shown here for guidance
regarding its form and content. 

         -- Average memory usage and processing time of \sex\ as a
function of image size, bit depth, and number of sources.  VmSize and
VmRSS show the average {\it maximum} memory used; in parenthesis is
this number as a fraction of the image size.  The average total
processing time and its RMS are also listed.
\end{table*}

\clearpage
\section{Appendix}

\subsection{The SDSS Photometric Pipeline: \photo \label{sec-alg_photo}}

The SDSS photometric pipeline \photo\ contains a complete suite of
data reduction tools that take the raw data stream, apply reduction
and calibration stages, and extract photometry from the calibrated
images.  Because the images we are using have been pre--processed by
\photo, we expect that \photo\ has a distinct advantage in the quality
of its photometric measurements.

In \photo, the data stream from each CCD (drift-scanning results in an
``infinitely'' long narrow image) is divided into an overlapping
series of 10\arcmin by 13\arcmin frames for ease of processing.  A
\photo\ module named {\tt frames} processes each of these separately.
However, in order to ensure continuity along the data stream, certain
quantities need to be determined on timescales up to the length of the
imaging run.  The astrometric and photometric calibrations certainly
fall into that category; in addition, a \photo\ module named the {\tt
postage-stamp pipeline (PSP)} calculates a global sky for a field,
flat-field vector, bias level, and the PSF.  Once these are provided,
a {\tt frames} run can be trivially parallelized.

\subsubsection{The Point Spread Function in \photo}

Even in the absence of atmospheric inhomogeneities the SDSS telescope
delivers images whose FWHMs vary by up to 15\% from one side of a CCD
to the other; the worst effects are seen in the chips furthest from
the optical axis.  Since the atmospheric seeing is not constant in
time, the delivered image quality is a complex two-dimensional
function.  The description of the PSF is critical for accurate PSF
photometry, for star/galaxy separation and for studies that measure
the shapes of non-stellar objects.

The SDSS imaging point spread function (PSF) is modeled heuristically
in each band using a Karhunen--Loeve (K--L) transform.  In particular,
using stars brighter than roughly 20th magnitude, the stellar images
from a series of five frames are expanded into eigenimages and the
first three terms are kept.  The variation of the coefficients that
multiply these terms with position across the chip is described by a
low-order polynomial.

The success of this K--L expansion depends critically on successful
selection of PSF stars.  In essence, to determine the PSF one needs to
select stars that look like the PSF, a requirement that results in
somewhat convoluted selection procedure.

The selection of PSF stars is done in two steps.  In the first crude
step stars that are grossly inadequate are rejected based on their
individual properties (i.e. without considering the overall sample
properties).  This category includes objects that are too faint, those
with saturated or cosmic ray pixels, objects with very close
neighbors, and significantly elongated objects (star/galaxy
information is not yet available at this processing stage).  In the
second step the distribution of image size and ellipticity is used to
reject stars that significantly deviate ($\sim3\sigma$ or more) from
the median.  Typically about 50\% of bright objects ($r<19$) survive
both rejection steps.

\subsubsection{Object Detection and Measurement in \photo}

Objects in the frame are detected and their properties measured in a
four-step process in each band.  First, an object finder is run to
detect bright objects.  In each band, the object finder detects pixels
that are more than 200$\sigma$ (corresponding roughly to $r = 17.5$)
above the sky noise; only a single pixel need be over this threshold
for an object to be detected at this stage.  These objects are flagged
as BRIGHT.  The extended power-law wings of BRIGHT objects that are
saturated are subtracted from the frame.  Such stars are marked
SUBTRACTED.  Then the sky level is estimated by median-smoothing the
frame image on a scale of approximately 100\arcsec; the resulting
``local'' sky image is subtracted from the frame (a global sky
determined on an entire frame has already been subtracted).

Third, objects are found by smoothing the image with a Gaussian fit to
the PSF and looking for 5$\sigma$ peaks over the (smoothed) sky in
each band.  After objects are detected, they are ``grown'' more or
less isotropically by an amount approximately equal to the radius of
the seeing disk.  An object is defined as a connected set of pixels
that are detected in at least one band.  All pixels in the object are
subsequently used in the analysis in every band, whether or not they
were originally detected in that band.  \photo\ never reports an upper
limit for the detection of an object but, rather, carries out a proper
measurement, with its error, for each of the varieties of flux listed
below.

Objects detected in a given band at this stage are flagged by setting
the mask bit BINNED1 in that band.  All pixel values in these BINNED1
objects are then replaced by the background level (with sky noise
added in), the frame is rebinned into a $2 \times 2$ pixel image, and
the object finder is run again.  The resulting sample is flagged in a
similar way with the BINNED2 mask, and pixel values in these objects
are replaced with the background level.  Finally, the original pixel
data is rebinned in a 4 $\times 4$ pixel image, and objects found at
this stage are flagged BINNED4.  The set of detected objects then
consists of all objects with pixels flagged BINNED1, BINNED2, or
BINNED4.

Fourth, the pipeline measures the properties of each object, including
the position, as well as several measures of flux and shape, described
more fully below.  It attempts to determine whether each object
actually consists of more than one object projected on the sky and, if
so, to deblend such a "parent" object into its constituent
``children'', self-consistently across the bands (thus, all children
have measurements in all bands).  Then it again measures the
properties of these individual children.  Bright objects are measured
twice: once with a global sky and no deblending run -- this detection
is flagged BRIGHT -- and a second time with a local sky.  For most
purposes, only the latter is useful, and thus one should reject all
objects flagged BRIGHT in compiling a sample of objects for study.

\subsubsection{Photometric Measurements in \photo}

There are several magnitude types provided by \photo, and all are
measured for all the detected sources.

\paragraph{PSF Magnitudes}

For isolated stars, which are well described by the PSF, the optimal
measure of the total flux is determined by fitting a PSF model to the
object.  In practice, this is done by sinc-shifting the image of a
star so that it is exactly centered on a pixel and then fitting a
Gaussian model of the PSF to it.  This fit is carried out on the local
PSF K--L model at each position as well; the difference between the
two is then a local aperture correction, which gives a corrected PSF
magnitude.  Finally, bright stars are used to determine a further
aperture correction to a radius of 7.4\arcsec as a function of seeing.
This involved procedure is necessary to take into account the full
variation of the PSF across the field, including the low
signal-to-noise ratio wings.  Empirically, this reduces the seeing
dependence of the photometry to below 0.02 mag for seeing as poor as
2\arcsec.

The PSF magnitude errors include contributions from photon statistics
and uncertainties in the PSF model and aperture correction.  Repeat
observations show that these errors are probably underestimated by
10\%.

\paragraph{Petrosian Magnitudes}

For galaxy photometry, measuring flux is more difficult than for
stars, because galaxies do not all have the same radial surface
brightness profile, and they have no sharp edges.  In order to avoid
biases, one wishes to measure a constant fraction of the total light,
independent of the position and distance of the object.  To satisfy
these requirements, the SDSS has adopted a modified form of the
\citet{Petro76} system, measuring galaxy fluxes within a circular
aperture whose radius is defined by the shape of the azimuthally
averaged light profile \citep[see][for more details]{Stou02}.

\paragraph{Model Magnitudes}

Just as the PSF magnitudes are optimal measures of the fluxes of
stars, the optimal measure of the flux of a galaxy would use a matched
galaxy model.  With this in mind, the code fits two models to the
two-dimensional image of each object in each band: a pure de
Vaucouleurs profile, and a pure exponential profile.  The models are
convolved with a double-Gaussian fit to the PSF.  Residuals between
the double-Gaussian and the full K--L PSF model are added on for just
the central PSF component of the image.

In order to measure unbiased colors of galaxies, their flux is
measured through equivalent apertures in all bands.  The model
(exponential or de Vaucouleurs) of higher likelihood in the $r$ filter
is applied (allowing only the amplitude to vary) in the other bands
after convolving with the appropriate PSF in each band.  The resulting
magnitudes are called model magnitudes.  The resulting estimate of
galaxy color is unbiased in the absence of color gradients.
Systematic differences from Petrosian colors are in fact often seen as
a result of color gradients, in which case the concept of a global
galaxy color is somewhat ambiguous.  For faint galaxies, the model
colors have appreciably higher signal-to-noise ratio than do the
Petrosian colors.

\subsubsection{Star/Galaxy Separation in \photo}

A simple star-galaxy separator, that works at the 95\% confidence
level to at least $r = 21$, is based on a difference between psf and
model magnitudes: ``unresolved'' objects are those with this
difference smaller than 0.145 mag.  This separation is done in each
band separately, and again globally based on the summed fluxes from
all bands in which the object is detected.

Experimentation has shown that simple variants on this scheme, such as
defining galaxies as those objects classified as such in any two of
the three high signal-to-noise ratio bands (namely, $g$, $r$, and
$i$), work better in some circumstances.  However, this scheme
occasionally fails to distinguish pairs of stars with separation small
enough ($<2$ \arcsec) that the deblender does not split them; it also
occasionally classifies Seyfert galaxies with particularly bright
nuclei as stars.

\subsubsection{Image Ellipticity}

While the model fits yield an estimate of the axis ratio and position
angle of each object, it is useful to have model-independent measures
of ellipticity.  Two further measures of ellipticity are computed by
{\tt frames}, one based on second moments, the other based on the
ellipticity of a particular isophot.  The model fits do correctly
account for the effect of the seeing, while these two methods do not.

The first method measures flux-weighted second moments, defined in
\citet{Stou02}.  This method is not ideal at low signal-to-noise
ratio.  A second measure of ellipticity is given by measuring the
ellipticity of the 25 mag per square arcsec isophot (in all bands).
In detail, {\tt frames} measures the radius of a particular isophot as
a function of angle and Fourier-expands this function.  It then
extracts from the coefficients the centroid, major and minor axes,
position angle, and average radius of the isophot in question.  It
also reports the derivative of each of these quantities with respect
to isophot level, necessary to recompute these quantities if the
photometric calibration changes.

\subsubsection{ The Deblender }

Once objects are detected, they are deblended by identifying
individual peaks within each object, merging the list of peaks across
bands, and adaptively determining the profile of images associated
with each peak, which sum to form the original image in each band.
The originally detected object is referred to as the ``parent'' object
and has the flag BLENDED set if multiple peaks are detected; the final
set of subimages of which the parent consists are referred to as the
``children'' and have the flag CHILD set.  All quantities are measured
for both parent and child.  For each child, parent gives the id of the
parent (for parents themselves or isolated objects, this is set to the
id of the BRIGHT counterpart if that exists; otherwise it is set to
$-1$); for each parent, nchild gives the number of children an object
has.  Children are assigned the id numbers immediately after the id of
the parent.  Thus, if an object with id 23 is set as BLENDED and has
nchild equal to 2, objects 24 and 25 will be set as CHILD and have
parent equal to 23.

The list of peaks in the parent is trimmed to combine peaks (from
different bands) that are too close to each other (if this happens,
the flag PEAKS\_TOO\_CLOSE is set in the parent).  If there are more
than 25 peaks, only the most significant are kept, and the flag
DEBLEND\_TOO\_MANY\_PEAKS is set in the parent.

In a number of situations, the deblender decides not to process a
BLENDED object; in this case the object is flagged as NODEBLEND.  Most
objects with EDGE set are not deblended.  The exceptions are when the
object is large enough (larger than roughly an arcminute) that it will
most likely not be completely included in the adjacent scan line
either; in this case, DEBLENDED\_AT\_EDGE is set, and the deblender
gives it its best shot.  When an object is larger than half a
frame,the deblender also gives up, and the object is flagged as
TOO\_LARGE.  Other intricacies of the deblending results are also
recorded in flags \citep[see][for more details]{Stou02}.

On average, about 15\%--20\% of all detected objects are blended, and
many of these are superpositions of galaxies that the deblender
successfully treats by separating the images of the nearby objects.
Thus, it is usually the childless (not BLENDED) objects that are of
most interest for science applications.

\subsubsection{ Astrometry }
\label{subsec-photo-astrom}

The SDSS astrometric pipeline, including treatment of chromatic
aberration and improved centroiding, is described in detail by
\cite{2003AJ....125.1559P}.  Of particular relevance here are
centroiding corrections that are similar in spirit to aperture
corrections for psf magnitudes.  A centroid correction (the difference
in position estimate between an approximate quartic method and true
centroid) is found using a high S/N PSF estimate, and then applied to
low S/N objects.  This correction may be as high as 1/4 of a pixel and
is applied {\it in situ}.  For this reason, it is expected that
photo's centroids will not perfectly agree with centroids determined
by other algorithms.

\subsection{\dao}

The \dao\ package contains a set of photometry algorithms primarily
designed to do stellar photometry and astrometry in crowded fields.
The tools are included as either subroutines in the executable {\tt
daophot} or as independent executable programs.  The programs are
typically used in the following groupings : {\tt daophot}\footnote{We
use the following conventions : when referring to \dao\ as a package
we will capitalize the name; when referring to the executable {\tt
daophot} we will use lower--case.}  and {\tt allstar}; and {\tt
daomatch}, {\tt daomaster}, {\tt montage2}, and {\tt allframe}.  These
programs are defined below.

\begin{itemize}

\item {\tt daophot} : Main executable program.  Typically used to
find stellar objects, perform aperture photometry, and derive a PSF
for the image from a selected set of stars.  The PSF--building task is
the most complex, and is highly iterative.  No accommodations are made
for the measurement of extended sources.

\item {\tt allstar} : Run in conjunction with {\tt daophot}.  Accepts
the results of {\tt daophot}'s photometry and PSF--building stages and
performs a multiple--profile PSF fit to stars in the image
simultaneously, optimally deblending neighbors and merging detections
if they are determined to be the same object.  {\tt Allstar} groups
objects for a joint fit based upon their proximity, thus does not
literally photometer the entire image at once.  This program
automatically undertakes an iterative process of merging stars in the
input star list based upon a signal--to--noise criterion, rejecting
bad objects, and re--fitting each group's centroids and brightnesses
until all objects have converged (or a certain number of iterations
are reached).  In practice, this package is used to yield the
``final'' photometry and astrometry for a single image.

\item {\tt daomatch} : If multiple images of a field have been
acquired (either in different filters or on different dates), {\tt
daomatch} may be used to determine a basic geometric transformation
(offset, scaling, and rotation) between the star lists.

\item {\tt daomaster} : Takes the output of {\tt daomatch} (an ensemble
of geometric transformations, one for each science image) and performs
a joint registration of the star lists, rejecting spurious matches and
enforcing a common list of stars in all images for the match.  The
transformations may be of higher order (up to cubic) than in {\tt
daomatch}.  {\tt Daomaster} also returns the list of common stars that
are present in a user--defined fraction of the images, up to a
user--defined matching radius, as well as the geometric
transformations derived from these stars.

\item {\tt montage2} : Takes the transformations from {\tt daomaster}
and makes a stacked image.  The user decides which percentile from the
ensemble of (sky-subtracted) input pixels yields the stacked image
(i.e. 0.5 = median).  The image weights scale as $({\rm Depth}/{\rm
FWHM})^{2}$.  Pixels are resampled using nearest-neighbor
interpolation, and the resulting images are {\it not} to be considered
``science grade''.  This step is typically done {\it after} {\tt
allframe} is run, where it is used to coadd star--subtracted images to
search for faint objects that were originally missed.

\item {\tt allframe} : Takes the master star list and geometric
transformations derived from {\tt daomaster} and performs simultaneous
PSF photometry on a given group of objects in the entire stack of
images.  This package is essentially a 3--dimensional version of {\tt
allstar}.  {\tt Allframe} mirrors in many aspects the envisioned LSST
Image Processing Pipeline (IPP) in regards to stellar photometry
\citep{2005AAS...207.2629B}.

\end{itemize}

The executable {\tt daophot} is designed to be command--line driven,
and in fact places the user in a small data processing environment.
For this reason, is has proven difficult to turn this into an
automated pipeline.  In particular, the generation of the
point--spread function in \dao\ is a highly iterative process
involving many stages.  We have chosen to use {\tt Perl}--language
scripts to automate this process (Section~\ref{dao-sec_perl}).

\subsubsection{How \dao\ is Written}

The \dao\ package is written in the language {\tt FORTRAN}. It
requires the {\tt cfitsio} libraries, as well various as {\tt IRAF}
libraries.  The code itself is very well documented, and in fact much
of what we have learned about how it operates at the algorithmic level
was derived straight from the {\tt FORTRAN} code.

However, the code also contains many hard--wired variables, and thus
is not flexible enough for LSST's needs as implemented.  Two prime
examples that caused us difficulties are the maximum number of PSF
stars allowed in the PSF model ({\tt MAXN}), which had to be changed
in two places in the file {\tt psf.f} (one apparent, one not), as well
as the maximum filename length allowed by \dao\ (including the
absolute path to the file), which was hard coded in enough places that
it was unfeasible to change them all.  As a workaround, during actual
\dao\ reductions we made a copy of each image in the {\tt /tmp/}
directory, operated on the file there, and then copied the derived
data products back into the pipeline workspace.  In addition, if one
wanted to change other variables such as the maximum image size or
maximum number of images to reduce, one has to edit a file and then
recompile the binaries.

\subsubsection{How \dao\ is Designed to be Used}

\dao\ is better described as a toolkit than a pipeline.  In fact, it
has been designed to be a user--interactive environment.  This is
particularly true for the generation of the PSF model, where the user
is encouraged to manually review each star that has been input to the
PSF generation section.  While this toolkit comprises many tools, we
only review the most relevant ones here.

\begin{itemize}

\item {\tt SKY} : \dao\ uses the following algorithm to estimate
the global sky value in the image for the purposes of object
detection: 10000 pixels are chosen uniformly distributed across the
image; the tails of this distribution are clipped; the mode is
estimated as $3
\times {\rm median} - 2 \times {\rm mean}$; and the RMS is derived from
the $1-\sigma$ width of the sky histogram about the mean.

\item {\tt FIND} : Based upon the user--input {\tt readnoise} and
{\tt gain} relevant for each image, and sky as derived above, \dao\
will compute the random error per pixel.  This value is normalized by
the inverse square root of : [sum of the squares of the values] - [the
square of the sum of the values] of a bivariate circular Gaussian
function with unit height and the user--supplied value of the
estimated FWHM.  This yields the estimated random noise in the
Gaussian--convolved background image.  A user--defined multiple of
this value is used as the star detection threshold.  This represents
the minimum {\it central} height above the local sky for an object to
be considered significant, {\it not} the integrated signal from the
entire detection.

\item {\tt PHOT} : This subroutine performs aperture photometry on a
list of stars.  In this process, {\it all} stars are subtracted from
the image (using the current PSF model), and each star is individually
added in turn to the image to estimate its aperture flux.  The user
chooses apertures for measurement, as well as an inner and outer
radius for local sky determination (determined in a manner similar to
{\tt SKY}).  A circular aperture is approximated by an irregular
polygon by only accepting fractions of the flux in each boundary
pixel, with a linear fractional flux scaling between 1 and 0 for
pixels within -0.5 and +0.5 of the aperture radius, respectively.  In
addition, {\tt PHOT} performs an azimuthal smoothing within each
annulus bounded by neighboring apertures to recognize hot pixels : if
a given pixel is discrepant relative to the mean and dispersion of
other pixels within the same annulus, the discrepant pixel value is
replaced by a weighted average of the pixel value and the mean value
for the annulus.  This is useful for ``curve of growth'' corrections
but not directly relevant to our analysis here.  If the photometry
process fails (e.g. the modal sky could not be determined, or there is
a bad pixel in the aperture) the magnitude error is set to 9.999.
Uncertainties in the magnitudes for good objects contain terms from :
random noise inside the star aperture, including readout noise and
contamination by other stars in the neighborhood, estimated by the
scatter in the sky values (this term increases as the square root of
the area of the aperture); the Poisson statistics of the observed star
brightness; and the uncertainty of the mean sky brightness (which
increases directly with the area of the aperture).

\item {\tt PICK} : This subroutine chooses good candidates for
PSF stars based upon their distance from the edge of the frame and
local crowding conditions.  In particular, stars near brighter stars
or within a user--defined threshold distance are rejected.  If at
least 3 apertures are specified in the {\tt PHOT} stage, {\tt PICK}
will use $M2 - M3$ as well as $M1 - M2$ to choose objects, under the
assumption that $M2 - M3$ will be larger for extended objects than for
stars.  In principle we could use \photo--selected stars for this
process, but have decided to allow \dao\ to select them.

\item {\tt PSF} : The use of this procedure is complex enough that
we address it in detail in Section~\ref{sec-dao_psf} and
Section~\ref{sec-dao_use}.  In summary, this routine takes a list of
objects (e.g. those selected by {\tt PICK}) and builds a model of the
point--spread function.

\item {\tt SUBSTAR} : This subroutine accepts an input list of
objects, scales and shifts the PSF according to
each star's magnitude and centroid, and subtracts them from the image.
This is useful when looking for faint neighbors which might
contaminate the PSF determination (in this mode, one subtracts off the
PSF stars and runs {\tt FIND}) or when undertaking additional rounds
of PSF fitting (where one subtracts off all faint neighbors and runs
{\tt PSF}).  The pattern of residuals left by {\tt SUBSTAR} is also a
critical diagnostic for determining the quality of the PSF.

\end{itemize}

\subsubsection{Star-Galaxy Separation}

As \dao\ is explicitly designed to do stellar photometry, {\tt
daophot} does not have the ability to do high confidence star--galaxy
discrimination.  The safeguards that have been built in are primarily
to discriminate against cosmic rays and instrumental artifacts, such
as bad pixels and CCD bleed from saturated pixels.  To reject
detections around these features, {\tt daophot} {\tt FIND} calculates
the following parameters per object :

\begin{itemize}

\item {\tt Sharp} : Ratio of : [the height of the best fit delta--function
that fits the data] divided by [the height of the best fit Gaussian
function that best fits the peak].  For cosmic rays, this should be
larger than one.  For bad negative going pixels, this should be close
to zero.  This statistic is primarily designed to filter against
cosmic rays and bad pixels.  The default tolerance for a good object
in \dao\ is a value between 0.2 and 1.0.

\item {\tt Round} : To calculate this value, the data are summed along each
dimension, and then fit with one--dimensional Gaussian functions along
both $x$ and $y$.  The {\tt round} parameter is the ratio : [the
difference between the heights of the Gaussians] divided by [the
average of the heights of the Gaussians].  An object elongated in the
$x$--direction will have {\tt round} $< 0$; in the $y$--direction,
{\tt round} $> 0$.  This is primarily designed to filter against
charge--overflow features.  The default tolerance for a good object in
\dao\ is a value between -1.0 and 1.0.  Note that objects elongated at
oblique angles will {\it not} be preferentially rejected, thus this
only marginally useful for star--galaxy separation.  An additional
roundness parameter is calculated that measures the four--fold
symmetry of the detection as a safeguard against diffraction spikes.

\end{itemize}

Clearly, neither of these statistics are optimal for doing
star--galaxy separation.  However, {\tt allstar} also calculates the
following parameters per object, which we ingest into our database as
{\tt PSFChiSq} and {\tt OrigClass}, respectively.

\begin{itemize}

\item {\tt Chi} : A weighted estimate of the standard deviation of the
residuals from the PSF fit.  This is derived from : [the ratio of the
observed pixel--to--pixel mean {\it absolute} deviation from the
profile fit] divided by [the value expected on the basis of the noise
properties].  The denominator is derived from the input gain and
readnoise, Poisson statistics, some fraction of the total measured
flux (input parameter {\tt PERCENT ERROR}, default $0.75\%$) to allow
for flat-fielding errors, plus an user supplied (input parameter {\tt
PROFILE ERROR}, default is $5\%$) estimated error of the fourth
derivative of the PSF at the peak of the profile to account for
uncertainties in interpolation.

\item {\tt Sharp} : A parameter with the same name but different
interpretation from {\tt daophot}'s {\tt Sharp} parameter.  This {\tt
Sharp} is a goodness--of--fit statistic describing how much broader
the actual profile of the object is compared to the profile of the
PSF.  Pixels within 6 half--widths of the PSF are included in
calculation of the quantity :

\begin{equation}
\frac{\sum e^{-r^2} * \Delta * (r^2 - 1) / \sigma^2}{\sum e^{-2r^2} *
  (r^2 - 1)^2 / \sigma^2} \nonumber
\end{equation}

where $\delta$ is the residual of the brightness of each pixel from
the PSF fit and $\sigma$ is the anticipated standard error of the
intensity of the pixel.  Objects less extended than the PSF (such as
cosmic rays) have {\tt Sharp} smaller than 1.0; objects more extended
than the PSF (such as galaxies) have {\tt Sharp} larger than 1.0.
This {\tt Sharp} parameter is an estimate of the intrinsic angular
size of a given object, and should tend to the same mean value
regardless of the seeing.

\end{itemize}

We also emphasize that \dao\ operates only with PSFs.  Any galaxy it
encounters (or any saturated star that has been interpolated by
\photo\ and thus does not follow exactly the image's PSF) tends to get
split up into multiple components.  {\tt FIND} will detect a peak at
the galaxy centroid, and after subtraction of a PSF at this position,
the remainder of the object flux is modeled as multiple additional
stellar objects.  For this reason, \dao\ photometry for galaxies and
the very brightest stars is not to be trusted.  This also causes
difficulties in the {\tt OPTICS} clustering runs
(Section~\ref{sec-cluster}), since a single galaxy may have multiple
components from \dao.

We also suspect that this is one reason \dao\ finds more objects than
\photo\ : it splits up galaxies (or saturated stars) into multiple
components, which then cluster with other \dao--reduced runs or
filters, but not with \photo.

\subsubsection{Deblending}

The process of object deblending is not strictly supported in \dao,
insomuch as the object detection ({\tt daophot}) and PSF photometry
({\tt allstar}) portions of the code are decoupled.  What happens in
practice to a blended pair is that the bright component is detected in
{\tt FIND}, photometered in {\tt allstar}, subtracted from an image
using {\tt SUBSTAR}, and its blended neighbor revealed in a call to
{\tt FIND} on the star--subtracted image.  This pair of detections is
then sent along with the original science image to {\tt allstar},
which then attempts to deblend them using the PSF.  This process does
not always succeed, and {\tt allstar} is able to {\it merge} stars
into a single detection if S/N criteria are not met.  It cannot
however {\it add} a component to the fit if it feels additional
deblending is required.

It is important to note that {\it all} objects are assumed to be
stellar.  This approach will fail in the general case where there are
significant numbers of background galaxies in the field, but should
succeed in the case of very crowded stellar fields, such as globular
clusters.

{\tt allstar} checks objects for merger if they are separated by 1
FWHM of the PSF.  Objects are considered merged if they are separated
by less than 0.375 the FWHM.  For neighbors with separation between
0.375 and 1.0 times the FWHM, {\tt allstar} will merge them into a
single detection if the signal--to--noise of the object with the
largest magnitude error is smaller than a given threshold.  This value
increases from 1.0 for iteration number 5 of {\tt allstar} up to 2.0
for iteration 15 and beyond.  An object is considered to have
converged once its determined to have a S/N $> 2.0$.

The process of merging objects yields a composite centroid from the
weighted means of the most recent centroid estimates of both stars,
and a composite brightness from the sum of brightnesses of both
elements.  This object is then marked for analysis in the next
iteration of {\tt allstar}.

The program {\tt allframe} uses a similar set of criteria for
deblending.  In this case, objects are considered critically blended
if they are within 0.375 times the FWHM of the best--sampled frame in
which they both appear.

\subsubsection{How We Married \dao\ to {\tt Perl}}
\label{dao-sec_perl}

Since the \dao\ package is more of a toolkit than a pipeline, to make
it into an automated pipeline we have chosen to use the {\tt Perl}
scripting language.  These scripts were derived from the thesis work
of \citet{AB}, and were designed to perform automated crowded field
photometry on Galactic bulge and LMC images taken on the CTIO 0.9m
telescope.

In {\tt Perl}, {\tt daophot} (and {\tt allstar}) is opened as a
filehandle to which commands may be written.  This is accomplished in
the following way

\begin{verbatim}
$daopid = open (DAOPHOT, ''|daophot >> $out_file'');
\end{verbatim}

The filehandle DAOPHOT is written to using simple {\tt print}
commands, such as

\begin{verbatim}
print DAOPHOT ''$re_dao\n'';
\end{verbatim}


where the variable {\tt \$re\_dao} contains the image readnoise.
In this way, we are able to send commands to the program as if
we were typing them on the command line.

Through trial--and--error, we have determined the sequence of prompts
requested by {\tt daophot} and {\tt allstar} for a given command
sequence, as well as the diversity of variations allowed.  Our {\tt
Perl} script is designed to itself recognize each possible fork (e.g.
if a file exists, do you overwrite it?) and send {\tt daophot} the
appropriate commands.  We are thus able to replicate an interactive
session with our automated scripts.

\subsubsection{The Point Spread Function in \dao \label{sec-dao_psf}}

\dao\ is very flexible in how it handles its PSF, and we believe
this flexibility is one of the main reasons that it performed so well
in our precision tests.

The \dao\ PSF model is a combination of two components : an analytic
approximation to the true PSF; and a pixel--wise look--up table
containing the average deviations of the true PSF from the analytic
model.  There are 6 analytic models for \dao\ to use\footnote{These
descriptions are lifted verbatim from the \dao\ manual} :

\begin{itemize}

\item A Gaussian function, having two free parameters:
half--width at half--maximum in $x$ and $y$.  The Gaussian function
may be elliptical, but the axes are aligned with the $x$ and $y$
directions in the image. This restriction allows for fast computation,
since the two--dimensional integral of the bivariate Gaussian over the
area of any given pixel may be evaluated as the product of two
one--dimensional integrals.

\item A Moffat function, having three free parameters:
half--width at half--maximum in $x$ and $y$, and (effectively) a
position angle for the major axis of the ellipse.  Since it's
necessary to compute the two--dimensional integral anyway, we may as
well let the ellipse be inclined with respect to the cardinal
directions.  In case you don't know it, a Moffat function is

$$ \propto{1\over{(1 + z^2)^\beta}}$$

\item{} where $z^2$ is something like ${x^2/\alpha_x^2} +
{y^2/\alpha_y^2} + {\alpha_{xy} x y}$ (Note: {\it not\/} $\ldots +
{xy/\alpha_{xy}}$ so $\alpha_{xy}$ can be zero).  In this case, $\beta
= 1.5$.

\item A Moffat function, having the same three parameters
free, but with $\beta = 2.5$.

\item A Lorentz function, having three free parameters: ditto.

\item A ``Penny'' function: the sum of a Gaussian and a
Lorentz function, having four free parameters.  (As always) half--width
at half--maximum in $x$ and $y$; the fractional amplitude of the
Gaussian function at the peak of the stellar profile; and the position
angle of the tilted elliptical Gaussian.  The Lorentz function may be
elongated, too, but its long axis is parallel to the $x$ or $y$
direction.

\item A ``Penny'' function with five free parameters.  This
time the Lorentz function may also be tilted, in a different direction
from the Gaussian.

\end{itemize}

It is perhaps worth noting that the data are not fit to an actual
analytic profile, but instead to the function as integrated over the
area of each pixel.

The look--up table is allowed to vary spatially in a constant, linear,
or quadratic fashion.  The table has a resolution of one half pixel,
centered on the centroid of the stars.  {\it It is necessary to both
cleanly subtract off all neighbors and accurately determine the
centroids of the objects for this mechanism to work optimally}.
High order terms of the look--up table have zero volume, so that the
volume of the PSF is constant across the image.

\dao\ has the option to automatically choose which analytic model best
fits the data, using as a metric the RMS of the residuals as a
fraction of the peak height of the analytic function.  In practice, we
allow \dao\ to fit all 6 models to the ensemble of data and select the
best fit profile.  This leads to significant computational overhead,
and is one culprit for the slowness of \dao\ relative to the other
algorithms.

After \dao\ has chosen the best model, it displays the star--by--star
RMS residuals, as well as indications that it thinks a particular star
is saturated, too near to the edge of the image, or has a RMS larger
than 3 times the average.  It is this list of RMS residuals that we
need to parse in {\tt Perl}.  We use this RMS distribution to reject
stars that fit the PSF model poorly, and then re--send the list of
acceptable stars to the {\tt PSF} stage.

\subsubsection{\dao\ in Practice \label{sec-dao_use}}

In our typical runs, we start with a high--threshold {\tt FIND}
command to locate bright stars.  We run {\tt PHOT} on the objects and
{\tt SELECT} the 800 brightest and most isolated objects in the image
to use as the inputs to the initial PSF generation stage.

In this first stage, we fit a pure analytic model with no lookup
table.  The program selects the best of the 6 analytic models, and
lists the resulting RMS values star by star.  We parse this list in
{\tt Perl} and reject those candidates that have more than 2.7 times
the median RMS.  The list of good objects is sent to {\tt allstar} to
determine positions, brightnesses, and local sky values.

At this point in time, we want to start building up the complexity of
the PSF by adding a look--up table.  We would ideally subtract off the
PSF stars, and run a {\tt FIND} on the residual image to detect faint
neighbors, subtract off only these objects using {\tt SUBSTAR}, and
re--run {\tt PSF} on the now--isolated PSF stars.  Blended neighbors
have a relatively small effect on the analytic model, but can
contaminate the look--up table significantly.

However, because of the complexities of the SDSS PSF, we encountered
problems with \dao\ finding incorrect initial centroids of the stars
(meaning the PSF model was not exactly and consistently centered on
the objects).  Since the PSF model is incomplete at this stage, and we
were not yet using a look--up table, the residuals between the
analytic model and the true PSF were being detected by \dao\ as
entirely new objects in {\tt FIND}.  Thus every bright star was split
in twain : the original detection, and the residual of this detection
from the initial PSF model.  \dao\ was not inclined to merge these
detections into a single object, and we ultimately ended up with an
incomplete PSF model and multiple detections per star.

We decided that we needed to first build a more complete model of the
PSF before doing neighbor detection.  This would allow {\tt allstar}
to successfully centroid each object, to allow {\tt PSF} to build a
more accurately centered model.  Essentially, we had to build up a
better approximation of the PSF so that we could generate a more
accurate PSF downstream.  This process of bootstrapping seemed to
solve the problem, but also slowed down the processing significantly.
It also required that we start the PSF modeling process with {\it
many} objects (we chose 800) since we wanted to beat down the
systematics in the initial look--up table due to un--subtracted
neighbors.

Therefore we first increase the complexity of the PSF to include a
look--up table without spatial variation, and re--run {\tt PSF} {\it
without} neighbor subtraction at this point.  Candidates with more
than 1.8 times the median RMS are rejected.  This culled list is
re--sent to {\tt PSF}.  We iterate this procedure until the list
converges or we reach 3 iterations, whichever comes first.  In
addition, we halt the sigma--clipping process if the number of PSF
stars falls below 100.  This culled list is then re--sent to {\tt
allstar} to yield an updated list of PSF stars.

We send this new list to {\tt daophot} and again increase the
complexity of the PSF look--up table to include linear variation
across the image and repeat the above loop, rejecting objects with
more than 1.5 times the median RMS, and sending the culled list to
{\tt allstar}.

At this point, we run a {\tt FIND} on the PSF--star--subtracted image
to find blended neighbors.  This list is appended to the PSF--star
list, and the ensemble is sent to {\tt allstar} for joint photometry.
{\tt Allstar} ideally deblends neighbors and merges spurious
detections, yielding accurate centroids.  We use {\tt SUBSTAR} to
remove only the neighbors from the image.  Finally, the PSF is
generated on the neighbor--subtracted image, using quadratic spatial
variation in the look--up table, and rejecting objects with more than
1.5 times the median RMS.  This yields our final PSF model.

We next detect {\it all} sources in the image by : calling {\tt FIND}
with the final FWHM as derived from the PSF; running {\tt allstar} to
photometer and subtract the objects; running {\tt FIND} on the
star--subtracted image to detect blended or dim objects; running {\tt
allstar} on the merged star list, yielding another star--subtracted
image; and a final run of {\tt FIND} and {\tt allstar} to produce the
final PSF photometry per image.  This list is sent to {\tt PHOT} to
produce aperture photometry results for the entire list of objects.

\subsubsection{{\tt Allframe} in Practice \label{sec-alf_use}}
We decided to produce {\tt allframe} results by hand for a subset of
our data because this algorithm is the closest existing piece of
software to the envisioned LSST Image Processing Pipeline and its
aggregate analysis of all images of a given sky patch.  We used the
field of globular cluster M2 (NGC 7089) for this analysis.  \photo\
frequently fails to reduce of this field due to its extreme crowding
conditions.  Thus it presents an opportunity to explore the parameter
space opened by \dao\ and {\tt allframe}.

We ran the standard \dao\ reductions of this field, and fed the
derived star lists from all 5 passbands and both runs into {\tt
daomatch}.  We used the $g$--band image from run 3437 as the reference
astrometric frame.  We next ran {\tt daomaster}, matching up all
objects in a 1--pixel (in the reference image) radius with quadratic
transformations.  This matching radius was monotonically decreased to
0.1 pixels, yielding an initial star list of $\sim~8000$ matches.  The
derived star list and transformations were fed to {\tt allframe},
which produces star--subtracted images for each input image.  These
images were co--added using {\tt montage2}, yielding an image
containing all objects {\it not} matched in the {\tt daomaster} stage.
We next ran {\tt FIND} on this image, and then {\tt allstar} using the
point--spread function of the reference image (a reasonable
approximation since we only want initial centroids, which will be
recalculated in subsequent calls to {\tt allframe}).  This starlist
was appended to the results of {\tt daomaster} and the images were
re--fed into {\tt allframe}.  We ran an additional {\tt FIND} and {\tt
allstar} on the co--added residuals of this second {\tt allframe} run.
The final star list was derived from a third and final {\tt allframe}
run on the images.

\subsubsection{Processing Time}

We found the preceding protocols sufficient to produce good results
from \dao\ and {\tt allframe}, but it is likely that not all of it was
necessary.  The amount of over--design in the construction of the PSF
is large, and this overhead can almost certainly be reduced.  We did
not test this parameter space, instead choosing to exercise the
algorithm with very conservative (and time--consuming) settings.

We address several points that affect the run--time of \dao:

\begin{itemize}

\item {\tt PSF} fits 6 models to the ensemble of data every time it is
called (up to 10 times per image).  This yields a factor of 60 in
run--time compared to the generation of a single PSF.  This could be
sped up by choosing a single analytic model to use, one that most
closely approximates the characteristics of your data.  With the
inclusion of a look--up table in the PSF, the overall differences when
using the different analytic models should ideally be minimal
(assuming you can build a high--fidelity look--up table).  In
practice, it is the case that you want capture as much of the PSF in
the analytic portion of the model.

\item We decided to use a large number of stars (800) to initially
feed to {\tt PSF}, assuming (rightly so) that many would be rejected
in our sigma clipping iterations.  This is 1 PSF star for every
100x100 pixel patch in the 2048x4083 image, perhaps a factor of 10
larger than is needed.  The final PSF model tends to be derived from
200--300 stars.

\item The executables {\tt daophot} and {\tt allstar} are run approximately
30 times in the normal mode where we generate the PSF and detect and
photometer all objects in the image.  Each of these calls loads the
image from disk.  Some processes write temporary files to disk.  And
for each call, the output stream is captured and parsed by the
controlling {\tt Perl} scripts.  This is clearly inefficient at the
system level.  A tighter integration between the processing software
and its various components (e.g. the individual executables {\tt
daophot} and {\tt allstar}) and the controlling software (middleware)
would yield a vast improvement in system load.

\end{itemize}

Overall, our automated implementation of \dao\ is very inefficient but
produces satisfactory results.  Our pipeline would benefit greatly
from tighter integration of the application and its controlling
middleware.  However, we feel that the most improvement to be gained
is in the generation of the PSF.  Had we known {\it a priori} the
locations of PSF stars and fed them directly to the PSF generation
stage, we could have sped up the processing dramatically.  We
recommend that LSST builds and then uses on a nightly basis a master
list of PSF stars to assist in this computation.

\subsection{\dop}

The \dop\ package \citep{Schechter93} is designed to robustly produce
a catalog of stellar positions, magnitudes and relatively crude
star/galaxy classifications for detections from astronomical images.
Like \sex\, \dop\ was designed to work on a large number of images
quickly with little to no interaction with the user.  According to
\citet{Schechter93} it was in fact, optimized to handle large numbers
of poorly sampled, low S/N images.  The major caveat made by the
authors states that \dop\ may not be the optimal program (sacrificing
completeness and accuracy) for use on datasets that differ
dramatically from the data it was originally designed to work on.

The version of \dop\ tested here is not the original software
implementation as designed by \citet{Schechter93}.  The original
FORTRAN source code was translated, using {\tt f2c}, into C--language
code by I. Bond of the MOA Microlensing Collaboration.  Much of the
elegance of the original source code was lost in translation, and the
resulting code is extremely difficult to interpret.  Many of the
subsequent changes to \dop\ were done in order to be able to do
photometry in difference imaging (forced photometry, photometry on
images with zero background, etc.).  Nevertheless, it has been
extensively modified to operate robustly in the {\tt Photpipe}
environment.  We emphasize that the original software should not be
implicated for any shortcomings in the analyses presented here.

Given the uniqueness of SDSS drift--scan data and the complexity of
the PSF for these images, we set out to investigate the usefulness of
\dop\ with respect to the other algorithms described in this section
with little expectation that \dop\ would measure up.  As demonstrated
below, the numerous input parameters and complicated implementation of
the source code have made a thorough investigation of \dop's
capabilities nearly impossible in the time frame given for this study.
We caution the reader that the results we quote in the following
sections for \dop\ may not be representative of the full capabilities
of \dop.
%

\subsubsection{An Overview}

To enable \dop\ to run within the {\tt Photpipe} framework, the {\tt
C} code version we used has been wrapped in an extensive amount of
{\tt Perl}.  For our study, several additional modifications to both
the {\tt Perl} code and {\tt C} code were necessary to accommodate the
SDSS images.  In particular, we added the second moments (sigx, sigy,
sigxy), the $\chi^{2}$ (chisqr), and PSF magnitudes and errors to the
default \dop\ output parameters.


\subsubsection{Object Detection and Measurement}

\dop\ returns both aperture magnitudes (again, using the optimal
\photo\ aperture of 37.17 pixels) and PSF magnitudes and respective
uncertainties.  The PSF is based on an analytic model, consisting of
similar ellipses of the form

\begin{equation}
I(x,y) = I_{o}(1 + z^{2} + 1/2 \beta_{4}(z^{2})^{2} + 1/6 \beta_{6}(z^{2})^{3})^{-1} + I_{s}
\end{equation}

where

\begin{equation}
z^{2} = [-1/2(\frac{x^{2}}{\sigma_{x}^{2}} + 2\sigma_{xy} xy + \frac{y^{2}}{\sigma_{y}^{2}})],
\end{equation}
\begin{equation}
x = (x^{'} - x_{o}); y = (y^{'} - y_{o}).
\end{equation}

This function is not allowed to vary spatially, putting this software
at an extreme disadvantage compared to \photo\ and \dao.  This is
particularly true for SDSS data, since temporal PSF changes (and the
PSF is always changing) in drift--scanned data translate into spatial
PSF variation in the images.

\dop\ uses the initial inputs (user defined) for the seeing, background
sky and the instrument to identify objects.  After this first pass
through the data \dop\ improves its initial estimate of the shape of
the object by fitting the model of a typical star to a number of
subrasters centered on a variety of detected objects.  It does this
until it finds the optimal model (star, galaxy, double star, cosmic
ray,) for each object (as described below).  In much the same fashion
as \dao, the detected objects are subtracted from the image and
another detection pass is performed and the object classification
routine is rerun to improve the model.

\dop\ produces a noise image which weights each pixel in its
non-linear least squares fitting routine.  This is also used to
determine if the detection is sufficiently above the background or
should be rejected \citep{Kor05}.

\subsubsection{Star/Galaxy Separation}

\dop\ makes a crude attempt at separating a potential star from a
double star or galaxy by comparing the shape parameters of the object
to the given initial guesses for a ``typical'' stellar shape in the
parameters file.  If these shapes differ significantly and are larger
than the specified footprint, \dop\ attempts to fit two typical
stellar profiles to the object.  If this too fails to meet a user
specified threshold, the object is then classified as a galaxy.
Discrimination between galaxies and double stars can be adjusted with
the {\tt STARGALKNOB} parameter .

\dop\ returns one of nine different object types : 1 = star, 2 =
galaxy, 3 = double star, and 4-9 flag the object for a variety
potential issues with the object and/or image that prevent a
definitive classification.

\subsubsection{Crowded Field Photometry Comparison}

\dop\ does a relatively good job on crowded fields.  \dop\ does better
than \sex\ under most circumstances but worse than \dao.  According to
the accompanying manual, tweaking the {\tt STARGALKNOB} parameter will
allow \dop\ to do better at discriminating double stars from galaxies
at low galactic latitudes.  \citet{Fer00} discuss the effect of using
\dop\ on cosmic ray--cleaned images and crowded fields.  They report
that \dop\ has the tendency to overestimate the sky brightness
significantly when cosmic rays are present.  We used fully reduced and
cosmic ray--cleaned SDSS images for our tests and were not sensitive
to this effect.

As in all packages, around bright stars residuals from the PSF
subtraction may trigger the false detection of new objects on the
residual flux.  To compensate for this, \dop\ adds noise to the noise
image it produces every time it subtracts a new detection from the
image.  However, this reduces the efficiency with which \dop\ can
detect faint sources near bright objects.



\subsection{\sex}

The \sex\
package\footnote{http://terapix.iap.fr/soft/sextractor/index.html} is
designed to quickly produce reliable aperture photometry catalogs on a
large number of detected sources from astronomical images.  
%
%
Aside from the ease of installation, \sex\ is also notable for its
speed and versatility.  Aside from \photo, it is one of the few
packages that promises to distinguish and photometer both stars and
galaxies.


\subsubsection{An Overview of the Software}

\sex\ uses {\tt autoconf} to configure the software to the particular
system it is being installed on, making it extremely portable and
flexible.  It comes with an ensemble of runtime configuration files,
including a list of default input and output parameters, neural
network weight files for star--galaxy separation, and convolution
masks to assist in object detection.  \sex\ is but one part of a
larger data processing environment that also includes {\tt EyE}
(Enhance Your Extraction,
\footnote{http://terapix.iap.fr/soft/eye/index.html}), which allows
you to generate non--linear filters that may be used for adaptive
filtering and feature detection in \sex.

\sex\ itself uses a custom FITS interface derived from the Leiden Data
Analysis Center (LDAC) toolset, and the
WCSLIB\footnote{http://www.atnf.csiro.au/people/mcalabre/WCS/} library
to perform pixel--to--sky transformations.

\subsubsection{Object Detection}

One of the most difficult issues in photometry is the accurate
determination of the sky background.  In \sex, the background is
determined locally in each mesh of a user--specified grid that covers
the image.  Sigma clipping of pixels occurs until convergence at $\pm 3
\sigma$ about the median.  If the sky estimate has changed less than
$20\%$ from the initial estimate, the mean of this clipped histogram
is considered the sky.  Otherwise the sky is estimated as the mode as
$2.5 \times {\rm median} - 1.5 \times {\rm mean}$.  Note that this is
different than \dao's definition of mode.  These values are median
filtered to avoid the influence of individual bright stars, and the
global background model is derived from a bicubic spline fit to the
mesh value.  

The background subtracted image is convolved with a filter optimized
to detect the objects of interest in the image.  This correctly
suggests that choice of filter is essential.  For example, the optimal
filter to detect stars is the PSF flipped about the $x$ and $y$ axes.
This occurs in practice by approximating this function with a
symmetric Gaussian whose full--width at half--maximum is similar to
the PSF FWHM.  However, this filter is {\it not} optimal for galaxy
detection, since galaxies are generally broader than the PSF, and
oriented arbitrarily.  In crowded fields, this convolution process
tends to blend neighboring objects together, and without a PSF model
makes it difficult to ``segment'' or ``deblend'' neighboring objects.
To assist in this problem, \sex\ provides filters to use under varying
seeing conditions and optimized to detect Gaussian functions (stars),
extended low surface brightness objects, or wavelet features designed
for crowded field detection.  Ideally, one should develop filters with
{\tt EyE} optimized for the features one wants to detect, and apply
these filters in \sex's filtering steps.

\subsubsection{Deblending}

\sex\ groups significant neighboring sets of pixels in the filtered
image into ``segments'', allowing connectivity at the sides or
corners.  The user sets the threshold above which pixels are
considered significant with parameter {\tt DETECT\_THRESH}.  Segments
must have at least {\tt DETECT\_MINAREA} pixels above this threshold to
be considered significant.  \sex\ attempts to deblend each segment by
building a model of how the segment bifurcates into different objects
as the detection threshold is diminished.  The decision to regard a
branch as distinct is based upon its relative integrated intensity.  If
the integrated pixel intensity of the branch is greater than a certain
fraction of the composite object, it is considered distinct.  The
default parameters allow a contrast of approximately 6 magnitudes in
blended objects.

\subsubsection{Object Measurement}

After detection and deblending, \sex\ characterizes each source.  Only
pixels above the detection threshold are considered.  In general, the
user requests a subset of desired characteristics from the longer list
of parameters \sex\ is able to measure.  However, some of the isophotal
measurements are required by \sex, and are performed even if not
requested by the user.  

As an example, the isophotal 2$^{\rm nd}$ order moments are calculated
from the image as follows :
\begin{eqnarray}
<x^2> &  =  & \frac{\sum_{i \in S}^{} I_{i} \ast x_{i}^2}{\sum_{i \in S}^{} I_{i}} - <x>^2 \nonumber \\
<y^2> &  =  & \frac{\sum_{i \in S}^{} I_{i} \ast y_{i}^2}{\sum_{i \in S}^{} I_{i}} - <y>^2 \nonumber \\
<xy>  &  =  & \frac{\sum_{i \in S}^{} I_{i} \ast x_{i} \ast y_{i}}{ sum_{i \in S}^{} I_{i}} - <x> \ast <y> \nonumber
\end{eqnarray}
However, isophotal measurements are not optimal, in that they are
sensitive to the thresholding level.  In \sex\ versions later than
2.4, ``windowed'' measurements of positions and shapes are allowed.
These include a Gaussian weighting, similar to the adaptive second
moments used by \photo.  While more robust than isophotal
measurements, they are derived iteratively, and thus more
computationally expensive.

\sex\ is capable of determining magnitudes in five different ways.
Each of these parameters is discussed in detail in the users guides
available on the TERAPIX site given above.  We have distilled the
information on these and other main features of this package here for
completeness but refer the reader to the manuals for further details.

\begin{itemize}

\item {\tt MAG\_ISO}: isophotal magnitudes - \sex\ uses a user defined
threshold for detection as the lowest isophot (pixels above the
threshold minus the background).  This uses the {\tt DETECT\_THRESH}
parameter in the setup file.

\item {\tt MAG\_ISOCOR}: corrected isophotal magnitudes - retrieves the
amount of flux in the wings of the isophotal (Gaussian) area.

\item {\tt MAG\_AUTO}: automatic aperture magnitudes - from Kron-like
elliptical apertures.

\item {\tt MAG\_BEST}: Choice between {\tt ISOCOR} and {\tt AUTO} -
typically AUTO unless nearest neighbors influences photometry by more 10\%.

\item {\tt MAG\_APER}: fixed-aperture magnitudes - user defined
circular apertures.

\item {\tt MAG\_PETRO}: petrosian aperture - similar to {\tt AUTO}'s
Kron-like aperture (as of version 2.4.4) with different radius but
similar position angle and ellipticity.

\end{itemize}

\subsubsection{Star-galaxy Classification}

\sex\ uses a neural--network--based star/galaxy classifier which
allows it to do a primitive classification of objects (returned as
{\tt CLASS\_STAR}).  This classifier may be augmented by using the
{\tt EyE} package\footnote{``Enhance Your Extraction'',
http://terapix.iap.fr/soft/eye} to design more complex classifiers.

The object classification in \sex\ is designed to detect and classify
both galaxies and stars using a neural network output.  \sex\ begins
its object classification with the pixel scale of the input image and
a user supplied estimate of the seeing FWHM.  The neural network uses
these values to make an initial rough guess about object shape and
size on the image.  The final classification for an object is
designated by the {\tt CLASS\_STAR} parameter and has a fractional
value between 0 and 1.  \sex\ considers a zero to be a galaxy and a
one to be a star.
In Section~\ref{sec_ana-stargal} we show exactly how easily the values
between 0 and 1 can be reliably interpreted as either a galaxy or a
star using \photo's galaxy/star classifications as ``truth'' for each
object and comparing the results.


Parameters for the detection and analysis thresholds ({\tt
DETECT\_THRESH, ANALYSIS\_THRESH}) and deblending ({\tt
DEBLEND\_MINCONT, DEBLEND\_NTHRESH}) can be set to improve the the
detection rate and quality.  Note however that much like \dao, if
given too fine a deblending \sex\ may deblend large galaxies into
several individual objects.

{\tt CLASS\_STAR} behaves as a sharply--tuned Bayesian classifier.
Results can become unreliable when the actual PSF shape is different
from what it was trained with (Moffat--like), or when the
user--provided {\tt SEEING\_FWHM} is inaccurate.  Asymmetric PSFs and
strong variations in the PSF across the field are additional factors
that limit the accuracy of the classifier.  These effects are
frequently seen in large--area CCD mosaics.  Because of these
shortcomings, using {\tt CLASS\_STAR} for star/galaxy separation is
generally not recommended in large surveys.  A preferred method is to
use {\tt FLUX\_RADIUS} (the radius of the disk which contains half of
the flux) as well as its variation across the image.



\subsubsection{Using a PSF Model}

Because of \sex's robust deblender, it does a reasonable job at
performing photometry in crowded fields.  The software will process
the images to completion, although the output catalog should be
closely inspected to verify the level of deblending was appropriate.
It is more robust than \photo\ in this regard, as \photo\ is known to
fail at the deblending stages in the most crowded of fields.  However,
the photometric accuracy of \sex\ in crowded fields, and for faint
sources, has generally been limited by the lack of a PSF model.

Contrary to most literature sources, \sex\ \emph{can} perform PSF
photometry and position measurements \citep[see][for
examples]{cfht1,cfht2,Bertin04}.  The {\tt PSFEx}\footnote{While {\tt
PSFEx} has not been officially released, the software may be
downloaded from the TERAPIX public repository at
http://terapix.iap.fr/wsvn/index/public/software/psfex/} package
provides this functionality.  This is accomplished in three steps: (a)
make an initial pass through \sex, and create a binary catalog
containing small images around each bright source; (b) pass this
catalog through {\tt PSFEx} to create a model describing the PSF and
its variations; (c) rerun \sex\ requesting parameters such as {\tt
MAG\_PSF}, {\tt MAGERR\_PSF}, etc.  At this stage, there are still
completeness issues in very crowded fields, which has prevented the
public release of the {\tt PSFEx} package.



\subsubsection{\sex\ In Practice}

Unlike \dop\ and \dao, \sex\ is relatively straightforward to use
within the framework of the {\tt Photpipe} pipeline, requiring little
initial setup and no modifications to the source code.

The parameters we used in our test runs with \sex\ from the setup file
(default.sex) and the requested output catalog parameters
(default\_sex.params) can be found in the Appendix.  In particular,
the parameter {\tt NUMBER} is a running number use for cross
identification and not recorded in the database.  {\tt X\_IMAGE}, {\tt
X2\_IMAGE}, {\tt Y\_IMAGE}, {\tt Y2\_IMAGE}, and {\tt XY\_IMAGE} have
been depreciated in the new version in favor of the new
Gaussian--windowed measurements.  As is demonstrated in the
photometric analysis, the windowed measures are vastly superior to the
old parameters, which were essentially isophotal quantities.  For
completeness, we requested the {\tt MAG\_APER} and {\tt MAGERR\_APER}
values in a 37.171 pixel aperture (7.36 arcsec at 0.396 arcsec/pixel),
which is the aperture we chose to use for \photo's aperture
photometry.

\subsubsection{Crowded Field Photometry Comparison}
\label{subsec-sex-crowd}

How well does \sex\ perform in crowded fields?  Relatively well if
deblend and threshold parameters are set at reasonable values for your
images.  
The unavoidable end result is that \sex's neural network breaks down
at the low magnitude end, especially when it comes to detecting faint
galaxies in crowded fields.  \citet[][and references therein]{Hol05}
suggest two novel approaches to detecting these faint galaxies using
\sex.

The first involves the use of \dao\ to first subtract all objects
\dao\ detects as stars in the crowded field and save the subtracted
image.  \dao\ is essentially optimized for such a task.  Without the
influence of the additional stars in the image, \sex\ does a better
job at finding faint galaxies, although we do not explore this claim
in our report.  
The second involves the use of two (or more) color images.
\citet{Gonz98} use $B-I$ images to detect sources instead of using the
single color images.  The major disadvantage of this is the increase
in noise associated with the image, which will in turn produce more
spurious \sex\ detections.



\subsection{Previous Tests Involving \dao, \dop, and \sex
\label{sec-previous_comparisons}}

There are a few noteworthy studies in the literature that investigate
the usefulness of the algorithms in this study. Most of the algorithm
comparisons found in the literature and on the web
\citep[e.g.][]{Alard00, Fer00, Neill05, Staude04, Kor05, Smolcic06}
use \sex\ version 2.3.2, \dop\ version 2.0, and/or \dao\ version II in
their analysis.  The latest versions used in this analysis of \sex\
(version 2.4.4), \dop\ (version 3.0) and \dao\ (version IV) include
significant upgrades and enhancements over their older, well used, and
well studied predecessors.  

\subsubsection{\citet{Smolcic06} : Assessment of \dop\ for Crowded
Field Photometry}

This study implements a new pipeline designed around a version of
\dop\ v2.0 that was wrapped in C and compiled under f2c by
E. Magnier.  They use this pipeline on crowded fields where \photo\
gives poor results.  Instead of determining the repeatability of their
photometric measurements or comparing their photometry to another
algorithm as we have done with \photo, the authors use \dop's PSF
model to generate synthetic stars and place them on an image through
Monte Carlo simulations.  They created both sparse and crowded fields
and quantify their completeness at different magnitudes as the ratio
of the number of artificial stars extracted by \dop\ to the number of
artificial stars on the frame, $n_{output}/n_{input}$.

Their completeness for sparse fields is comparable to that of \photo\
at the bright end ($\sim$95\%--99\%) and falls below 90\% at
magnitudes fainter that 20--21 (filter dependent).  \photo\ is quoted
as having 95\% completeness for magnitudes between 21.3--22.2 (for
$g,r,i$).  For magnitudes brighter than 21 ($g,r,i$) our recovery of
stars as compared to \photo\ is 83\%($i$)--93\%($g$) for Run 3437 and
87\%($i$)--96\%($g$) for Run 4207 (refer to Table~\ref{tab-sg1}).

For crowded fields \cite{Smolcic06} find that in regions of high
stellar density (center of Leo I) there is no appreciable effect on
the number of synthetic stars recovered to a magnitude limit of $\sim
20$.  At fainter magnitudes and stellar densities of $\sim 200$
stars/arcmin$^{2}$ their completeness suffers a 10\%--30\% decrease in
the number of stars recovered by \dop.


The success of the \citet{Smolcic06} \dop\ pipeline in crowded fields
is likely due to their attention to the background sky model.  We used
the simple uniform gradient model which is supposed to give a
reasonable description of the background sky.  The \citet{Smolcic06}
pipeline uses the modified Hubble profile model and estimates the
seeing and background sky directly from each image. They claim this
gives them a better detection rate in crowded fields by a factor of
$\sim 3$.

\citet{Smolcic06} were most concerned with detecting sources in the
crowded field SDSS images of the dwarf spheroidal galaxy Leo I and
apparently were less concerned with detecting faint galaxy sources and
the accuracy of their astrometry as they do not discuss any analysis
or fine-tuning of their pipeline to accommodate these techniques.


\subsubsection{\citet{Fer00} : Comparison of \dop\ and \dao/\all\ 
on Crowded Stellar Fields}

This study tests both \dop\ and \all\ using artificial star
simulations with a variety of complex backgrounds and stellar
densities for crowded fields observed with HST/WFPC2.  Their goal was
to determine the distances to Cepheid variables and investigate the
effect, if any, these two packages had on the distance determinations.
The authors find that when using \dop\ it is crucial the frames have
cosmic rays removed, otherwise \dop\ tends to overestimate the sky
brightness.  \all\ photometers all frames simultaneously which allows
it to easily flag and ignore cosmic rays.  Our frames were cleaned of
cosmic rays prior to using \dop, and therefore not significantly
affected by this bias.

\dop\ photometry on their artificial frames was found to be more
complete that \all.  \dop\ and \all\ agree to within 0.05 magnitudes
(within uncertainties for aperture corrections).  In crowded field
regions, confusion noise and rapidly varying background contribution
resulted in stars being measured consistently too bright $\sim 25$\%
for \dop\ and $\sim5-10$\% for \all. This effects the photometry for
single-epoch observations significantly.  For \dop\ the effect can be
as little as 0.05 magnitudes in moderately crowded fields and as large
as 0.2 magnitudes for the most crowded of their observed fields.
Surprisingly, \citet{Fer00} find that this bias is worse when \all\
photometry is used.

Their overall conclusion was that both packages are equally suited to
determining the distances to Cepheid variables with \all\
underestimating the distances by 1\% and only slightly larger for \dop\
(2\%). 


\subsubsection{Other \dop\ Studies}

\citet{Bel04} use a version of \dop\ modified by P. Montegriffo
(Bologna Observatory) to read images in double precision format.  Like
\citet{Smolcic06}, they use images seeded with synthetic stars to
confirm that their photometric uncertainties are small and that
blended sources do not impact their analysis in any significant way.
They report a completeness of over 80\% over the range in magnitudes
for their sample.

A similar analysis is performed by \citet{RM91} using \dop\ \citep[see
also][]{Vogt95,Gal99}.  They also perform a limited $i$--band
comparison between \dop\ and \dao\ where they find that \dop\ does a
better job at estimating the sky background in the crowded field
images.  \dop\ systematically finds faint stars to be brighter in
magnitude than \dao, and attributes this to \dop\ determining the sky
background from the fully subtracted frame, whereas \dao\ computes the
background before star subtraction resulting in a difference of less
than 1\% in the computed sky backgrounds (\dop's is lower).

%

\end{document}